\documentclass{aa}

\usepackage[]{natbib}
\bibpunct{(}{)}{;}{a}{}{,} 
\usepackage{graphics}   
\usepackage{graphicx}   
\usepackage{epstopdf}
\usepackage{longtable}   
\usepackage{url}      
\usepackage{bm}        
\usepackage[varg]{txfonts}
\usepackage{pdflscape}
\usepackage{supertabular}
\usepackage{hyperref}   
\usepackage{wasysym}
\usepackage[svgnames]{xcolor}
\usepackage{amsmath}
\usepackage{amssymb}  
\usepackage{nccmath}

\hypersetup{
    unicode=false,          
    colorlinks=true,       
    linkcolor=blue,          
    citecolor=blue,        
    filecolor=blue,      
    urlcolor=blue           
}
\usepackage{color}

\usepackage[utf8]{inputenc}
\usepackage{lmodern}
\usepackage{tocloft}
\usepackage{enumitem}
\usepackage{enumerate}
\usepackage{graphicx}
\usepackage{appendix}
\usepackage{multirow}
\usepackage{array}
\usepackage{makecell}
\usepackage{cancel}
\usepackage{fancyhdr}
\usepackage{tikz}
\usetikzlibrary{decorations.markings,calc,patterns,decorations.pathmorphing}

\usepackage{float}
\usepackage{indentfirst}

\usetikzlibrary{shapes}
\usepackage{afterpage}

\usepackage{soul}

\newcounter{maacounter}
\newenvironment{maaenvironment}[1][]{\refstepcounter{maacounter} \nobreakspace 
   {\color{Sienna}{MAA(\themaacounter):~{#1}}} \rmfamily}{}

\newcommand{\modl}[1]{\textcolor{black}{\texttt{#1}}} 
\newcommand{\secref}[1]{Sect.~\ref{#1}}
\newcommand{\figureref}[1]{Fig.~\ref{#1}}

\defcitealias{Wheeler_2015}{WKC15}

\begin{document}

\title{The magneto-rotational instability in massive stars}

\author{Adam Griffiths\inst{1,2}, 
Patrick Eggenberger\inst{2},
Georges Meynet\inst{2},
Facundo Moyano\inst{2},
Miguel-Á. Aloy\inst{1,3}
}

 \authorrunning{Griffiths et al.}

 \institute{Departament d'Astronomia i Astrofísica, Universitat de València, 46100 Burjassot, Spain
 \\
 \email{adam.griffiths@uv.es}; \email{miguel.a.aloy@uv.es}
 \and
 Geneva Observatory, University of Geneva, Chemin Pegasi 51, CH-1290 Sauverny, Switzerland
\and
Observatori Astronòmic, Universitat de València, 46980 Paterna, Spain}

\date{Received /
Accepted}
\abstract
{The magneto-rotational instability (MRI) has been proposed as a mechanism to transport angular momentum (AM) and chemical elements in theoretical stellar models.} 
{Using as a prototype a massive star of $15\,M_\odot$  with solar metallicity, we explore the effects of the MRI on 
the evolution of massive stars.}
{We used the Geneva Stellar Evolution Code to simulate the evolution of various models, up to the end of oxygen burning, including the MRI through effective, one-dimensional, diffusion coefficients. We consider different trigger conditions (depending on the weighting of chemical gradients through an arbitrary but commonly used factor) and different treatments of meridional circulation as either advective or diffusive. We also compare the MRI with the Tayler-Spruit (TS) dynamo in models that included both instabilities.}
{The MRI triggers throughout stellar evolution. Its activation is highly sensitive to the treatment of meridional circulation and the existence of chemical gradients. The MRI is very efficient at transporting both matter and AM, leading to noticeable differences in rotation rates and chemical structure, which may be observable in young main sequence stars. While the TS dynamo is the dominant 
mechanism for transferring AM, the MRI remains relevant in models where both instabilities are included. Extrapolation of our results suggests that models that include the MRI tend to develop more compact cores, which likely produce failed explosions and black holes, than models where only the TS dynamo is included (where explosions and neutron stars may be more frequent).}
{The MRI can be an important factor in massive star evolution but is very sensitive to the implementation of other processes in the model. The transport of AM and chemical elements due to the MRI
alters the rotation rates and the chemical makeup of the star from the core to the surface and may change the explodability properties of massive stars.}

\keywords{instabilities - stars: rotation - stars: magnetic field - stars: abundances- stars : evolution }

\maketitle

\titlerunning{MRI in stars}
\authorrunning{Griffiths et al.}

\noindent
\section{Introduction}

The effects of rotation in stellar evolution have been studied for a long time and by many authors \citep[e.g.][]{Zeipel1924, Eddington1933, Sweet1950,Mestel1966, Goldreich1967, 1969Fricke, tassoul1978, 1981Endal,Maeder_2004A&A...422..225, Maeder2009}. However, only in the last two decades have extensive grids of rotating models for both single stars and binaries appeared \citep[e.g.][]{Heger_2000, Brott2011, Ekstrom2012, Choi2017,Limongi2018}. These models explore the impact of axial rotation on different aspects, such as nucleosynthesis \citep[e.g.][]{Choplin2018, Prantzos2020}, re-ionisation \citep[e.g.][]{Yoon2012, Murphy2021}, the nature of supernova progenitors \citep[e.g.][]{Georgy2012,Groh2013}, the origin of long soft gamma ray bursts \citep[e.g.][]{Levan2016, MM2017}, and the nature and properties of stellar remnants, especially those in close binary systems that are at the origin of gravitational wave detections \citep[e.g.][]{Qin2019, Bavera2020, Bel2020}.
These works find that rotation plays an important role, underlining the need to better understand its effect on the evolution of stars\footnote{ The term `rotating models' actually hides many types of models whose results, for a given initial mass, metallicity, and rotation, can differ very substantially. A given physical process can be numerically implemented in such a different way in different models that the outputs are significantly different.}.

Rotating models have not yet reached a state of robustness and reliability that would allow them to be used universally to obtain consistent predictions. Among the challenges these models face we highlight two: first, the production of internal rotation rates that are compatible with what is deduced from asteroseismology of low mass stars \citep[e.g.][]{Egg2019a, Deheuvels2020}; and second, explaining the low rotation rates of white dwarfs and pulsars \citep[][]{Suijs2008, Heger2005, Hart2019}. \citet{Fuller2019c} discuss the impact of an efficient angular momentum (AM) transport mechanism on the rotation rates of black holes (BHs) and find that most BHs should be born with very low spin. However, the formalism proposed cannot simultaneously reproduce the constraints on the internal rotation rates of subgiants and red giants as deduced from asteroseismology \citep{Egg2019b}. In brief, these constraints show that present-day stellar models lack additional physics to describe the transport of the AM. This justifies new studies that hopefully will unveil the nature of the missing processes. Magnetic fields seem to be part of these missing processes, as their inclusion in rotating models predicts different rotation profiles than models that discount magnetic fields \citep[][]{2019A&A...626L...1EggenbergerSun, 2014ApJ...793..123Maederbraking}.

The magneto-rotational instability (MRI) was first considered in the context of accretion disks \citep{1991BandH}. However, it also plays a role in stellar objects 
\citep[e.g.][]{Balbus_1995ApJ...453..380, Menou_2004ApJ...607..564,  Kagan_2014ApJ...787...21,Jouve_2015A&A...575A.106} and, therefore, will most likely have an impact on stellar evolution, as well as on the post-collapse evolution of massive stellar cores  \citep[e.g.][]{Akiyama_2003ApJ...584..954, Masada_2006ApJ...641..447, Masada_2012ApJ...759..110, Rembiasz_et_al__2016__mnras__Onthemaximummagneticfieldamplificationbythemagnetorotationalinstabilityincore-collapsesupernovae, Rembiasz_et_al__2016__mnras__Terminationofthemagnetorotationalinstabilityviaparasiticinstabilitiesincore-collapsesupernovae, Reboul-Salze_2021A&A...645A.109}.
The MRI has been mostly ignored in terms of stellar evolution, under the assumption that other magnetic instabilities may trigger more easily under the conditions of thermal and chemical stratification  common in stellar interiors. 
\cite{Wheeler_2015} \citepalias[][hereafter]{Wheeler_2015}, however, studied the MRI in massive stars ranging from $7M_{\odot}$ to $20M_{\odot}$ using the Modules for Experiments in Stellar Astrophysics code \citep[MESA]{2011ApJS..192....3Paxton, 2013ApJS..208....4Paxton,2019zndo...3473377Paxton}. They found the MRI to be active during stellar evolution and showed that it impacted the course of a star's lifetime with respect to both transport of AM and transport of chemical elements. Nonetheless, they conclude that instabilities such as the Tayler-Spruit \citep[TS;][]{Spruit2002} dynamo are likely the dominant factors impacting the fate of a star during its evolution.  

In this work we aim to assess the impact of the MRI on the evolution of massive stars using the Geneva Stellar Evolution Code \citep[GENEC;][]{2008Ap&SS.316...43Eggenberger}. This code takes a very different approach to the modelling of the effects of rotation than the one used by 
\citetalias{Wheeler_2015}. We aim to benchmark our results with those of \citetalias{Wheeler_2015}, rather than to fit any specific observation. Although we give more details in \secref{sec:physics}, here we briefly mention two main differences. 

First, the GENEC code accounts for the effects of meridional currents and shear instabilities along the theory proposed by \citet{JPZahn1992} \citep{2008Ap&SS.316...43Eggenberger, Ekstrom2012}. The AM transport due to meridional currents is treated through the resolution of an advective equation.
In contrast, \citetalias{Wheeler_2015} employ a diffusive equation for this process. It should be noted that transport of AM by meridional currents is an advective process and, in principle, cannot be optimally modelled by a diffusion equation. Here we show that the advective treatment yields a large impact. We anticipate that an advective treatment of meridional currents may build up a velocity shear off which the MRI can feed. The previous treatment of \citetalias{Wheeler_2015} excludes this possibility.

Second, the gradients of chemical composition are extremely important quantities in rotating models. They represent barriers for any mixing in the radial direction and, thus, if strong enough, may completely suppress any mixing \citep[see e.g. the discussion in][]{MM1997}. In \citetalias{Wheeler_2015} the authors artificially reduce the barriers of the chemical gradients by a factor of $f_\mu^{-1}=20$ (see Sec.\,\ref{sec:MRI_description}).
    This factor appears in the instability criterion used to decide whether the MRI instability develops, and this blurs the picture of whether or not the MRI has an impact.
     Here we assess how not degrading the chemical gradients in the triggering condition affects the results.

This paper is organised as follows.
In Section \S\ref{sec:physics} we present the physics that we apply in our models for the transport of AM and for magnetic instabilities. Section \S\ref{sec:models} gives a full description of the set of models that we present here, listing their differences and how we expect these differences to impact the MRI. In Sections \S\ref{sec:main_sec} and \S\ref{sec:late_stages} we discuss the results of our models, focusing first on the main sequence (MS) and second on the later stages of stellar evolution. In Section \S\ref{sec:extrapolation} we extrapolate our results to shed light on what may happen to our models at collapse and explore their explodability properties. We give our concluding remarks in Section \S\ref{sec:conclusions}.

\section{Physics of the models}
\label{sec:physics}

The models presented in this paper use the same implementation of mass loss, overshooting, opacities and nuclear reactions as in \citet{Ekstrom2012}. We describe in this section only the physics of the magnetic and hydrodynamic instabilities.

 We denote the angular velocity, constant on isobars\footnote{The angular velocity as well as all physical quantities are averaged over isobars \citep{Zahn1992}.}
under the assumption of shellular rotation, by $\Omega$, and $q=\frac{\partial\ln \Omega}{\partial \ln r}$ refers to the radial shear. The thermal and composition components of the Brunt-Väisälä frequency, $N_T$ and $N_{\mu}$ respectively, are given by

\begin{equation}
\centering
    N_{T}^2=\frac{g\delta}{H_p}(\nabla_{\rm ad}-\nabla_{\rm rad}),
    \label{eqn:NT2}
\end{equation}
and
\begin{equation}
\label{eqn:Nmu}
    N_{\mu}^2=g\phi\left|\frac{\partial \ln \mu}{\partial r}\right|,
\end{equation}
with $\nabla_{\rm ad}$ and $\nabla_{\rm rad}$ the adiabatic and radiative gradients with the usual definitions; $\nabla_{\rm ad}=(\partial \ln T / \partial \ln P)_{S}$ and $\nabla_{\rm rad}=3\kappa \tilde{l} P/(64\pi\sigma G M_r T^4)$, where $S$ denotes the entropy, $T$ the temperature, $P$ the pressure, and $\tilde{l}=4\pi r^2 F$ where $F$ is the diffusive flux of radiative energy. The pressure scale height is $H_p=P/(\rho g)$, with $g$ the local gravity and $\mu$ the mean molecular weight. As usual, we set $\delta=-(\partial \ln \rho/\partial \ln T)_{P,\mu}$  and $\phi= (\partial \ln \rho)/(\partial \ln \mu)_{P,T}$ with $\rho$ the local density.

In all of our models, we take the magnetic resistivity as \citep{Spitzer2006}

\begin{equation}
\label{eqn:resistivity}
        \eta \approx 5.2 \times 10^{11}\frac{\ln \Lambda}{T^{3/2}}\ {\rm cm}^2 \ {\rm s}^{-1},
\end{equation}
where $\ln\Lambda$ is the Coulomb logarithm,
\begin{equation}
    \ln \Lambda \approx \left\{ \begin{array}{ll } 
        -17.4 + 1.5\ln T - 0.5\ln \rho & T< 1.1\times 10^5 {\rm K},  \\
         -12.7 + \ln T - 0.5\ln \rho & T> 1.1\times 10^5 {\rm K}. 
    \end{array}\right.
\end{equation}

Finally, for the thermal diffusivity we use
\begin{equation}
\label{eqn:kappa}
\kappa=\frac{16\sigma T^3}{3\kappa_R\rho^2 c_P},
\end{equation}
where $c_P$ is the specific heat at constant pressure, $\kappa_R$ is the Rosseland mean radiative opacity, and $\sigma$ the Stefan-Boltzmann constant.
The values of $\eta$ and $\kappa$ vary significantly during evolution and also within the star. In the MS $\eta$ ranges from $10\, \text{cm}^2\, \text{s}^{-1}$ in the core to  $10^3\,\text{cm}^2\, \text{s}^{-1}$ at the surface, while $\kappa$ goes from $10^9\, \text{cm}^2\, \text{s}^{-1}$ in the core to  $10^{13}\, \text{cm}^2\, \text{s}^{-1}$ at the surface. By the time the end of oxygen burning is reached the ranges of these values are quite different ($10^{-3}\, \text{cm}^2\, \text{s}^{-1} < \eta< 10^6\, \text{cm}^2\, \text{s}^{-1}$, $10^4\, \text{cm}^2\, \text{s}^{-1} < \kappa < 10^{18}\,  \text{cm}^2\, \text{s}^{-1}$). We note that the ratio $\eta/\kappa\ll 1$ and it stays roughly in the same range of $10^{-7} - 10^{-10}$ throughout the star's life. 

In this work, we suppose that magnetic instabilities operate only in radiative zones. We assume that the transport of AM and chemical species is sufficiently efficient in convective regions to impose solid body rotation and a homogeneous composition. Hence, instabilities feeding off of the differential rotation of the star (the case of MRI or the TS dynamo) are surmised to not operate in convective stellar layers.
\subsection{Description of MRI}
\label{sec:MRI_description}

We apply the same description for the MRI as in \citetalias{Wheeler_2015} and references within.
The MRI is active once the local shear is above the trigger condition,

\begin{equation}
\label{eqn:Instab}
    -q > q_{\rm min,MRI}:=\frac{\Big(\frac{\eta}{\kappa}N^2_T+f_{\mu}N^2_{\mu}\Big)}{2 \Omega^2},
\end{equation}
with $q_{\rm min,MRI}$ the minimum shear required to trigger the instability.
 The instability can only be activated in regions where the angular velocity decreases outwards (so $-q$ is positive)
and also overcomes the  stabilising action of the chemical and temperature gradients. In radiative regions,  $N_T^2>0$ and thus acts as a stabilising term\footnote{In convective regions $N_T^2<0$ and would promote the MRI however we exclude these regions from the stability criterion.} in Eq.~\eqref{eqn:Instab}. The mean molecular weight is almost always decreasing monotonically within the star, so $N_{\mu}^2$  is nearly always stabilising with respect to the MRI. During the MS the chemical gradient is very strong at the 
boundary between the hydrogen dominated envelope and the convective core thus inhibits the MRI at the core--envelope boundary. This is replicated at later phases where different chemical elements are burnt (H, He, C, etc.), causing strong chemical gradients at the borders of the burning layers.

In the instability condition we use the reduced thermal Brunt-Väisälä frequency term, $\Big(\frac{\eta}{\kappa}\Big)N^2_T$, which accounts for diffusive effects that diminish the thermal buoyancy term as is done in  \cite{1999Spruit} in his discussion of  magnetic instabilities induced by differential rotation.
Thermal diffusivity helps a displaced fluid element reach thermal equilibrium faster, therefore reducing the stabilising influence of thermal buoyancy. The instability condition Eq.~\eqref{eqn:Instab} also features the parameter $f_{\mu}$, which weights the chemical gradients. 

A small value of $f_{\mu}$ lessens the strength of the chemical gradients, thus allowing the MRI to trigger more easily within the star. \citetalias{Wheeler_2015} set this parameter to $f_{\mu}=0.05$ following  \citet{Heger_2000}, where $f_{\mu}$ was introduced to calibrate the transport of chemical elements not based on a physical mechanism. In GENEC, a very different implementation of the hydrodynamical instabilities due to rotation is in place compared to MESA. Hence, in the absence of any physical justification to reduce $f_\mu$, we employ $f_\mu=1$ by default. Nevertheless, we compute certain models with $f_{\mu}=0.05$ to compare with \citetalias{Wheeler_2015}.

When the shear inside the star verifies Eq.~\eqref{eqn:Instab} the MRI is active and an
effective viscosity, $\nu_{\rm mag,MRI}$, begins to act \citepalias{Wheeler_2015},
\begin{equation}
     \label{eqn:Viscosity-MRI-2}
    \nu_{\rm mag,MRI}=\alpha|q|\Omega r^2,
\end{equation}
where the definition of the stress efficiency parameter, $\alpha$, 
is
\begin{equation}
    \label{eqn:alpha}
    \alpha := \frac{\langle B_r B_\phi \rangle}{4\pi P_0}
,\end{equation}
where $P_0=\rho v_A^2 \sim \rho(q\Omega r)^2$ ($v_A=B/\sqrt{4\pi\rho}$ is the Alfvén speed) is the maximum magnetic pressure produced by the MRI, and $\langle B_r B_\phi \rangle/(4\pi)$ the space and time average of the azimuthal Maxwell stress once the MRI-generated, small-scale magnetic field components $B_r$ (in the poloidal plane) and $B_\phi$ (toroidal component) reach saturation (see below). The length-scale of the MRI-generated fields approximately corresponds to the wavelength of the fastest growing MRI mode, which can be roughly estimated as in \citetalias{Wheeler_2015}, 
$\lambda_{\rm MRI, fgm} \sim v_A/\Omega \sim B\rho^{-1/2}\Omega^{-1}$. More quantitatively  \citep{Obergaulinger_2009A&A...498..241},
\begin{align}
\lambda_{\rm MRI, fgm} \sim 0.7\,\text{km}&\left(\frac{B}{100\,\text{G}}\right)\times\nonumber
\left(\frac{\rho}{10^5\,\text{gr\,cm}^{-3}}\right)^{-1/2}\\& \left(\frac{\Omega}{3\times10^{-5}\,\text{s}^{-1}}\right)^{-1},
\end{align}
where $B$ is the initial magnetic field strength.

In order to estimate suitable values of $\alpha$, we resort to results obtained in local and global simulations of the MRI. Most of the existing literature considers conditions characteristic of accretion disks, which cannot be directly extrapolated to stellar environments. In such cases, the ratio $\alpha_\mathrm{g}:=\langle B_r B_\phi\rangle/(4\pi P_{\rm gas})$ (note the different pressure used for normalisation of the time and volume averaged Maxwell stress; here the gas pressure ($P_{\rm gas}$))  is, for example, $\alpha_\mathrm{g} \approx 0.015- 0.016$ in \cite{Hirose_2009ApJ...691...16} and \cite{Shi_2010ApJ...708.1716}, $\approx 0.01-0.05$ in \cite{Hawley2011}, or
$0.0014-0.018$ in \cite{Shi_2016MNRAS.456.2273}. Using conditions more (geometrically) similar to the ones we may meet in stellar interiors, the MRI saturation has been studied in the context of nascent proto-neutron stars (PNSs), that is, at the end of stellar evolution. For instance, \cite{Masada_2012ApJ...759..110} find
$\alpha_\mathrm{g} \approx 0.002 - 0.01$ depending on the shear rate $q$ (they show a monotonic increase of this ratio for $0.2<|q|<1.8$), while the semi-global models of \cite{Rembiasz_et_al__2016__mnras__Terminationofthemagnetorotationalinstabilityviaparasiticinstabilitiesincore-collapsesupernovae} provide $\alpha_\mathrm{g} \approx 0.004$ for $|q|=1.25$.
 We finally note that $\alpha=(P_\mathrm{gas}/P_0)\alpha_\mathrm{g}$, and hence the actual value of the $\alpha$ parameter may differ from $\alpha_\mathrm{g}$ unless $P_\mathrm{gas}\simeq P_0$. As argued in \cite{Braithwaite_2006A&A...449..451}, the source of the magnetic field growth is the shear energy, that is to say, the difference in kinetic energy between a uniformly rotating star with the same AM and the actual (differentially rotating) star, which is often (much) smaller than the thermal energy. While in the models of, for example \cite{Rembiasz_et_al__2016__mnras__Terminationofthemagnetorotationalinstabilityviaparasiticinstabilitiesincore-collapsesupernovae} $P_0\lesssim P_\mathrm{gas}$, 
the setup of \cite{Masada_2012ApJ...759..110} implies that $P_0< P_\mathrm{gas}$. 
 This means that the value of $\alpha$ could be sensitively larger than $\alpha_\mathrm{g}$.
We note that in local or semi-global simulations the boundary conditions, geometry, stratification, local shear, and so on may influence the value of $\alpha_\mathrm{g}$ (and, thus, of $\alpha$); in order to ease the comparison with \citetalias{Wheeler_2015}, we take $\alpha=0.02$ as a reference value. This value may still be larger than the typical values of $\alpha$ at the end of the stellar evolution found in local simulations. Nevertheless, given all the uncertainties in the estimation of $\alpha$, the fact that the MRI trigger condition is independent of $\alpha$, and the parametric use we make of it in our models, it seems adequate for our purposes. Besides, we further explore other values of $\alpha$ modifying the reference value within a factor of 5.

The physical nature of the MRI means that the effective viscosity impacts transport of both AM and of chemical elements within the star\footnote{The MRI behaves in this respect similarly as a pure hydrodynamical shear instability.}. GENEC implements these two effects through two diffusion coefficients. One accounts for AM transport, $D_{\mathrm{mag,O}}$, and the other for chemical element transport, $D_{\mathrm{mag,X}}$. Once the MRI is active (i.e. in mass shells where the criterion of Eq.~\eqref{eqn:Instab} holds) these diffusion coefficients become equal to the effective viscosity related to the MRI,
\begin{equation}
    \label{eqn:DmagO-MRI}
    D_{\mathrm{mag,O}}=D_\mathrm{mag,X}=\nu_\mathrm{mag,MRI}
    , \quad (-q>q_\mathrm{min,MRI}).
\end{equation}
\noindent
After its activation, the MRI exponentially rapidly amplifies the azimuthal magnetic field up to its saturation value.
The growth rate of the fastest growing MRI is, approximately, $\gamma_{\textrm{MRI}} \approx  |q|\Omega$ \citepalias{Wheeler_2015}\footnote{Compare with e.g. \cite{Masada_2006ApJ...641..447}, who find $\gamma_{\textrm{MRI}} =  |q|\Omega/2$.}, which for typical conditions over the MS is $\sim  10^{-5}-10^{-4}\,$s$^{-1}$, and $\sim 10^{-3} - 10^{-1}$s$^{-1}$ at the end of core oxygen burning. Thus, if the MRI triggers, it grows on a timescale significantly smaller than any other in the system. This is only not the case when either the action of MRI or the expansion after the MS slows down the rotational frequency of the outer envelope. Then, both $q\sim 0$ and $\Omega \sim 10^{-10}\,\text{rad\,s}^{-1}$ are very small. Excluding these outer regions, which are fully mixed and where the transport of AM by MRI is not effective, we find that, if $\Delta t$ is the time step our $15M_\odot$ stellar model computed with GENEC,  $10^{13}> \Delta t \gamma_{\textrm{MRI}}>10^3$ until the end of oxygen burning.  Hence, we assume that the MRI results in an instantaneous growth of the local (sub-grid scale) seed magnetic field up to its saturation value. A similar assumption is customary done for the TS dynamo (see \secref{sec:TS_description}). This instantaneous action of the magnetic field growth by the MRI also drives an equally rapid chemical mixing, something that is analogous to assuming instantaneous mixing in convective layers.
We consider as in \citetalias{Wheeler_2015} that $\lambda_{\rm MRI,fgm}$ can be arbitrarily small, since differently from the TS dynamo, the MRI triggering condition (Eq.~\ref{eqn:Instab}) does not depend on the (seed) magnetic field strength. This means that, for sufficiently small magnetic field, $\lambda_{\rm MRI,fgm}$ is much smaller than the typical radial length scale over which the thermal diffusivity is effective, $\ell_r\sim \sqrt{\kappa/N}$. While the MRI field growth could cease once $\lambda_{\rm MRI,fgm}\approx \ell_r$, it is more likely that saturation happens when
$\omega_A \sim q\Omega$ \citep{RevModPhys.70.1}, where $\omega_A=B/\sqrt{4\pi\rho r^2}$ is the Alfvén frequency, and thus the dominant toroidal magnetic field is

\begin{equation}
    B_{\phi} \sim q\Omega r \sqrt{4\pi\rho}.
    \label{eqn:MRI-magnetic-field}
\end{equation}
This field strength develops over length scales $\sim \lambda_{\rm MRI,fgm}$, typically much smaller than the radial size, $\Delta r$, of the mass shells used in GENEC. Hence, it is more significant to provide an estimation of the time and volume average saturation magnetic field. For that, we again resort to semi-global models performed for the saturation of the MRI in PNSs. \cite{Masada_2012ApJ...759..110} report values $\beta_{\phi}^{-1}:=\langle B_\phi^2\rangle/(8\pi P_{\rm gas})\approx 0.04$ for $|q|=0.2$, monotonically increasing up to $\approx 0.3$ for $|q|=1.8$. For comparison, in shearing-box accretion disk simulations, \cite{Hawley2011} shows $\langle B_\phi^2\rangle/(8\pi P_{\rm gas})
\approx 0.08$. 
Under the aforementioned assumption that $P_{\rm gas}$ can be replaced by $P_0$ in stellar models, one finds
\begin{align}
\label{eq:Bphi-saturation}
    \overline{B_\phi}:=\sqrt{\langle B_\phi^2\rangle} \sim \sqrt{2\beta_\phi^{-1} P_0} =  \sqrt{2\beta_\phi^{-1}}B_\phi\sim 0.40B_\phi,
\end{align}
where the last numerical factor results for $\beta_\phi^{-1}\approx 0.08$.

To estimate the time and volume averaged poloidal magnetic field,
$\overline{B_r}:=\sqrt{\langle B_r^2\rangle}$, we reason as follows.%
\footnote{Strictly speaking, $\overline{B_r}$ is the time and volume averaged radial component of the magnetic field at saturation. Since we are not computing the consistent magnetohydrodynamic evolution of the three components of the magnetic field, we assume that  $\overline B_r$ is representative of the poloidal component of the magnetic field within an order of magnitude.} 
\cite{Reboul-Salze_2021A&A...645A.109} in a similar context as \cite{Masada_2012ApJ...759..110}, but using also global models for PNSs find that $\zeta_{r\phi} := \langle B_r^2 \rangle/\langle B_\phi^2 \rangle\simeq 0.04-0.06$, a range of values significantly narrower than in local and global models of accretion disks 
\citep[where $0.0015 \lesssim \zeta_{r\phi} \lesssim 0.2$;][] {Hawley2011,Shi_2010ApJ...708.1716,Shi_2016MNRAS.456.2273}. Thus, we may estimate
\begin{align}
 \label{eq:Br-saturation}
   \overline{B_r}\sim \sqrt{2\beta_\phi^{-1} \zeta_{r\phi} } B_\phi \sim 0.09 B_\phi,
\end{align}
where the last numerical factor corresponds with taking $\zeta_{r\phi}\approx 0.05$. We note that different from the TS dynamo (see next section), the poloidal and toroidal (time and volume averaged) magnetic field components are proportional to each other.
   
We remark that even at their saturation values, the magnetic fields are dynamically negligible. Once the driving MRI condition \eqref{eqn:Instab} does not hold, magnetic reconnection reduces the magnetic field strength. Following \citetalias{Wheeler_2015}, the reconnection timescale is $\tau_{\rm recon}\sim \sqrt{\tau_{\rm diff} \tau_{A}}\sim 1/(\sqrt{\alpha} |q| \Omega)$, where we have estimated the diffusion timescale as $\tau_{\rm diff}=r^2/\nu_{\rm mag,MRI}$, and the Alfvén timescale at saturation as $\tau_{\rm A}=1/\omega_{\rm A}\sim 1/(|q|\Omega)$. Hence, the decay time from the saturated state is, approximately, a factor  of $1/\sqrt{\alpha}\sim 7$ times longer than the exponential growth time of the MRI, which remains significantly shorter than the evolutionary timescale. As a consequence, we also assume that once a certain mass shell does not fulfil the MRI triggering condition, the magnetic field is instantaneously destroyed. This assumption may not hold in the nearly non-rotating envelope developing after the MS, and is poor after the end of core oxygen burning. In such advanced stages of stellar evolution, the contraction of the core spins it up, raising $\Omega$ and, at the same time, the shear in the transition layer may grow (hence rising $|q|$). Both factors increase the growth and decay timescales of MRI. We note that our models stop after core oxygen burning. Thereby, our assumption that $\tau_{\rm recon}\ll \Delta t$ is adequate.

\subsection{Description of TS dynamo}
\label{sec:TS_description}

We also included the effects of the TS dynamo in a subset of models. The TS dynamo is a more widely studied instability already proven to impact significantly the transport of AM in stellar evolution \citep[e.g.][]{Maeder_2004A&A...422..225, Petrovic_2005A&A...435.1013, Heger_et_al__2005__apj__Presupernova_Evolution_of_Differentially_Rotating_Massive_Stars_Including_Magnetic_Fields,Yoon_2006A&A...460..199,Hartogh2019, Egg2019b, Bel2020,Marchant2020}. As in \citetalias{Wheeler_2015} the inclusion of the TS dynamo alongside the MRI serves the purpose of comparing the effects of each magnetic instability and how they interact together when included in the same model.
We implement, as in \citetalias{Wheeler_2015}, the original version of the TS dynamo as described by \cite{Spruit2002}. Other implementations of the instability such as those proposed by \cite{Fuller2019b} are not explored in this work.
According to \cite{1999Spruit}, the TS dynamo is more efficient at transporting AM than the MRI. For the TS dynamo to trigger, the presence of a small magnetic field is sufficient if there is a significant amount of shear, similarly to the MRI. We define the trigger condition of the TS dynamo as in \cite{Spruit2002},
\begin{equation}
    \label{eqn:TSinstab}
    |q|>q_{\rm min, TS}:=\left(\frac{N_{\rm eff}}{\Omega}\right)^{3/2}\left(\frac{\eta}{r^2\Omega}\right)^{1/4},
\end{equation}
where the square of the effective buoyancy frequency reads
\begin{equation}
    \label{eqn:N}
    N_{\rm eff}^2=\left(\frac{\eta}{\kappa}\right)N^2_T+N^2_{\mu}.
\end{equation}
In these expressions $\eta$ is the same as given by Eq.~\eqref{eqn:resistivity}. 
Different from the MRI, it is only the amplitude of the shear that is important to trigger the TS dynamo and not the sign of said shear. This is because the winding up process of the poloidal magnetic field contributing to the TS dynamo occurs whenever there is a gradient of rotational velocity, regardless of whether it is positive or negative. 

The effective viscosity associated with the TS dynamo follows from \citet{1999Spruit},
\begin{equation}
    \label{eqn:TSvisco}
    \nu_\mathrm{mag,TS}=r^2\Omega q^2\left(\frac{\Omega}{N_{\rm eff}}\right)^4,
\end{equation}
with $N_{\rm eff}^2$ given by Eq.~\eqref{eqn:N}.

In the same way as the MRI, we encode the effects of the TS dynamo on AM by means of the diffusion coefficient $D_\mathrm{mag,O}$. When the dynamo is active we have
\begin{equation}
\label{eqn:DmagO-TS}
    D_\mathrm{mag,O}=\nu_\mathrm{mag,TS}.
\end{equation}
Unlike the MRI, the transport of chemical elements is very inefficient for the TS dynamo.

As reasoned in \cite{Spruit2002}, the same displacements that produce the effective magnetic diffusivity affect the mixing of chemical elements in the radial direction. Thus, we implement the following diffusion coefficient,
\begin{equation}
\label{eqn:DmagX-TS}
    D_\mathrm{mag,X}=\eta.
\end{equation}
The expressions for the diffusion coefficients, $D_\mathrm{mag,O}$ and $D_\mathrm{mag,X}$, given so far are valid when either the MRI or the TS dynamo is active alone. When we include both instabilities in the same model, the magnetic diffusion coefficients are simply a linear combination of the two (\secref{sec:MRI+TS_MS} includes a rough justification of this choice). 

The growth rate of the magnetic field generated by the TS dynamo is, approximately, $\gamma_{\textrm{TS}} \approx  \omega_{\rm A}^2/\Omega$ if $\omega_{\rm A}\ll \Omega$ \citep{Spruit2002}, which for typical conditions of stellar interiors lets the field grow on a timescale significantly smaller than any other in the system. As stated in \secref{sec:MRI_description} for the MRI, our treatment assumes that the TS dynamo instantaneously amplifies the local seed magnetic field up to its saturation value.
The azimuthal and poloidal magnetic fields, in areas where the TS dynamo is active, are computed directly from the effective viscosity, $\nu_{\rm mag,TS}$, using \cite{Heger_et_al__2005__apj__Presupernova_Evolution_of_Differentially_Rotating_Massive_Stars_Including_Magnetic_Fields}:
\begin{eqnarray}
    B_{\phi}^4&=&16\pi^2\rho^2\nu_{\rm mag,TS}q^2\Omega^3r^2
    \label{eqn:Bphi-TS}\\
    B_{r}^4 &=& B_{\phi}^4\left(\frac{\nu_{\rm mag,TS}}{\Omega r^2}\right)^2
    \label{eqn:Br-TS}
.\end{eqnarray}
For the TS dynamo, the ratio between poloidal and azimuthal magnetic fields is not constant like the MRI but depends on the rotation profile and the chemical structure, this can lead to differences between these two field components of several orders of magnitude (see~\secref{sec:B-field_topology}). 
We have not found in the literature suitable estimates of the time and volume averaged magnetic fields at saturation due to the TS dynamo (cf. the contradicting results of \cite{Braithwaite_2006A&A...449..451} and \cite{Zahn_2007A&A...474..145} regarding whether a dynamo operates or not after the Tayler instability). These values should be smaller than the (small-scale) estimates in Eqs.~(\ref{eqn:Bphi-TS}, \ref{eqn:Br-TS}). Guided by the MRI case, where the factor relating the small-scale saturation fields and their time and volume averages is of order one, we assume that $\overline{B_\phi}\sim B_\phi$ and $\overline{B_r}\sim B_r$.

For the decay timescale of the fields generated by the TS dynamo, we use the estimate of \citetalias{Wheeler_2015}, based on the reconnection timescale of the generated fields. They find $\tau_{\rm recon}\sim 100\,\text{yr}H_{p,9}^{3/2}T_8^{3/4} \rho^{1/4} B_{8}^{-1/2}$, where $H_{p,9}$, $T_8$, and $B_8$ are the pressure scale height in units of $10^9\,$cm, the temperature in units of $10^8\,$K and the magnetic field in units of $10^8\,$G. This timescale is very short compared with the typical timescales of evolution through the MS. Thus, assuming that the magnetic field instantaneously decays once the triggering conditions do not hold is well justified. However, it may not be the case late in the evolution, once the density and temperature grow.

In the model \modl{MTS100}\footnote{See \secref{sec:models} for nomenclature details.}, which implements both the MRI and TS dynamo, we chose to compute the magnetic field as the maximum of the saturation values given by each instability. We suppose that the dominant instability is the one with the largest growth rate, which we surmise reaches saturation first. A more accurate determination of the magnetic field growth would require a precise tracing of the magnetic field, but such a method is not implemented in GENEC.

\subsection{Equations of transport}
In addition to implementing the MRI, we wish to see how magnetic instabilities interact with other physical processes due to rotation, specifically the circulation by meridional currents. The latter is an advective process and thus can lead to a buildup of differential rotation within the star \citep[see e.g.][]{MZ1998, MandM2000}, helping the MRI to trigger. Often in stellar evolution models, the circulation by meridional currents is treated as a diffusive process, with which there is no possibility of a buildup of differential rotation. We aim to study how this choice impacts the MRI. 

\subsubsection{Transport with an advective meridional circulation}
\begin{table*}
\caption{List of models computed for $M_{\rm ZAMS}=15 M_{\odot}$, $V_{\rm ini}$ = 206 km s$^{-1}$, and $Z$=0.012. The first column provides the model name. The second column, the treatment of the AM transport by meridional currents, either `advective' or `diffusive'. The third and fourth columns show whether the MRI and/or the TS dynamo are included. Columns five and six provide the values of $f_\mu$ (Eq.~\eqref{eqn:Instab}) and of $\alpha$ (Eq.~\eqref{eqn:alpha}). The seventh and eighth columns provide the equation number in the text used to compute the magnetic diffusion coefficients of AM, $D_\mathrm{mag,O}$ , and of chemical elements, $D_{\mathrm{mag,X}}$, respectively, induced by either the MRI, the TS dynamo or both. The final column gives the ratio between the core and surface angular velocity, given half-way through the MS for each model.}

\centering
    \begin{tabular}{||c c c c c c c c c ||}
        \hline \\[-2.4ex]
 Model  & \hspace*{2.2mm} Transport \hspace*{2.2mm}     & \hspace*{2.2mm} MRI   \hspace*{2.2mm}      & \hspace*{2.2mm} TS \hspace*{2.2mm} &  \hspace*{2.2mm} $f_{\mu}$ \hspace*{2.2mm}        & \hspace*{2.2mm}$\alpha$   \hspace*{2.2mm}       & \hspace*{2.2mm} $D_{\rm mag,O}$ \hspace*{1.9mm} & \hspace*{2.2mm} $D_{\rm mag,X}$ \hspace*{2.2mm}&  $\Omega_{\rm surf}/\Omega_{\rm core}$ \\[0.5ex] 
        \hline 
        \hline
        \modl{AN100} & \textit{advective}     & N & N & -    &  -   & - & -  & 0.64 \\
        \hline   
         \modl{AM100} & \textit{advective}       & Y  & N & 1.00 & 0.02 & \ref{eqn:Viscosity-MRI-2} & \ref{eqn:Viscosity-MRI-2} & 0.74\\
        \modl{AM005} & \textit{advective}        & Y & N & 0.05 & 0.02 & \ref{eqn:Viscosity-MRI-2} & \ref{eqn:Viscosity-MRI-2} & 0.91 \\
         \modl{AM100L}& \textit{advective}       & Y & N  & 1.00 & 0.01& \ref{eqn:Viscosity-MRI-2} & \ref{eqn:Viscosity-MRI-2} & 0.74 \\ 
         \modl{AM100H} &\textit{advective}      & Y & N & 1.00 & 0.05 & \ref{eqn:Viscosity-MRI-2} & \ref{eqn:Viscosity-MRI-2} & 0.74  \\      
        \hline
        \modl{TS100} & \textit{-}   & N & Y & 1.00   &  -  & \ref{eqn:TSvisco} & \ref{eqn:DmagX-TS}    & 0.97 \\
        \modl{MTS100} & \textit{-}   & Y & Y & 1.00 & 0.02 & \ref{eqn:Viscosity-MRI-2}+\ref{eqn:TSvisco} & \ref{eqn:Viscosity-MRI-2}+\ref{eqn:DmagX-TS} & 0.97 \\
        \hline
       \modl{DN100} & \textit{diffusive} & N & N & - & - & - & - & 0.91 \\ 
         \modl{DM100}& \textit{diffusive}  & Y & N & 1.00 & 0.02 & \ref{eqn:Viscosity-MRI-2} & \ref{eqn:Viscosity-MRI-2} &0.92 \\[0.4ex]  
        \hline
        \hline
    \end{tabular}
\label{table:nomenclature}
\end{table*}
\label{sec:advection}

In the case where we implement an advective circulation by meridional currents, we solve the following transport of AM equation within GENEC,

\begin{align}
    \rho \frac{d}{dt}(r^2\Omega)_{M_r}=\:&\frac{1}{5r^2}\frac{\partial}{\partial r}\left(\rho r^4 \Omega U(r)\right) \nonumber \\
    &+\frac{1}{r^2}\frac{\partial}{\partial r}\left(\rho D r^4 \frac{\partial \Omega}{\partial r}\right).
    \label{eqn:TransportAng}
\end{align}
On the left-hand side of this equation, is the Lagrangian derivative of the specific AM taken at a mass coordinate $M_r$.
The first term in the right-hand side of this equation is the divergence of the advective flux of AM, while the second term is the divergence of the diffusion flux. $U(r)$ expresses the dependence with $r$  of the vertical component of the meridional circulation velocity \citep{MZ1998}. The coefficient $D$ is the total diffusion coefficient, taking into account the various instabilities that transport AM diffusively. These instabilities here are convection, secular-shear, and magnetic instabilities,
\begin{equation}
\label{eqn:D}
    D=D_{\rm shear}+D_{\rm conv}+D_{\rm mag,O}
,\end{equation}
\noindent
with $D_{\rm shear}$ defined as in \citet{Maeder1997-2} assuming a very small Peclet number\footnote{This expression corresponds to Eq.~5.32 in \citet{Maeder1997-2} neglecting the term multiplied by $\Gamma/(\Gamma-1)$, where $\Gamma$ is equal to the Peclet number multiplied by 1/6. 
},
\begin{equation}
    \label{eqn:Dshear}
    D_{\rm shear}=\frac{H_p}{g}\frac{\kappa}{\left[\nabla_{\mu}+\delta(\nabla_{\rm ad}-\nabla_{\rm rad})\right]}\left(\frac{9\pi}{32}\Omega q\right)^2
,\end{equation}
 with $\kappa$ given by Eq.~\eqref{eqn:kappa} and $\nabla_{\mu}=(\partial \ln \mu / \partial \ln P)_{T,\mu}$.
The convection coefficient $D_{\rm conv}$ is derived from mixing length theory.
The magnetic diffusion coefficient $D_{\rm mag,O}$ is given either by Eq.~\eqref{eqn:DmagO-MRI}, for the MRI, Eq.~\eqref{eqn:DmagO-TS}, for the TS-dynamo or the sum of the two.

The transport of chemical species is described 
as the sum of the transport by the shear (same diffusion coefficient as for the AM), $D_{\rm shear}$, by a diffusion coefficient $D_{\rm eff}$, describing the interaction of a strong horizontal turbulence with the meridional currents \citep{1992Chaboyer-Zahn}, and by the magnetic instabilities. $D_{\rm eff}$ reads
\begin{equation}
    \label{eqn:Deff}
    D_{\rm eff}=\frac{1}{30}\frac{|rU(r)|^2}{D_{\rm h}}.
\end{equation}
In this expression  $D_{\rm h}$ denotes the horizontal turbulence coefficient \citep{Zahn1992},
\begin{equation}
    \label{eqn:Dh}
    D_{\rm h}=r\left|2V(r)- \left(1+\frac{1}{2}q\right) U(r)\right|,
\end{equation}
where $V(r)$ describes the radial dependence of the azimuthal component of the meridional circulation velocity.
To evolve the chemical species distribution we then solve\begin{align}
       \rho \frac{d }{dt}\left(X_i\right)_{M_r}=\:&\frac{1}{r^2}\frac{\partial}{\partial r}\left(\rho r^2 \left[ D +D_\mathrm{eff} \right]\frac{\partial X_i}{\partial r} \right) \nonumber \\
       &+ \rho\left(\frac{d X_i}{dt} \right)_\mathrm{nucl},           
       \label{eqn:TransportChem}
\end{align}
where $\left(\frac{d X_i}{dt} \right)_\mathrm{nucl}$ accounts for the changes in the mass fractions of the species due to nuclear reactions, and
\begin{equation}
    \label{eqn:D2}
    D=D_\mathrm{shear}+D_\mathrm{conv}+D_\mathrm{mag,X},
\end{equation}
$D_\mathrm{mag,X}$ being equal to either Eq.~\eqref{eqn:DmagO-MRI}, for the MRI, Eq.~\eqref{eqn:DmagX-TS} for the TS dynamo or a combination of the two.

\subsubsection{Transport with a diffusive meridional circulation}
\label{sec:diffusive_transport}

In GENEC the AM transport due to meridional currents is treated as an advective process, by default. However, many other stellar evolution codes model this process through a diffusion equation. Thus, we also employed  a diffusive treatment of the meridional circulation for certain models. 
In this case, we drop the advective term that described the circulation and instead add another diffusive coefficient to model this process. Therefore, Eq.~\eqref{eqn:TransportAng} reduces to
\begin{equation}
\label{eqn:TransportAng-Diff}
    \rho \frac{d}{dt}(r^2\Omega)_{M_r}=\frac{1}{r^2}\frac{\partial}{\partial r}\left(\rho D r^4 \frac{\partial \Omega}{\partial r}\right),
\end{equation}
with\begin{equation}
 \label{eqn:D-circ}
    D=D_{\rm shear}+D_{\rm conv}+D_{\rm mag,O}+D_{\rm circ}.
\end{equation}
The diffusion coefficient of the meridional circulation, $D_{\rm circ}$, is given by
\begin{equation}
    \label{eqn:Dcirc}
    D_\mathrm{circ}=\left|rU(r)\right|.
\end{equation}
The equation describing the transport of chemical elements  Eq.~\eqref{eqn:TransportChem} remains unchanged under the diffusive treatment. However, $D_{\rm h}$ has an approximate expression in this case (thus changing $D_{\rm eff}$), which we obtain by taking $V(r) \approx U(r);$ therefore,

\begin{equation}
    \label{eqn:Dh_diffusive}
    D_{\rm h}=r\left|\left(1-\frac{1}{2}q\right) U(r)\right|.
\end{equation}

\section{Models computed}
\begin{figure*}[!]
    \centering
    \begin{tikzpicture}
    \pgftext{%
        \hspace{-1.8cm}\includegraphics[width=0.51\textwidth]{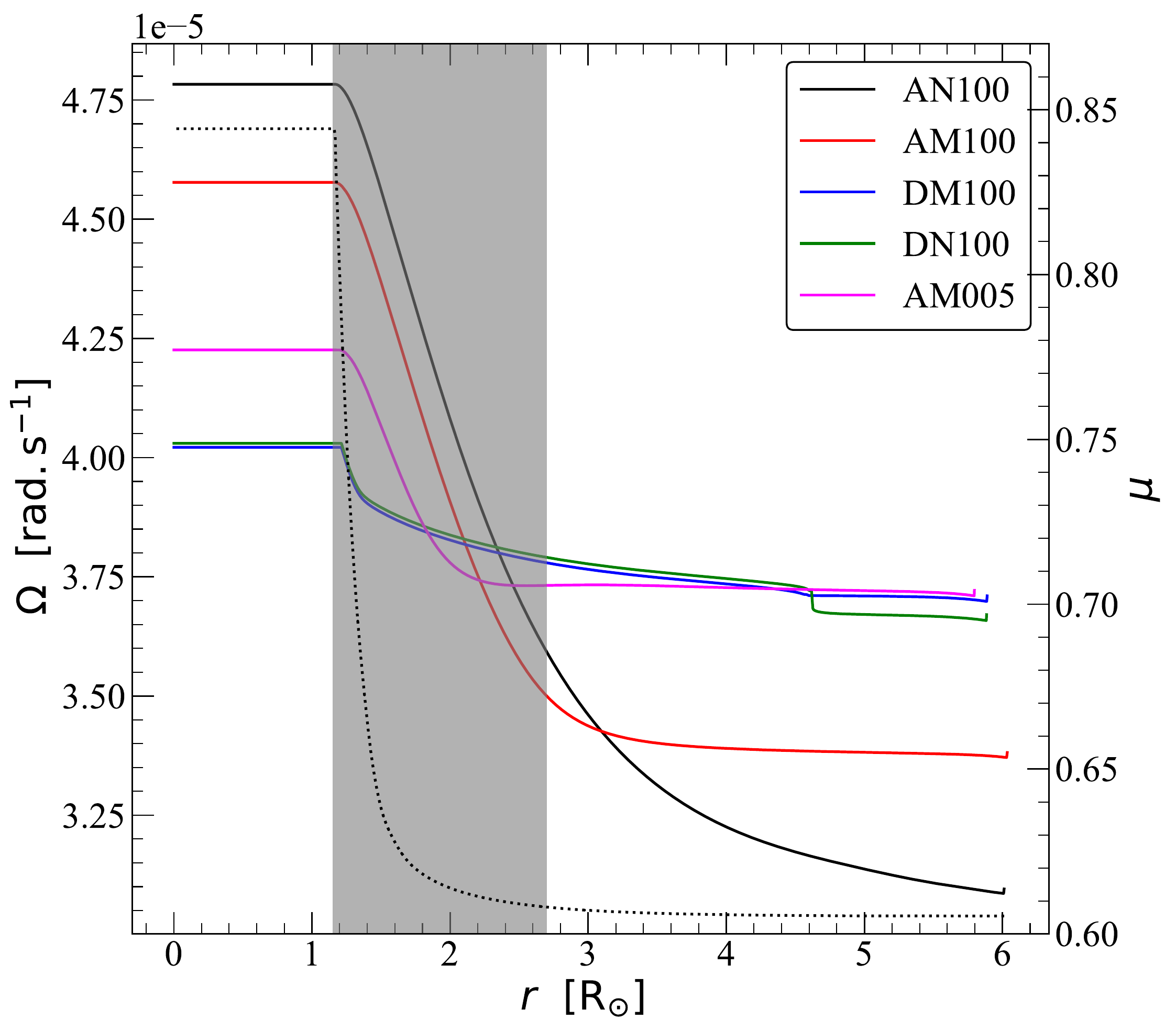} \includegraphics[width=0.465\textwidth]{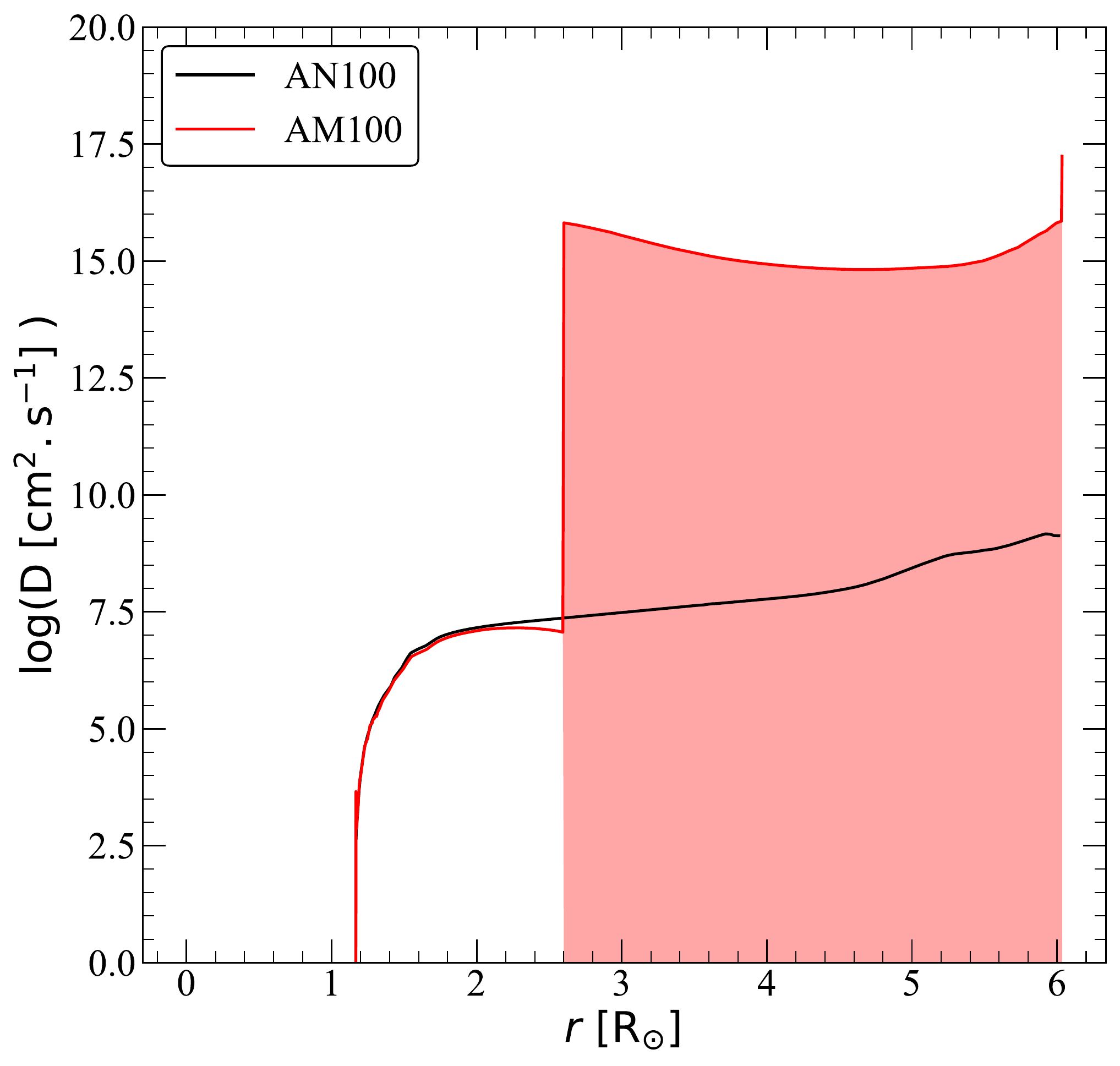}
            }%
    \node at (-9.5,-3.2) {\large (a)};
    \node at (-0.2,-3.2) {\large (b)};
    \end{tikzpicture}
\caption{ Effects of the MRI during the MS {\it Left panel:} 
Distribution of angular velocity, $\Omega$, when the core hydrogen abundance reaches $X_\mathrm{c}=0.35$ for different models. For model \modl{AN100}, we show with a dotted line the mean molecular weight, $\mu$. The area in grey indicates where the $\mu$ gradient is strong enough to suppress MRI activation in model \modl{AM100}. 
   {\it Right panel:} Comparison at $X_\mathrm{c}=0.35$ of the unrestricted AM diffusion coefficient $D=D_{\rm shear}+D_{\rm MRI}$, given by Eqs.\eqref{eqn:Dshear} and \eqref{eqn:DmagO-MRI}, for model \modl{AN000} (black) and  model \modl{AM100} (red).
 The zone in shaded red shows the radial extension of the region where the MRI is active in model \modl{AM100}.}
    \label{fig:MRI-none-omega-30}
\end{figure*}
\label{sec:models}

Our models have an initial mass of 15 M$_\odot$, a moderate initial equatorial rotation velocity at the time of zero age main sequence (ZAMS) of 206\,km\,s$^{-1}$ (roughly 0.32 of critical velocity), and an initial metallicity $Z=0.012 \approx Z_\odot$. 
These values have been chosen to be the same initial conditions as in \citetalias{Wheeler_2015} to allow comparisons.

Our models perform a parametric scan of different physical properties (see Table~\ref{table:nomenclature}), and their naming convention is the following: models names contain a prefix part with two or three letters that indicate the physical processes included, followed by three numbers that account for the value of the parameter $f_\mu$ (see below). Further suffixes account for variations on the default $\alpha$ parameter (Eq.~\eqref{eqn:alpha}). More precisely: (i) models with either an advective or a diffusive scheme for the AM transport include the letter `A' or `D' in their names, respectively; (ii) models with and without  MRI, are prefixed with  an `M' or an `N', respectively; (iii) models that include the TS dynamo are annotated with `TS', while models simultaneously including the TS dynamo and the MRI are named with `MTS'; (iv) models with different values of $f_{\mu}$, say $f_\mu=0.05$ and 1.00 contain in their name, respectively, 005, and 100; and (v) in models that include the MRI, values of $\alpha=0.01$ and $0.05$ are distinguished from the default $\alpha=0.02$ case with suffixes `L' and `H', respectively.

A comparison between models \modl{AM}, including MRI,  and model 
\modl{AN100}, without MRI,  facilitates assessing the impact of the MRI in rotating models, where the AM transport by meridional currents is accounted as an advective process. This requires the resolution of a fourth order equation in $\Omega$ \citep[see e.g.][]{Meynet2009}. Numerically, this equation is decomposed in four first order equations with four unknowns resolved using a Henyey method.
This complex treatment of the AM transport by meridional currents is applied only during the MS phase. During the post-MS phase, the AM variation inside the star is mainly dominated by the local conservation of AM.

Models \modl{AM100}, \modl{TS100,} and \modl{MTS100} allow us to compare the effects of the MRI acting alone versus the action of the TS dynamo as a standalone process or in combination with the MRI. Comparing model \modl{DM100} with model \modl{DN100} will assess the impact of the MRI when `the AM transport by meridional currents is treated through a diffusive equation. To calibrate the effects of the treatment of meridional currents without the additional interplay of magnetic instabilities, we compute models \modl{AN100} and \modl{DN100}.

\begin{figure*}
    \centering
        \begin{tikzpicture}
    \pgftext{%
        \includegraphics[width=0.5\textwidth]{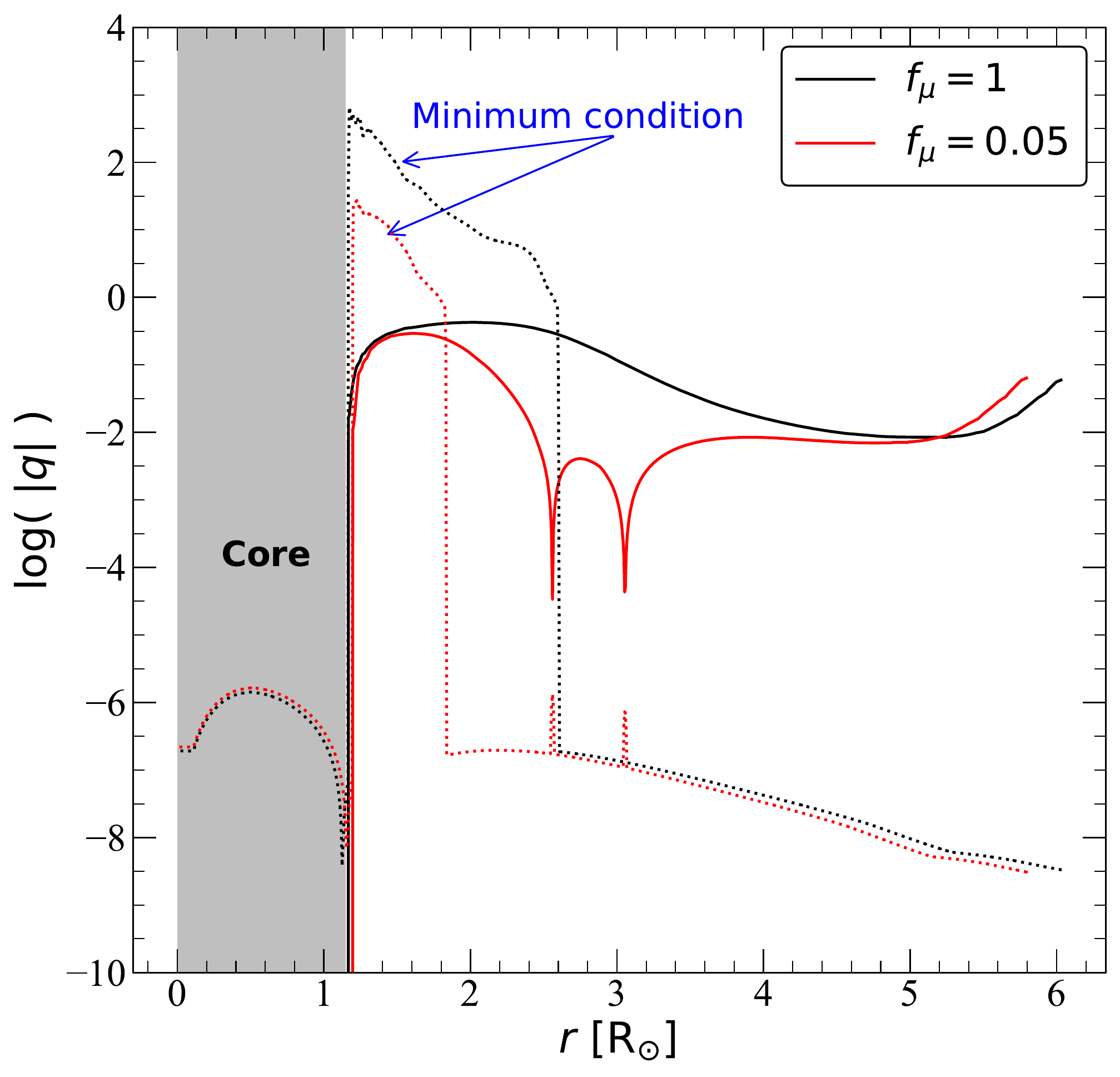}
        \hspace{0.2cm}
        \includegraphics[width=0.495\textwidth]{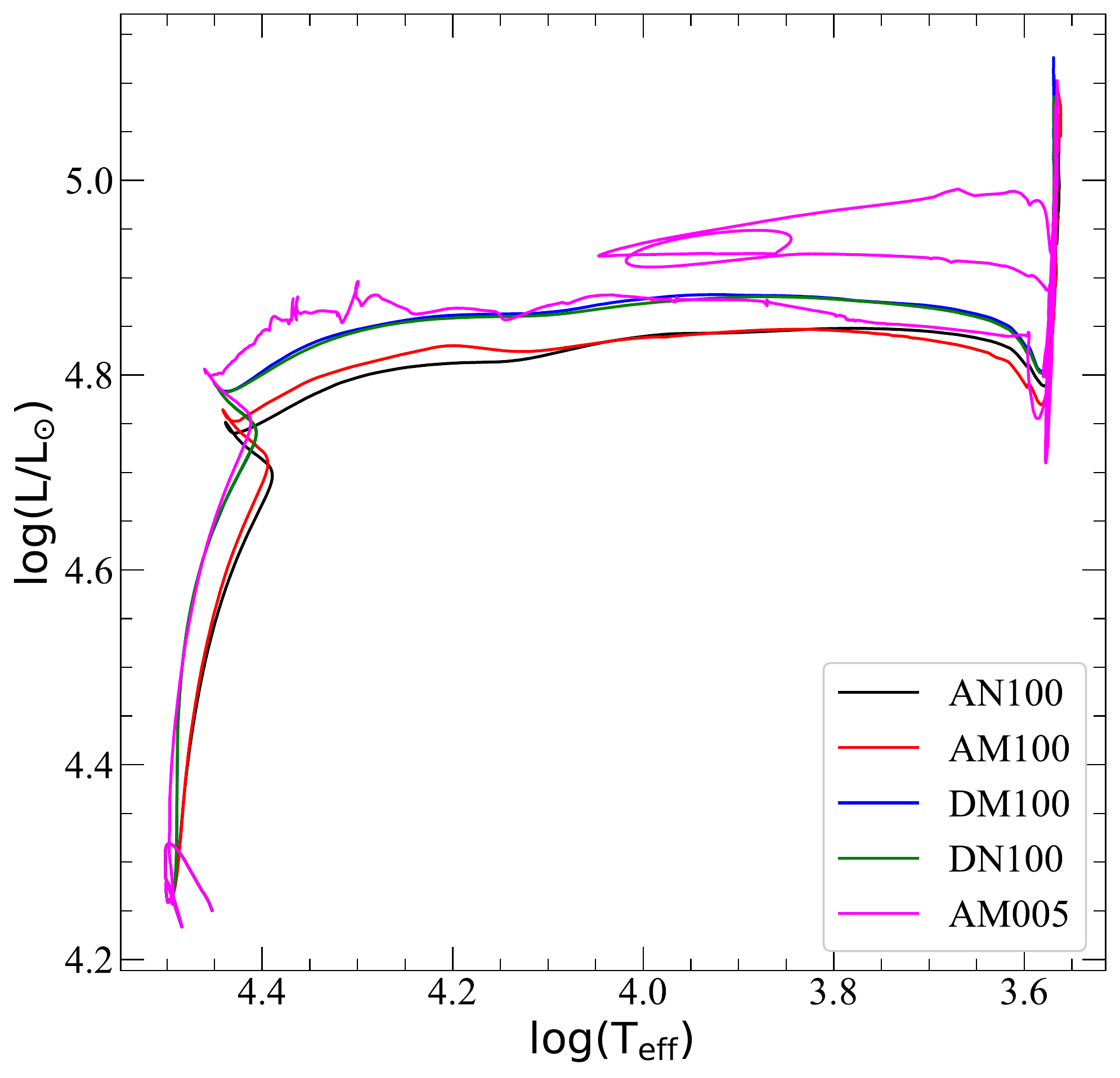}
            }%
    \node at (-9.2,-3.7) {\large (a)};
    \node at (+0.2,-3.7) {\large (b)};
    \end{tikzpicture}
    \caption{%
    Minimum shear condition for the MRI and evolutionary tracks of our models.
    \textit{.Left panel: }Logarithm of the absolute value of the shear, $q$, as a function of the radius when $X_\mathrm{c}=0.35$ for model \modl{AM100} (black) and model \modl{AM005} (red). The values of $q_\mathrm{min,MRI}$ given by the expression in Equation~\eqref{eqn:Instab} are plotted as dotted lines. The grey zone covers the convective core of the stellar models.
    \textit{Right panel:} HR diagram showing the evolutionary tracks during the MS for various models (see legends).}
    \label{fig:q-qmin+HR}
\end{figure*}
\section{Massive star evolution with MRI during the MS}
\label{sec:main_sec}

We investigate here the impact of MRI during the MS. We consider in \secref{sec:MRI_MS_ang} and \secref{sec:MRI_MS_chem} the stellar evolution of models in which the MRI instability may trigger, and the meridional currents are treated as an advective process. Then, in \secref{sec:MRI+TS_MS} we compare the effects of the MRI and the TS dynamo, either acting separately (and exclusively) or working together.
Finally, we discuss the action of the MRI in the frame of models that treat the transport of the AM as a diffusive effect in \secref{sec:MRI+Diffusion_MS}.

\subsection{Transport of angular momentum due to the MRI}
\label{sec:MRI_MS_ang}

During the MS the MRI triggers and subsequently transports AM within the star. Figure~\ref{fig:MRI-none-omega-30}a shows the variation of angular velocity as a function of radius in various models, among which we find \modl{AN100} (in black) without any magnetic field and, hence, without including the action of the MRI and \modl{AM100} (in red) with MRI. The models in this figure are shown half-way through the MS, estimated to be when central hydrogen abundance, $X_c$, equals 0.35.
With the MRI included, the profile of $\Omega$ is flattened (i.e. $|q|$ decreases), in the region where the $\mu$-gradients are not strong enough to inhibit the instability (from $r\gtrsim 2.6R_\odot$ for \modl{AM100}).
As  a result of the AM transport due to MRI the time averaged surface velocity increases, though not by much. The surface of model \modl{AM100} rotates $\sim 14\%$ faster than \modl{AN100}.

In Fig.~\ref{fig:MRI-none-omega-30}b, we plot the sum of the diffusion coefficients for the secular shear instability and the MRI, $D=D_\mathrm{shear}+D_\mathrm{mag,O}$, for models \modl{AN100} and \modl{AM100}. From $r\approx1.1R_{\odot}$ up to the surface is the radiative region of the star, the MRI does not trigger below $r\approx2.6R_{\odot}$. Thus, $D$ is the same for both \modl{AN000} and \modl{AM100} below $r=2.6 R_{\odot}$ because $D_{\rm mag,O}=0$. Farther out in the star, the MRI triggers because the strong chemical gradient that is present close to the core boundary drops off as we enter the more homogeneous outer layer (see the dotted line showing the variation of the mean molecular weight for \modl{AN000} 
in Fig.~\ref{fig:MRI-none-omega-30}a). This region is shaded in red in Fig.~\ref{fig:MRI-none-omega-30}b. There the MRI is extremely strong and the combined diffusion coefficient reaches values of $10^{15}\,{\rm cm}^2\,{\rm s}^{-1}$.%
\footnote{For numerical reasons we restrict the transport coefficient so that $D\le 10^{12}\,{\rm cm}^2\,{\rm s}^{-1}$. For values $D>10^{12} \,{\rm cm}^2\,{\rm s}^{-1}$ the same evolution is achieved, driven by the nearly instantaneous transport of AM. In Fig.~\ref{fig:MRI-none-omega-30}b we display the unrestricted value of $D$.} \label{foot:saturation}
This large coefficient allows for a fast AM transport, therefore flattening $\Omega$ in that region. 

Model \modl{AM005} incorporates a less stringent criterion to activate the MRI ($f_{\mu}=0.05$; see Sec.~\ref{sec:MRI_description}), in line with those used by other practitioners in the field.
 \modl{AM005} shows a flatter $\Omega$-profile with a lower contrast between the core and envelope rotation than model \modl{AM100} (Fig.~\ref{fig:MRI-none-omega-30}a). Indeed, the ratio of the surface ($\Omega_{\rm surf}$) to the core angular velocity ($\Omega_{\rm core}$) is $0.74$ for \modl{AM100} while this ratio is $0.91$ for \modl{AM005}.
A lower $f_{\mu}$ reduces the stabilising impact of chemical gradients, which in practice leads to a smaller value for $q_{\rm min,MRI}$ (Eq.~\eqref{eqn:Instab}), allowing the MRI to trigger more easily. In Fig.~\ref{fig:q-qmin+HR}a the values of $q_{\rm min,MRI}$ are shown in dotted lines alongside the actual values of $|q|$, solid lines, for  models \modl{AM100} and \modl{AM005}. The value of $q_{\rm min,MRI}$ maximises just at the core--envelope interface, where the $\mu$-gradients are the strongest.
Beyond that region the MRI has triggered and $q_{\rm min,MRI}$ drops off to very small values due to the efficient transport of the MRI.

As expected, a low value of $f_\mu$ significantly enlarges the region where the MRI triggers (diminishing the lower boundary of this region from 2.7 R$_\odot$ when $f_\mu=1$ to 1.8 R$_\odot$ when $f_\mu=0.05$). Therefore, in model \modl{AM005} the MRI transports more AM from the core (and also nuclear species as we see in Sect.~\ref{sec:MRI_MS_chem}), than \modl{AM100}. 
The efficient AM transport over a larger zone for model \modl{AM005} leads to a lower value of $|q|$ in the star,  Fig.~\ref{fig:q-qmin+HR}a,
 and also the more efficient mixing is responsible for the radius of \modl{AM005} being slightly smaller than in model \modl{AM100}.

We computed two other models that include just the MRI with an advective transport, \modl{AM100L} and \modl{AM100H} that differ with respect to our reference model \modl{AM100} only by the value of $\alpha$ (see Eq.~\eqref{eqn:alpha}), equal to $0.01$ and $0.05$, respectively. During the MS the shear is strong throughout most of the stellar envelope, and thus the amplitude of $|q|\Omega r^2$ in Eq.~\eqref{eqn:Viscosity-MRI-2} is very large. As seen in Fig.~\ref{fig:MRI-none-omega-30}b the unrestricted diffusion coefficient for the transport of AM and chemical elements saturates at \textit{D} $\simeq 10^{15} \, {\rm cm}^2\,{\rm s}^{-1}$, where the MRI is active. For these values of $D$, the transport is so strong that it is instantaneous. Therefore, the changes in $\alpha$ within a factor of 5 do not yield noticeable variations in the amount of AM and elements transported during the MS.

\begin{figure*}
    \centering
    \begin{tikzpicture}
    \pgftext{%
        \includegraphics[width=0.44\textwidth]{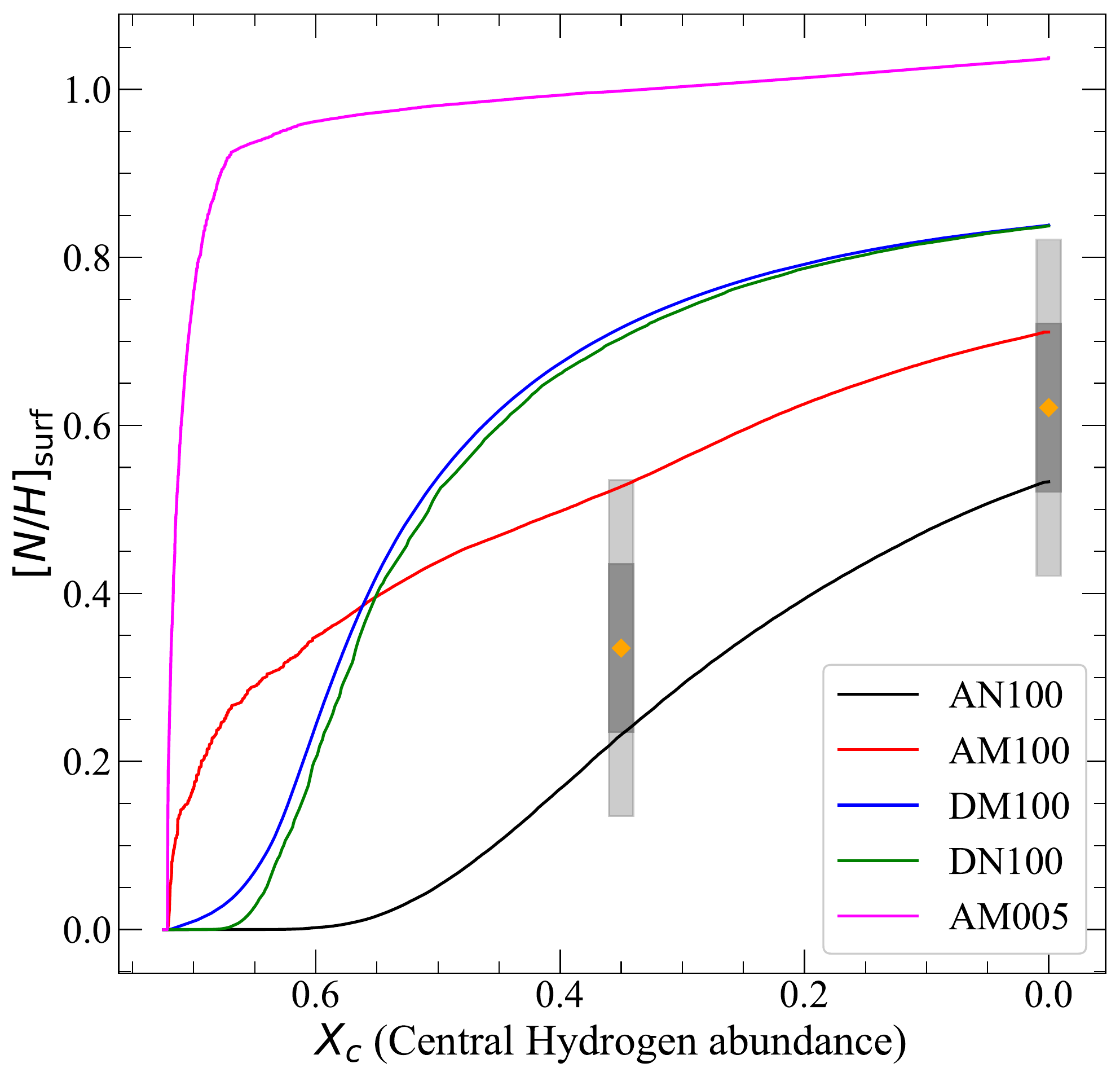}
        \hspace{0.2cm}
        \includegraphics[width=0.49\textwidth]{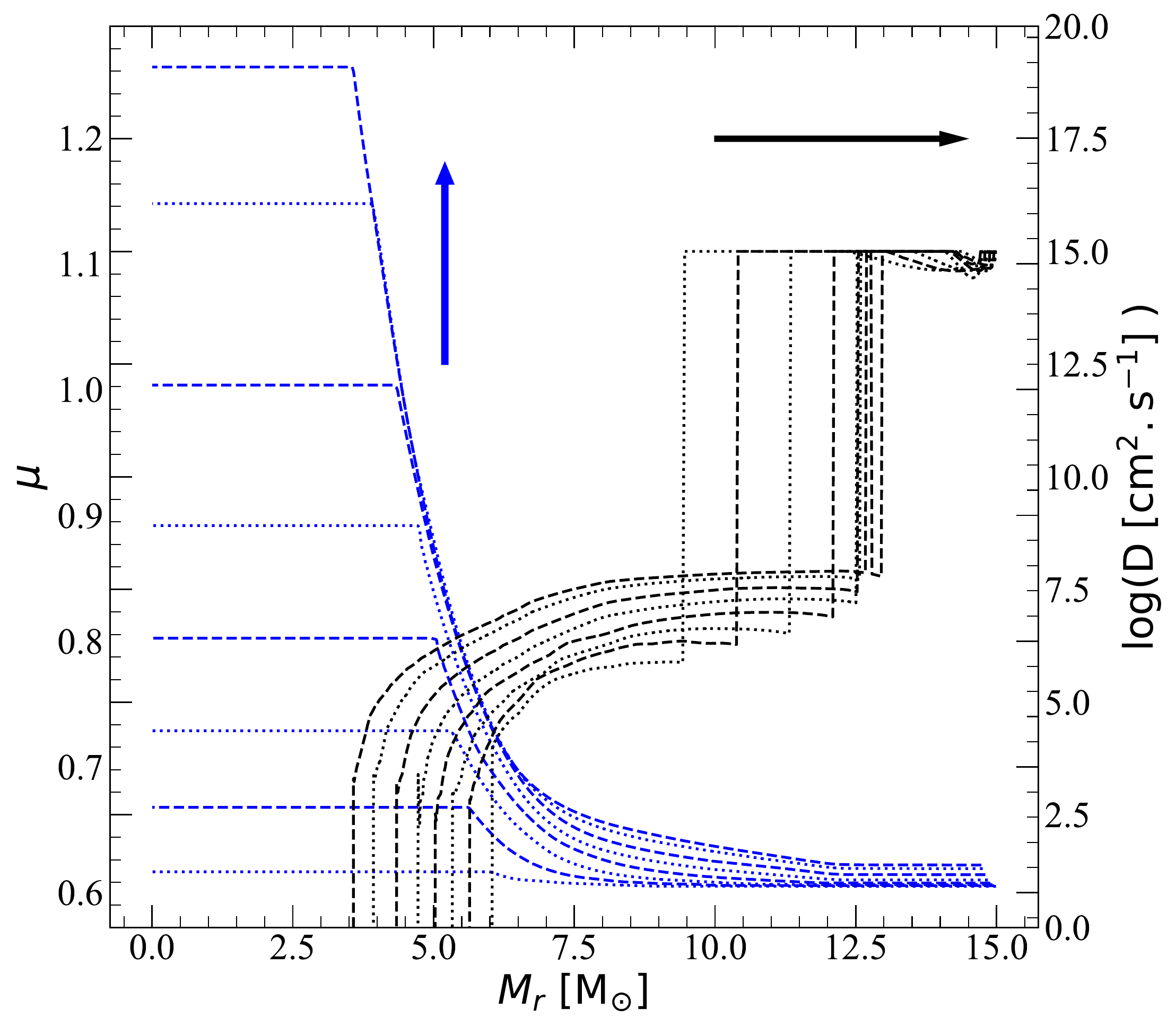}
            }%
    \node at (-9.2,-3.7) {\large (a)};
    \node at (+0.2,-3.7) {\large (b)};
    \end{tikzpicture}
     \caption{Impact of MRI on chemical transport during the MS. {\it Left panel:} Evolution of the ratio between the surface nitrogen abundance and surface hydrogen abundance, $\left[N/H\right]_\mathrm{surf}$, as defined in Eq.~\eqref{eqn:abundance-ratio}
     for selected models (see plot legends) as a function of the central hydrogen abundance, $X_\mathrm{c}$.
  The yellow diamonds are the mean of the observed surface values found in Fig.~11 of \citet{Ekstrom2012}, who compute models with rotation. The grey boxes encompass observed values of [N/H] for stars at the middle and end of the MS band, which are also from Fig.~11 of \citet{Ekstrom2012}. The light grey bars extend the values  of \citet{Ekstrom2012} to account for the large errors of the enrichment of individual stars with respect to the averaged values.  {\it Right panel:} Evolution of the mean molecular weight (in blue) and of $D=D_\mathrm{shear}+D_\mathrm{mag,O}$ (in black) for model \modl{AM100} as a function of the Lagrangian mass coordinate, $M_r$, during the MS. The curves shown with different line styles are for $X_\mathrm{c}=0.7,0.6,0.5,0.4,0.3,0.2,0.1,0.04$, where the arrows, blue and black respectively, show the direction of evolution going from $X_\mathrm{c}=0.7$ to $X_\mathrm{c}=0.04,$ increasing $\mu$ and restricting the region with large values of $D$ towards larger mass coordinates with decreasing values of $X_\mathrm{c}$ (equivalently, increasing time).  }
    \label{fig:Surface-Abund}%
\end{figure*}
\subsection{Transport of chemical elements due to the MRI}
\label{sec:MRI_MS_chem}

The MRI is akin to a hydrodynamical instability in that it transports AM and matter in equal measure. Here we focus on the impact of the transport of chemical elements during the MS.
Figure~\ref{fig:q-qmin+HR}b  shows the evolutionary tracks on a Hertzsprung-Russell (HR) diagram for our models. 
The difference between advective MRI (red curve) and non-MRI (black curve) models is modest.
Despite the fact that the mixing is more efficient in the MRI model, there is not much change in global surface properties.  Of course, this is based on models with a moderate initial rotation. 
We expect larger differences on the HR track for models with larger initial rotational rates.

In our MRI models the more virulent chemical mixing brings more helium to the outer envelope, which decreases the opacity, making the star over-luminous and more compact. Figure~\ref{fig:q-qmin+HR}b shows that the model, \modl{AM005} (magenta curve), is  more luminous than model \modl{AM100} due to the enhanced MRI induced chemical mixing. We note also that the \modl{AM005} model develops a blue loop during the core helium burning phase, while the other models do not, which reflects some significant differences in the chemical structure of this model.

Although the HR track is not substantially modified by the action of the MRI for model \modl{AM100}, the surface composition presents notable changes.
 In Fig.~\ref{fig:Surface-Abund}a we display the evolution of the normalised nitrogen to hydrogen surface ratio,
\begin{equation}
\label{eqn:abundance-ratio}
[N/H]_\mathrm{surf}:=\log\left(\frac{N_\mathrm{surf}}{H_\mathrm{surf}}\right) -\log\left(\frac{N_\mathrm{surf,ini}}{H_\mathrm{surf,ini}}\right)
,\end{equation}
as a function of central hydrogen mass fraction, $X_\mathrm{c}$. Here $N_{\rm surf}$ and $H_{\rm surf}$ are the surface number densities
of nitrogen and hydrogen and the subscript `ini' denotes the initial quantities. The quantity $X_\mathrm{c}$ is a proxy for time during the MS. The model with no MRI, \modl{AN100} (black curve), has no nitrogen surface enrichment above 0.2\,dex until halfway through the MS, which occurs around 9\,Myr, while the model with MRI, \modl{AM100} (red curve), reaches that same level of enrichment very early during the MS, at $X_\mathrm{c}=0.65$, roughly after 1\,Myr. Not only initially, but also during the whole MS phase, the model with MRI presents a larger surface enrichment than the one without it. The difference between the two models becomes smaller as evolution proceeds, reaching values around $0.2\,$dex towards the end of the MS.
This is mainly due to the fact that, during the MS, the gradient of $\mu$ becomes larger as time goes on, this inhibits the MRI more in the final stages of the MS, to the extent where the MRI can no longer trigger towards the end of the MS. In Fig.~\ref{fig:Surface-Abund}b we show the diffusion coefficient $D=D_\mathrm{shear}+D_\mathrm{mag,O}$, in black, inside the star at different time steps during the MS for model \modl{AM100} alongside the mean molecular weight, in blue. The chemical gradient increases with time at the boundary between the envelope and the core. Therefore, there is a reduction of the region where MRI can trigger (identified in Fig.~\ref{fig:Surface-Abund}b by the regions where $D\gtrsim 10^8\,\text{cm}^2\,\text{s}^{-1}$). As the MS progresses, the MRI is active over a progressively smaller area, so the MRI can only have an important effect on  chemical element transport for the first few million years of the MS. The MRI enriches the surface with heavy elements rapidly and early in the MS, this highlights a potential observable characteristic for detecting magneto-rotational instabilities in massive stars. Observing high surface enrichments for young massive stars may suggest that the MRI has developed in their radiative zones. We point out that the observational ranges displayed in \cite{Ekstrom2012} have here been enlarged in Fig.\ref{fig:Surface-Abund}a (light grey bands) due to various reasons. First, we considered also more recent samples of massive galactic stars from \cite{Aerts_2014}. Second, we aim at accounting for the fact that quite large errors in the determination of the nitrogen enrichment may be found for individual stars (singularly, when the number of nitrogen lines detected is small). Third, most samples include both rotating and non-rotating cases to compute their respective averaged values. Our models are rotating, and thus, a direct comparison with the averaged values of samples including also non-rotating stars may be too restrictive. In any case we should be cautious about the conclusions we may draw from the comparisons presented in Fig.~\ref{fig:Surface-Abund}a. Rotating models of any kind, the models of this paper included, hide free parameters whose values are specifically chosen in order to reproduce an observed constraint. In general, rotating models are calibrated so that a typical 15 $M_{\odot}$ mass model, with an initial rotation chosen to fit with an observed distribution of rotation velocities, then reproduces a mean observed  surface nitrogen enrichment.
The model with no MRI (see the black line in Fig.~\ref{fig:Surface-Abund}a), goes through the grey boxes (although a bit below their central regions), not by chance but because the physics used is the same as in \cite{Ekstrom2012}, and  the calibration is set in order to reproduce this enrichment.
Our model is slightly under what would produce a good fit because it has a smaller initial rotation than the model by \cite{Ekstrom2012},
and it has a slightly smaller initial metallicity. Using the same calibration, we note that the MRI model (see the red line in Fig.~\ref{fig:Surface-Abund}a) predicts larger surface enrichment at a given evolutionary stage, most notably in the very early MS. Thus, for such a model to go through the centre of the grey regions, we would need to change the value of the free parameters.

In the case of \modl{AM005} the suppression of chemical gradients through $f_{\mu}$ allows the MRI to be much more efficient. The surface enrichment of \modl{AM005} shown in (Fig.~\ref{fig:Surface-Abund}a) is extremely rapid and reaches significantly higher values than all other models. It is so fast that after just a few million years the amount of nitrogen at the surface goes far beyond the enrichment reached by all other models at the end of their respective MSs. This extreme enrichment surpasses any expected observations (Fig.~\ref{fig:Surface-Abund}a). Hence, it suggests that the use of $f_{\mu}<1$ with the GENEC code (including an advective treatment of the meridional circulation) is not necessary.

\begin{figure*}
    \centering
        \begin{tikzpicture}
    \pgftext{%
        \hspace{-0.9cm}\includegraphics[width=0.52\textwidth]{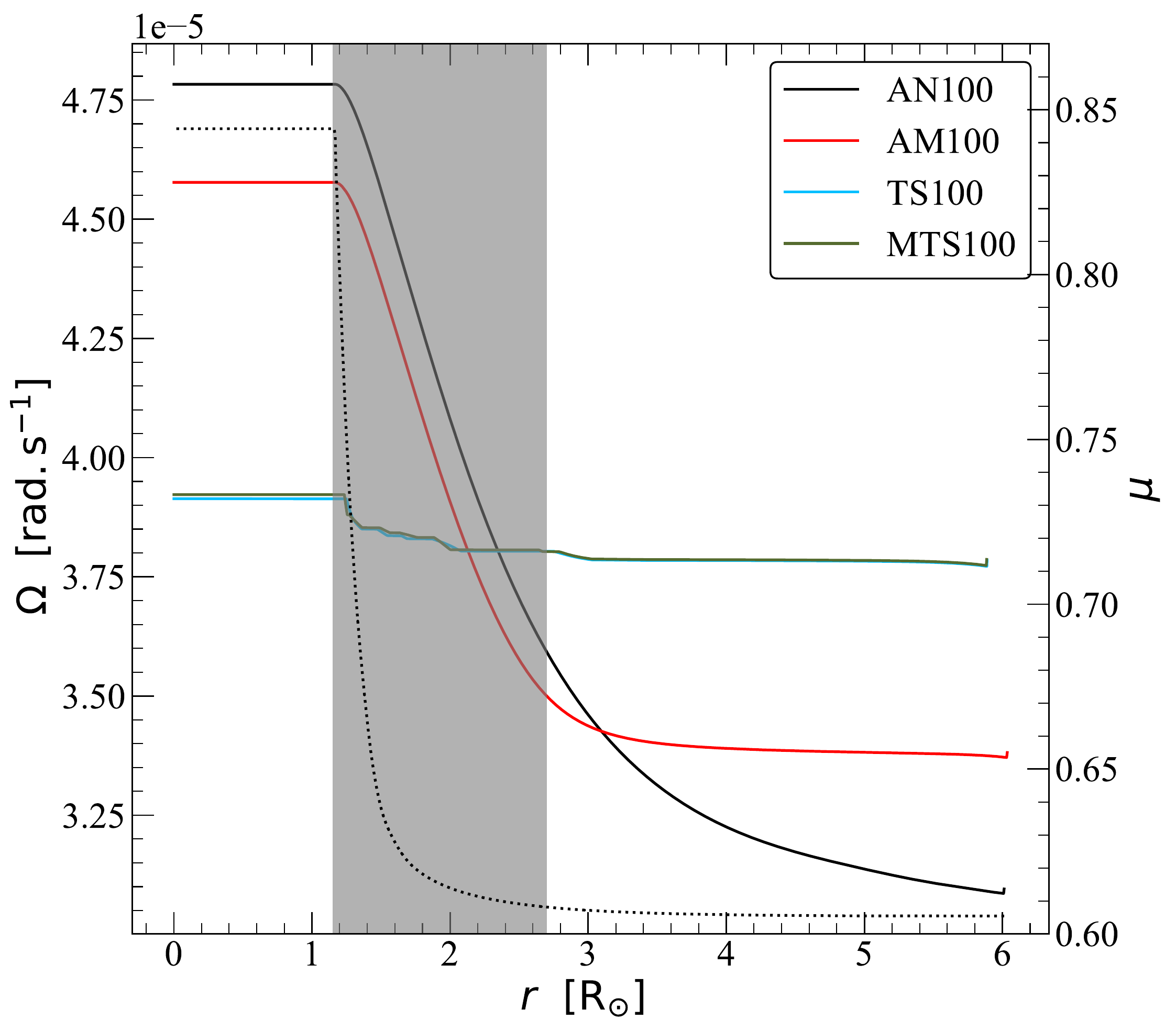}
        \includegraphics[width=0.465\textwidth]{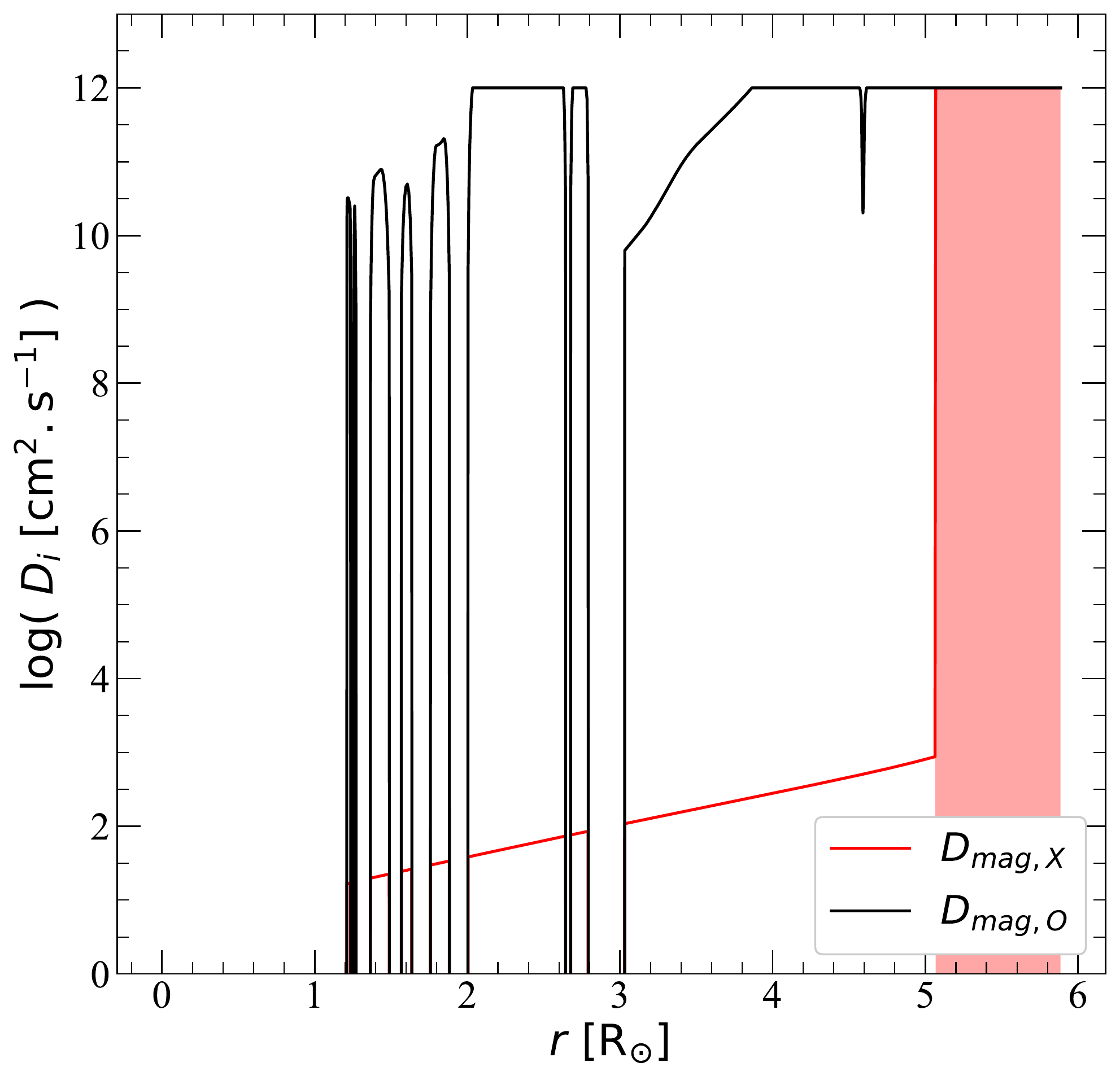}
            }%
    \node at (-9.0,-3.2) {\large (a)};
    \node at (+0.4,-3.2) {\large (b)};
    \end{tikzpicture}
   \caption{Effects of the MRI and the TS dynamo during the MS. {\it Left panel:} Same as in Fig.~\ref{fig:MRI-none-omega-30} but for the set of models discussed in \secref{sec:MRI+TS_MS}.
   {\it Right panel:}  (Restricted) diffusion coefficients for the transport of AM, $D_\mathrm{mag,O}$ (black), and for the transport of chemical elements, $D_\mathrm{mag,X}$ (red), for the sum of both magnetic instabilities in model \modl{MTS100} when the central hydrogen abundance reaches $X_\mathrm{c}=0.35$. The area highlighted in red shows where the MRI is active in the star.
   }
    \label{fig:Omega-35-TS}
\end{figure*}
\subsection{MRI with the TS dynamo during the MS}
\label{sec:MRI+TS_MS}

We computed models including the effects of the TS dynamo alongside the MRI in order to allow a closer comparison to \citetalias{Wheeler_2015}. 
Since differential rotation triggers both instabilities, their interaction may involve non-linear effects of an order difficult to quantify without thorough theoretical work beyond the scope of this paper. The  linear combination we apply here may hold in most cases if the conditions required for the two instabilities do not occur simultaneously, or if they do, the instability that triggers first grows exponentially and dominates the other. 
The TS dynamo triggers before the MRI because Eq.~\eqref{eqn:TSinstab} is less restrictive than Eq.~\eqref{eqn:Instab}. Once activated, the TS, grows at a rate $\gamma_\mathrm{TS}\simeq \omega_A^2/\Omega$.
The magnetic field strength, $B$, builds up so that $\omega_A$ grows eventually making $\gamma_\mathrm{TS}>\gamma_\mathrm{MRI}$ hold. Under such conditions, the TS dynamo is very efficient at transporting AM and decreasing the shear $q$, which limits in turn the possibility for the MRI to trigger afterwards. We will see in the next subsection that approximating the AM transport due to meridional currents by a diffusive equation has qualitatively the same effect.

In this subsection we compare model \modl{TS100},  which includes just the effects of the TS dynamo, with model \modl{MTS100}, which combines the effects of both the TS dynamo and the MRI. In these models we employ a special treatment for meridional circulation. For massive stars, an advective treatment of the circulation alongside the TS dynamo is not straightforward and would make for a cumbersome model to compute. To circumvent this problem, we exclude the effect of meridional currents on the AM transport, preferring to not account for it rather than account for the transport but in the wrong direction. We then only include the impact of meridional currents on chemical elements. As the TS dynamo is already very efficient at transporting AM this choice should not have a strong impact. 

In Fig.~\ref{fig:Omega-35-TS}a we show, as in \secref{sec:MRI_MS_ang}, the rotation profiles of the models halfway through the MS. The models that include the TS dynamo, \modl{MTS100} and \modl{TS100} have flat profiles  and are close to that of solid body rotation. The transport of AM due to the TS dynamo is extremely efficient during the MS. 

As the TS dynamo creates an almost entirely flat rotation profile, the MRI has little impact, so \modl{MTS100} and \modl{TS100} have qualitatively the same rotation profile (lying nearly on top of each other). In Fig.~\ref{fig:Omega-35-TS}b, we show the diffusion coefficients for the AM transport , $D_{\rm mag,O}$, and the transport of chemical elements, $D_{\rm mag,X}$ for model \modl{MTS100}. 
When both instabilities are active, each coefficient is the sum of the individual coefficients associated with each instability. In Fig.~\ref{fig:Omega-35-TS}b we see that, $D_{\rm mag,O}$ reaches large values%
\footnote{As stated in footnote \ref{foot:saturation}, the saturation limit for the diffusion coefficients is set at $10^{12}\,{\rm cm}^2\,{\rm s}^{-1}$}
$\sim 10^{12}\,{\rm cm}^2\,{\rm s}^{-1}$ in almost the entirety of the radiative envelope including up to the core--envelope boundary. The transport of AM is dominated by the TS dynamo as the only area where the MRI is active, shown in red, is very small and restricted to the surface. 

\begin{figure*}
    \centering
    \centering
    \begin{tikzpicture}
    \pgftext{%
        \includegraphics[width=0.5\textwidth]{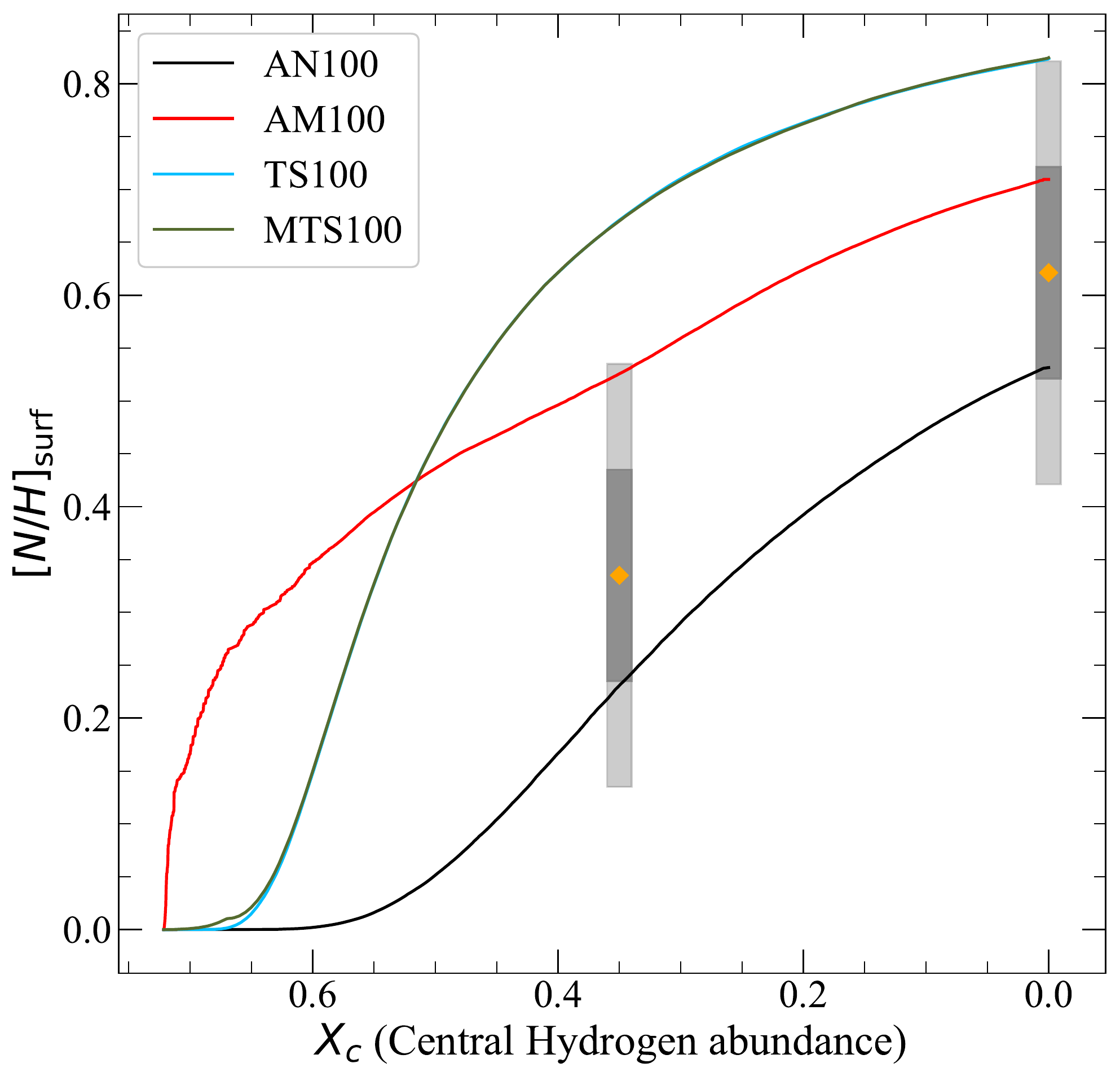}
        \includegraphics[width=0.5\textwidth]{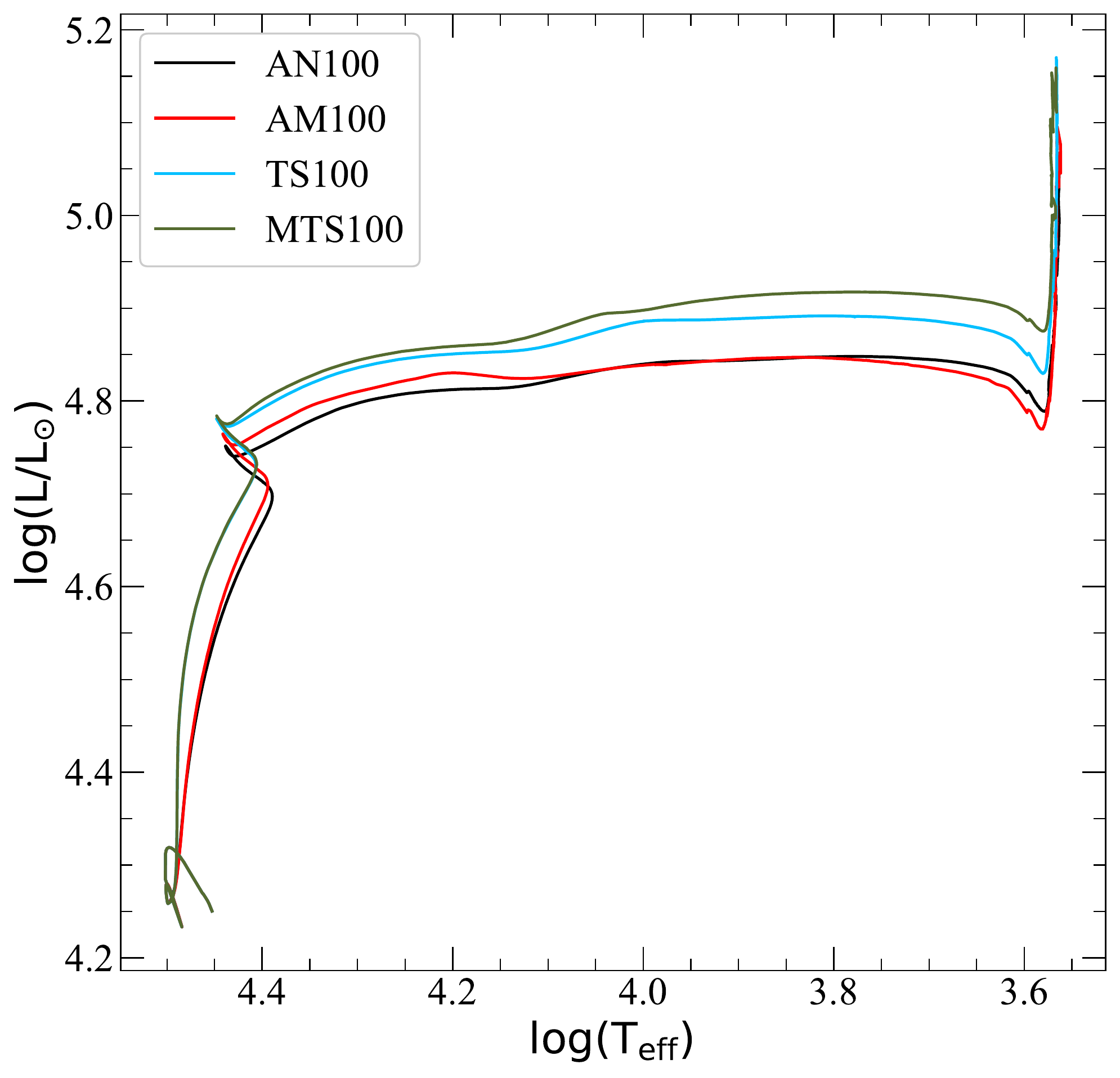}
            }%
    \node at (-9.0,-3.5) {\large (a)};
    \node at (+0.2,-3.5) {\large (b)};
    \end{tikzpicture}
   \caption{Impact of MRI and the TS dynamo on chemical transport during the MS. {\it Left panel:} Same as the right panel of Fig.~\ref{fig:Surface-Abund} but comparing our reference models to those where the TS dynamo is included. {\it Right panel:}  Same as in Fig.~\ref{fig:q-qmin+HR}b but for the models discussed in Sect.~\ref{sec:MRI+TS_MS} (see plot legends). 
   }
    \label{fig:Chemical-TS}
\end{figure*}

The TS dynamo is inefficient at transporting chemical elements. The diffusion coefficient for chemical transport due to TS is equal to the magnetic diffusivity $\eta$. During the MS, this is of order $10^3\,{\rm cm}^2\,{\rm s}^{-1}$, much smaller than the contribution due to the MRI. Figure~\ref{fig:Omega-35-TS}b shows the combined $D_{\rm mag,X}$ of both instabilities. The MRI is active in the red highlighted region, here chemical transport is very efficient but in most of the star the TS dynamo alone is active thus $D_{\rm mag,X} = \eta$ and is negligible compared to other processes. 
In Fig.~\ref{fig:Chemical-TS}a, we again compare the nitrogen enrichment at the stellar surface. $[N/H]_\mathrm{surf}$ grows monotonically in models \modl{TS100} and \modl{MTS100}, both models are identical here due to the fact that the MRI triggers only at the surface in model \modl{MTS100} and thus is not sufficient to dredge up the heavier elements, there is no rapid enrichment caused by the MRI as we saw for model \modl{AM100}. However, the final value reached is larger than in models with MRI alone or no magnetic effects. This is an indirect effect of the TS dynamo.  As the degree of differential rotation in the radiative region decreases significantly, the meridional circulation velocities increase (on average), over the whole radiative region, this leads to a more efficient mixing caused by the circulation. At the chosen initial rotation rate magnetic models do not fit the observational constraints used by \citet[see references therein for observations]{Ekstrom2012},  but as small variations in initial rotation or metallicity change this final value we would need to study a grid of magnetic models to draw effective comparisons with observations. 

In Fig.~\ref{fig:Chemical-TS}b we show the theoretical HR diagram for the models discussed in this section. There are little differences in the global surface properties between models, only that the TS models are slightly more luminous during the MS (roughly 2\%).

\subsection{MRI in purely diffusive models during the MS phase}
\label{sec:MRI+Diffusion_MS}

In \citetalias{Wheeler_2015}, the circulation of meridional currents is treated as a diffusive process. The circulation by meridional currents is treated advectively for the models presented in \secref{sec:MRI_MS_ang}. This leads to a buildup of differential rotation in the star, which is favourable for the MRI to trigger. The diffusive treatment unavoidably destroys the $\Omega$-gradient and is, therefore, unfavourable for the MRI. We compute models \modl{DN100} and \modl{DM100}, which use a diffusive treatment using $D_\mathrm{circ}$ given by Eq.~\eqref{eqn:Dcirc}.

In Fig.~\ref{fig:MRI-none-omega-30}a the diffusive models, \modl{DN100} and \modl{DM100} are shown by the blue and green curves.  
The rotation profiles are almost the same, except close to the surface, due to the fact that the MRI can only trigger close to the surface of the star. These profiles are very flat due to the strong action of meridional currents in the diffusive treatment. Comparing models \modl{AN100} (with an advective treatment) and \modl{DN100} (with a diffusive treatment) we see that the diffusive process yields a significant decrease of $q$ everywhere in the star, with a corresponding slowdown of the core rotating $\sim 20\%$ slower. The very efficient AM transport  due to diffusive meridional currents decreases $q$ fast enough to stop the MRI from triggering during the MS. This behaviour is very similar to the TS models presented in \secref{sec:MRI+TS_MS} where a buildup of differential rotation was suppressed by the activation of TS dynamo. We note that, even in the case where we use a diffusive treatment and set $f_{\mu}=0.05$ (corresponding to a model \modl{DM005} not presented in this paper), the reduction in chemical gradient is not enough for the MRI to significantly trigger as it did in models \modl{AM100} and \modl{AM005}.

In the diffusive models, the transport of chemical elements shows similarities with the models that include the TS dynamo.
The stellar surface is enriched with nitrogen (Fig.~\ref{fig:Surface-Abund}b) due to the flat rotation profile, increasing the efficiency of transport of the chemical elements as for model \modl{TS100} in Sect.~\ref{sec:MRI+TS_MS}. The MRI cannot develop significantly, so we see little impact in \modl{DM100}.
Throughout most of the MS, models employing diffusive meridional currents are slightly more luminous than their advective counterparts (Fig.~\ref{fig:q-qmin+HR}b). However, the HR evolutionary tracks are pretty much the same in both cases. 

In the models of \citetalias{Wheeler_2015}, no significant impact was noted by the MRI during the MS, a fact that we confirm here. Indeed, with the diffusive treatment, the circulation by meridional currents is so efficient that the MRI instability is suppressed almost completely. A reminder that this is not the case for model \modl{AM100} where the impact of the MRI is visible during the MS.

\begin{figure*}
    \centering
    \begin{tikzpicture}
    \pgftext{%
    \includegraphics[width=0.8\textwidth]{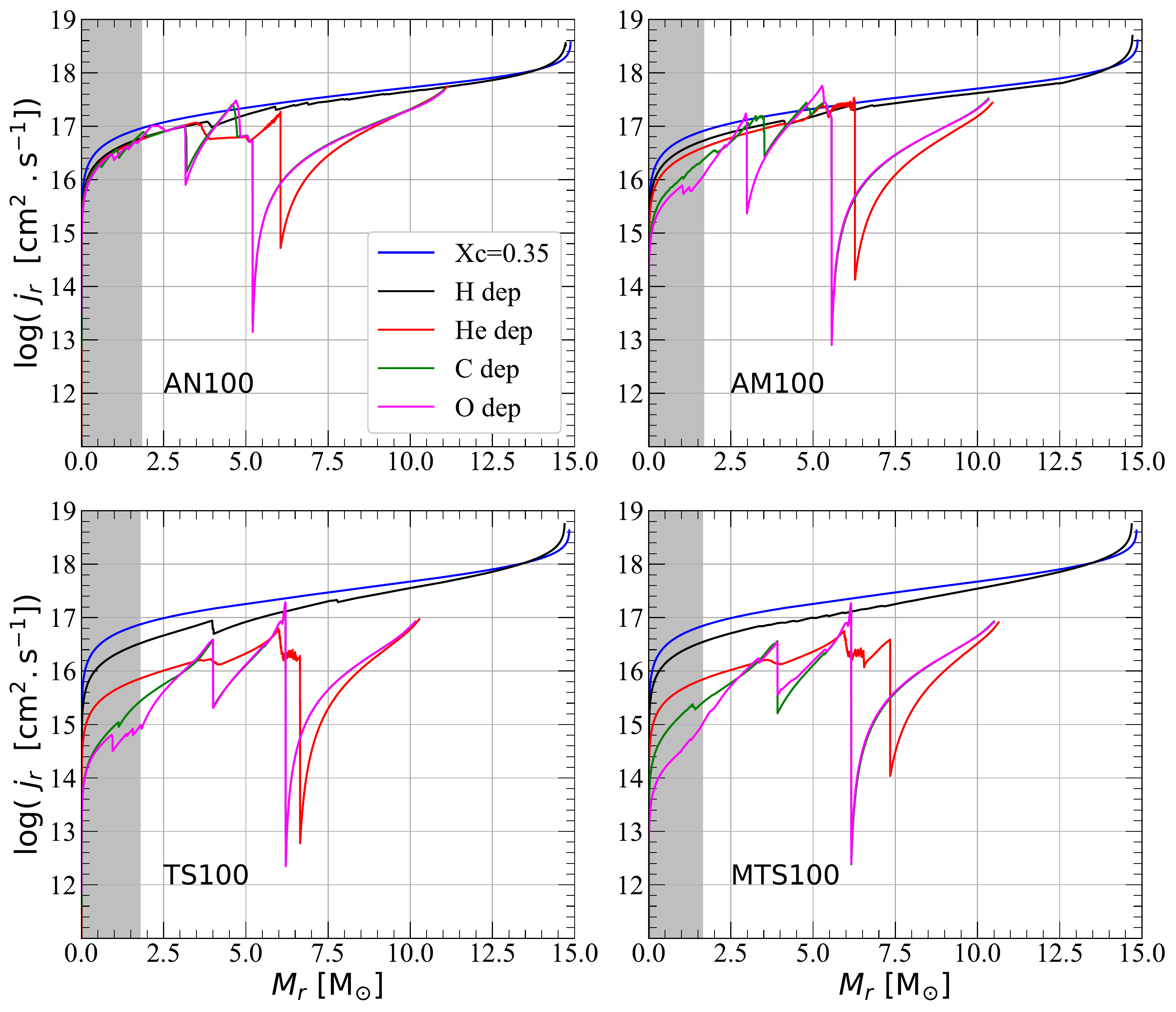}
            }%
    \node at (-7.5,0.5) {\large (a)};
    \node at (+7.5,0.5) {\large (b)};
    \node at (-7.5,-5.0) {\large (c)};
    \node at (+7.5,-5.0) {\large (d)};
    \end{tikzpicture}        
    \caption{Distribution with respect to mass of the specific AM, $j_r$, at different burning stages. Shown are the specific AM midway through the MS, $X_c=0.35$, at the depletion of hydrogen, helium, carbon, and finally oxygen in the core. Models shown are \modl{AN100} (top left), \modl{AM100} (top right), \modl{TS100} (bottom left), and \modl{MTS100} (bottom right). We show a  greyed-out area that highlights the inner region up to mass coordinate $M_4$ as defined in Eq.~\eqref{eqn:M4} and listed in Table.~\ref{table:magnetic}.
}
     \label{fig:j-and-D}
\end{figure*}
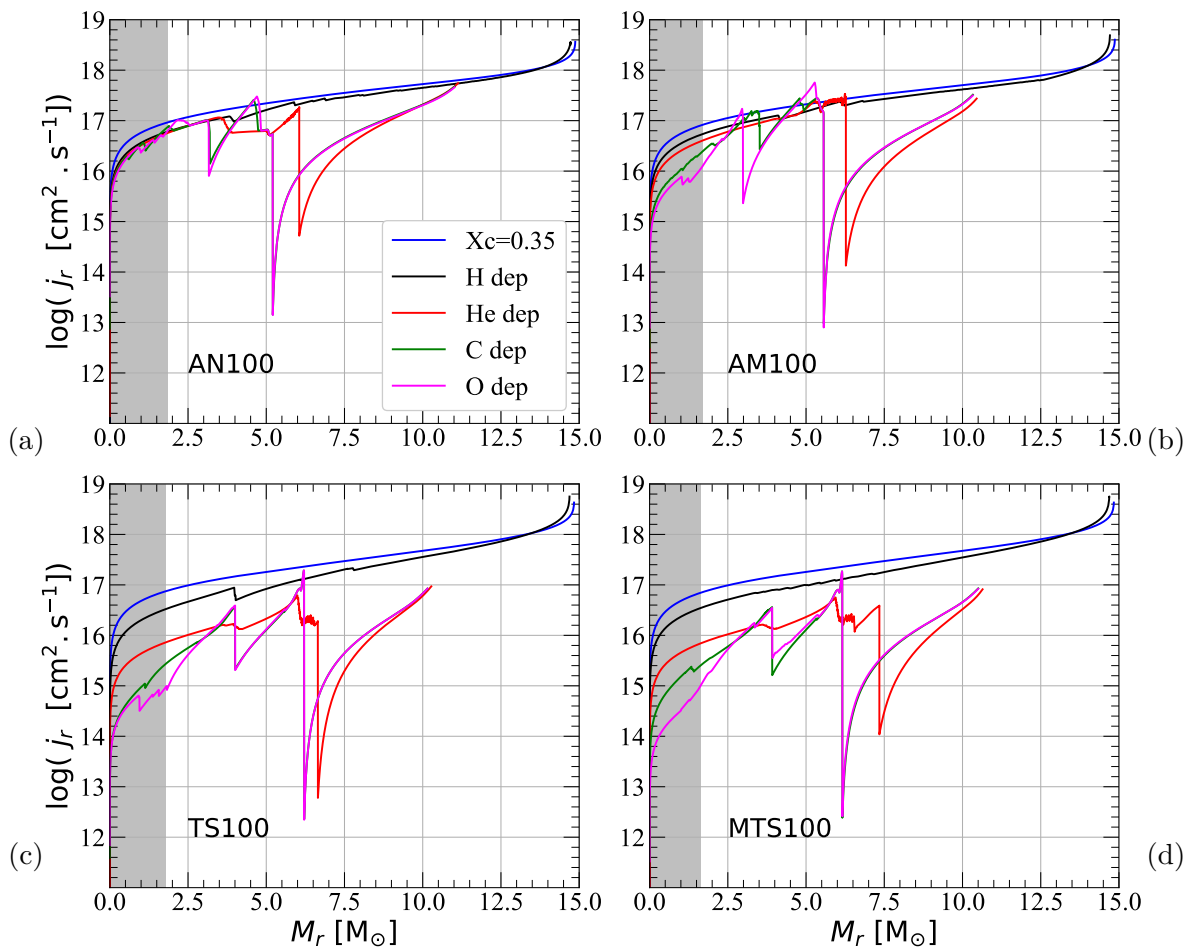

\section{Massive star evolution with MRI in the final stages of evolution}
\label{sec:late_stages}

Now we turn our attention to the effects of MRI in the post-MS heading towards the final stages of stellar evolution. All of our models are run to the end of oxygen burning in the stellar core. In these final stages, the MRI may occur more often because the chemical gradient between shells within the star are not as strong as that between the hydrogen-helium border during the MS. 
Indeed, the jump in the mean molecular weight, is smaller at the boundary between a Carbon shell with an oxygen shell than between a hydrogen shell and a helium one. In \secref{sec:main_sec}, we saw that the chemical gradient was the main inhibitor of the MRI. This gradient decreases the heavier are the fusing elements, and thus the leading term of Eq.\eqref{eqn:N}, $N_{\mu}^2$, decreases significantly. Furthermore, as we proceed to the later stages of evolution, there is a spin up of the inner regions of the star due to contraction increasing $\Omega$. Both the increase in $\Omega$ and the decrease in $N_{\mu}^2$ contribute to the reduction of $q_{\rm min,MRI}$ in the late stages of evolution (see Eq.\eqref{eqn:Instab}). At $M_r = 5 M_{\odot}$ for example $q_{\rm min,MRI}$ goes from order of 100 during the MS to order 0.1 during oxygen burning making it much easier for the MRI to trigger in such regions.
This effect, combined with the fact that compared to other hydro-dynamical instabilities, the MRI acts on a shorter timescale, will culminate in the MRI impacting significantly the post-MS evolution. The MRI affects the chemical structure and the distribution of AM in our models, and both of these factors may change the final fate of the stars, deciding whether either a BH or a neutron star (NS) forms. In this section we look at how exactly the MRI modifies the distribution of AM and chemical elements, and how the magnetic field is structured in the evolved star.

\subsection{Impact of the MRI on the distribution of angular momentum and chemical structure}
\label{sec:MRI-j_post-MS}

In Fig.~\ref{fig:j-and-D} we show the distribution of specific AM, $j_r=r^2\Omega$, with respect to the mass coordinate $M_r$ at different burning stages. The burning stages shown are, midway through the MS, and then at the end of main burning stages in the core (hydrogen, helium, carbon and oxygen). The end of a specific burning stage is defined arbitrarily by the central abundance of the given element going below a threshold value set for each phase as ($X_{\rm end,H}<10^{-5},X_{\rm end,He}<10^{-5},X_{\rm end,C}<10^{-4},X_{\rm end,O}<10^{-3}$).
For the model without magnetic fields (\modl{AN100}; Fig.~\ref{fig:j-and-D}a), in the innermost part of the core (grey shaded zone), which extends to $M_r=M_4$ (see Sect.~\ref{sec:extrapolation} for details on this value), the specific AM varies the most during the MS. After this phase, it remains close to constant in the core. This is not the case for the magnetic models - shown in the 3 other panels - where the specific AM continues to vary in core of the star during the whole the post-MS.
 \begin{figure*}
    \centering
        \begin{tikzpicture}
    \pgftext{%
    \includegraphics[width=0.88\textwidth]{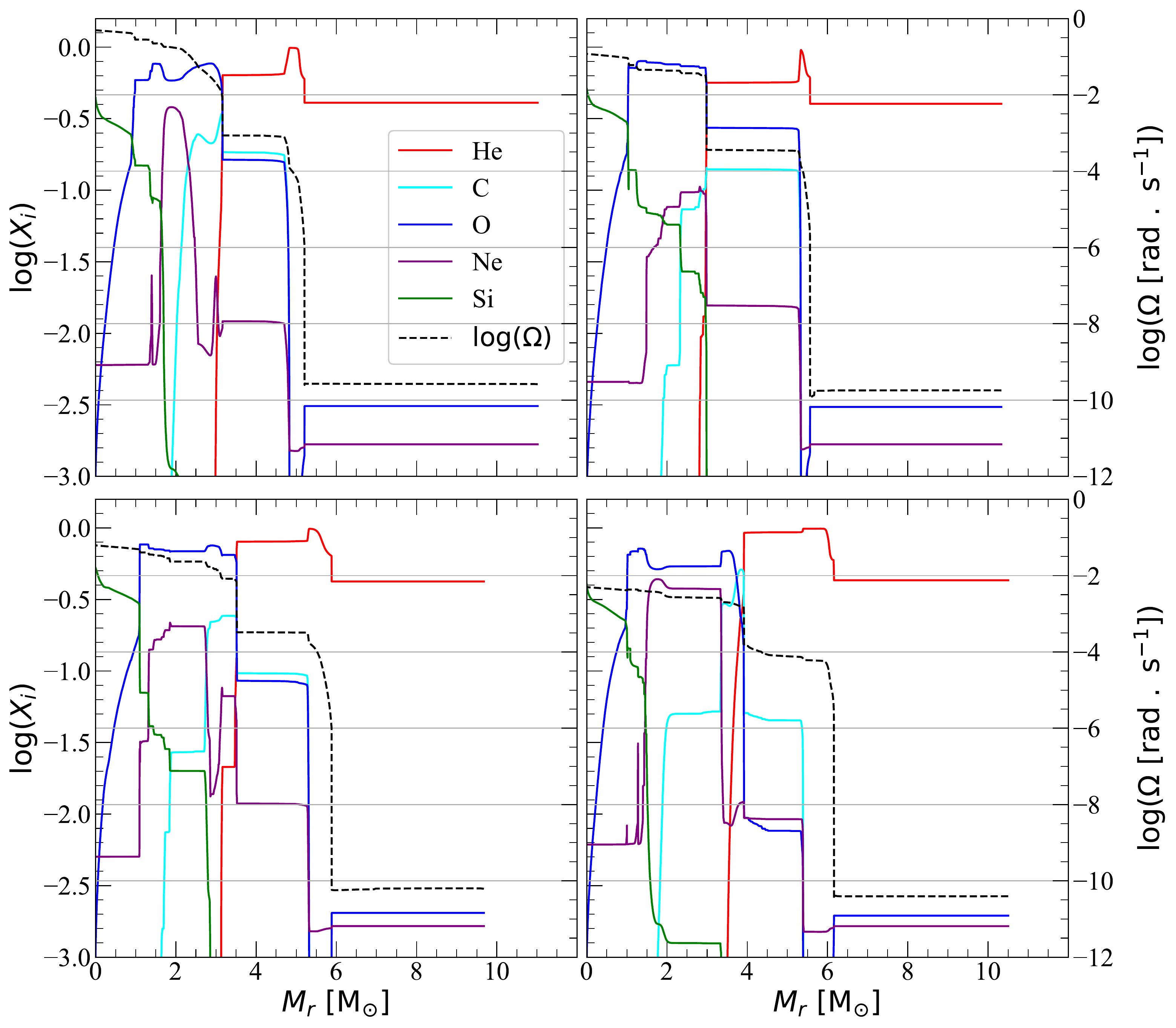}
            }%
    \node at (-8.9,0.4) {\large (a)};
    \node at (+8.8,0.4) {\large (b)};
    \node at (-8.9,-4.4) {\large (c)};
    \node at (+8.8,-4.4) {\large (d)};
    \end{tikzpicture}     
    \caption{Distribution with respect to mass of the composition reached at the end of oxygen burning for models \modl{AN100} (top left), \modl{AM100} (top right), \modl{DM100} (bottom left), and \modl{MTS100} (bottom right). The
    logarithm of the angular velocity, $\Omega$, is shown in dotted lines, and the grid lines shown are those of the axis for $\rm log(\Omega)$}
    \label{fig:Composition-oxygen-burning}
\end{figure*}
 
 For model \modl{AM100}, the most significant reduction of AM in the core occurs between He and C depletion. After that, the specific AM in the inner $\sim 1M_\odot$ remains nearly unchanged, and significantly above $\sim 10^{15}\,\text{cm}^2\,\text{s}^{-1}$. Direct comparison between models \modl{AN100} and \modl{AM100} shows that the MRI has extracted AM from the core into the outer shells, reducing the specific AM by roughly an order of magnitude. In Fig.~\ref{fig:j-and-D}c we show model \modl{TS100}, where we included only the effects of the TS dynamo. This model can also be compared with models existing in the literature, which only incorporate the TS dynamo. For instance, the model 16SG of \cite{Woosley_2006ApJ...637..914} consisting of a solar metallicity star with $M_{\rm ZAMS}=16M_\odot$ initially rotating at $215\,\text{km\,s}^{-1}$ at the equator. This model displays a qualitatively similar reduction of $j_r$ after the MS and the magnitude of $j_r$ is roughly the same in the core (i.e. $10^{14}\,{\rm cm}^2\,{\rm s}^{-1}$). Comparing our models to Fig. 11 of \citetalias{Wheeler_2015}, we find they are broadly in line given the differences between the two codes GENEC and MESA. However, in models such as \modl{AM100}, where the circulation by meridional currents is treated as advective with GENEC, we notice slightly higher values of specific AM in the core at the end of oxygen burning when compared to \citetalias{Wheeler_2015}. This demonstrates that transport of AM by meridional currents is much more efficient if treated as diffusive as the constructive advective motions cannot develop, so shear is always destroyed. Although true that the MRI will trigger more in model \modl{AM100} than in \modl{DM100} due to these developing $\Omega$-gradients, the overall effect of the larger transport of AM in diffusive models still yields slightly slower rotating cores.
 When we add the effects of the TS dynamo, the slowdown of the core is significantly larger \citepalias[in agreement with][]{Wheeler_2015}. The TS dynamo dominates the transport of AM due to its less stringent instability condition (see \secref{sec:MRI+TS_MS}).
 In our case, model \modl{MTS100} shows the largest slowdown of the core, consistent with the literature.%
 \footnote{We also directly compare \modl{MTS100} with \citetalias{Wheeler_2015}. At the end of oxygen burning, the specific AM for the Wheeler model reaches values of $10^{13} - 10^{14}\,\text{cm}^2\,\text{s}^{-1}$ as does \modl{MTS100}.}
 The MRI contributes to slower rotation rates, even when added alongside the TS dynamo.  We have roughly half an order of magnitude drop of the specific AM inside the core (in $m\lesssim 1 M_\odot$) and a smoother profile predicted in \modl{MTS100} (with a rough dependence $j_r\propto M_r^{2.2}$ in the interval $0.5\lesssim M_r/M_\odot \lesssim 3.4$, Fig.~\ref{fig:j-and-D}d) compared with \modl{TS100} (Fig.~\ref{fig:j-and-D}c) at the end of the oxygen burning phase.
Despite TS largely dominating AM transport, as seen during the MS, the MRI continues to trigger during advanced burning stages, specifically in the core and contributes to the total AM transport throughout the star. Thereby, the MRI should not be disregarded in favour of TS, since both may be important to correctly predict the specific AM distribution.

We anticipate that the relative change in $j_r$ as different burning phases occur may depend on metallicity, in as much as $Z$ is connected to the mass-loss rate of the star \citep{Woosley_2006ApJ...637..914}, since additional mass loss at the surface decreases the total AM of the star. In the models that include the TS dynamo, the AM transport is so efficient that there is a strong connection between surface and core in the first burning phases, (hydrogen and helium),  and loss of AM at the surface would be rapidly felt in the core. During the last burning stages a disconnection between core and envelope occurs and mass loss rates there should not impact significantly the core rotation rates.

In Fig.~\ref{fig:Composition-oxygen-burning}, we show the chemical structure of the models \modl{AN100}, \modl{AM100,} \modl{DM100,} and \modl{MTS100,} which we overlay with the rotation profile, $\Omega$ (dashed black line). There are a number of subtle differences in the chemical structure between these models, which depend on the details of the evolution.
 For example, we can compare the oxygen shell, extending from $1$ to $3M_{\odot}$ in \modl{AM100} ({Fig.~\ref{fig:Composition-oxygen-burning}b}) or up to $\sim 3.5M_\odot$ in model \modl{DM100}.
Models including only the MRI, tend to show piecewise monotonic abundances, with sharp discontinuous composition transitions. The inclusion of the TS dynamo indirectly impacts mixing by flattening the rotation profile and enhancing other hydro-dynamical mixing processes. But, as we have discussed in \secref{sec:MRI+TS_MS}, the TS dynamo has a weak effect on chemical transport directly. In Fig.~\ref{fig:Composition-oxygen-burning}d, we show model \modl{MTS100} and, while not hosting a very different structure than the other models, we do notice a larger oxygen shell and a greater quantity of neon and carbon mixed homogeneously. 

Overlapped over the chemical composition is  log ( $\Omega$ ) where we can see the rotation rates of each of the shells. Model \modl{AN100} displays the largest differential rotation of all models. The inclusion of the MRI reduces significantly the differential rotation, and the addition of the TS dynamo leads to an almost flat rotation profile in the inner 4 $M_{\odot}$ at the end of oxygen burning.  In Table.~\ref{table:Wheeler_comparaison} we compare the core rotation rates of some of our models (without an advective treatment of the meridional circulation) with those published by \citetalias{Wheeler_2015}. With the caveat that our models are computed only to the end of oxygen burning whereas the values of \citetalias{Wheeler_2015} are at the pre-supernova (pre-SN) link. That difference is important since, we expect that in between the end of oxygen burning and the pre-SN link two counteracting effects modify $\Omega_{\rm core}$. On the one hand, the  transport of AM will induce the decrease of $\Omega_{\rm core}$. On the other hand, further core compression stimulates the growth of $\Omega_{\rm core}$. Nonetheless, the central rotation rates of our models including magnetic instabilities are relatively close, but systematically smaller than in \citetalias{Wheeler_2015}. A moderate core contraction from the end of the oxygen burning to the pre-SN link may yield values closer those of \citetalias{Wheeler_2015}. Qualitatively, we find the same trend for $\Omega_{\rm core}$ as \citetalias{Wheeler_2015}. The lowest rotation rate is with MRI and TS dynamo implemented simultaneously, while the largest $\Omega_{\rm core}$ corresponds to the model without the TS dynamo. 
\begin{table}
\centering
    \begin{tabular}{||c c c  ||}
        \hline
        Model & \small{\citetalias{Wheeler_2015}} & \small{This paper} \\ [0.5ex] 
        \hline 
        \hline
        \small\modl{DM100} & -0.7 & -1.2 \\
       \small \modl{TS100} & -1.7 & -1.9  \\
        \small\modl{MTS100} & -1.9 & -2.3 \\[0.5ex] 

        \hline
        \hline
    \end{tabular}
\caption{Comparison of the values of $\log(\Omega_{\rm core})$ in $\rm s^{-1}$  between models computed in this paper at the end of oxygen burning and the equivalent model from \citetalias{Wheeler_2015} (corresponding to the pre-supernova link).
$\Omega_{core}$ corresponds to the value of $\Omega$ in the central point, which is representative of the whole core, since cores are nearly rigidly rotating at this evolutionary stage.} 
\label{table:Wheeler_comparaison}
\end{table}

\subsection{Magnetic field structure and strength}
\label{sec:B-field_topology}
 \begin{table*}[ht!]
\centering
  \begingroup
  \setlength{\tabcolsep}{4pt} 
    \begin{tabular}{||c  c c c c c c c c c c c c c ||}
        \hline 
Model &  $\mathcal{B}_\mathrm{t}$ &  $\mathcal{T}_\mathrm{t}$ &$\mathcal{B}_\mathrm{t}/\mathcal{T}_\mathrm{t}$ & $M_\mathrm{mag}$ & $M_\mathrm{t}$ &$\mathcal{P}_{\rm rm}$ & $\lambda_{2.5}$ & $\xi_{2.5}$ & $M_4$ & $\mu_4$ &  $\mathcal{T}_{4}$ & $a_{\rm rm}$ & remn.\\ 
 & [$10^{45}$\,erg] & [$10^{48}$\,erg] & $[\%]$ & $[M_\odot]$&  $[M_\odot]$& $[\text{ms}]$  & & & & & [$10^{46}$\,erg] & &\\ [0.5ex]
        \hline 
        \hline
\modl{ AN100} &      0 &     25.1 &     0 &  0 &  11.0 & 0.18 &    0 &  0.18 &    1.85 &   0.117 &    240 & (3.8) & BH\\
\modl{AM100H} &      92.7 &     2.87 &     3.2 &  1.56 &  10.4 & 0.82 &  0.024 &  0.16 &    1.69 &   0.104 &    8.79 & 0.75  & BH\\
\modl{ AM100} &      99.4 &     3.41 &     2.9 &  1.36 &  10.3 & 0.76 &  0.040 &  0.17 &    1.69 &   0.119 &    10.6 & 0.81 & BH\\
\modl{AM100L} &      77.2 &     3.01 &     2.6 &  2.09 &  10.2 & 0.75 &  0.027 &  0.17 &    1.74 &   0.125 &    9.79 & 0.85 & BH\\
\modl{ DM100} &      9.96 &     1.54 &    0.65 & 0.60 &  9.68 & 0.92 &  0.015 &  0.17 &    1.77 &   0.106 &    5.90 & 0.70 & BH\\
\modl{ AM005} &      2.08 &   0.10 &     2.2 & 0.80 &  10.5 &  2.51 & 0.004 & 0.01 &    1.78 & 0.005 &   0.97 & 0.26  & NS\\
\modl{ DN100} &      0 &     4.77 &     0 &  0 &  10.2 & 0.38 &    0 &  0.17 &    1.86 &  0.096 &    46.3 & (1.8) & BH\\
\modl{ TS100} &   0.005 &   0.05 &  0.01 &  4.05 &  10.2 &  9.96 &  0.068 &  0.14 &    1.80 &  0.070 &  0.07 & 0.07 & NS\\
\modl{MTS100} &    0.012 &   0.06 &   0.02 &  5.92 &  10.5 &  11.8 &   0.120 &  0.15 &    1.65 &  0.090 &  0.04 & 0.05 & BH \\
        \hline
        \hline
    \end{tabular}
    \endgroup
\caption{Properties of our models at the end of oxygen burning. For each model we list at the end of oxygen burning the total magnetic energy, $\mathcal{B}_t$, the total kinetic energy, $\mathcal{T}_t$, and the ratio between magnetic and kinetic energy. We also list the mass of the star that is magnetised, i.e. one or two magnetic instabilities are active, $M_{\rm mag}$, the total mass, $M_{\rm t}$, the rotational period of the hypothetical NS produced from the baryonic mass $\mathcal{P}_{\rm rm}$ (see Eq.~\eqref{eqn:PNS}), $\lambda_{2.5}$ the ratio of the average coherence length of the poloidal magnetic field within the inner $2.5M_\odot$. The ensuing five columns show the compactness ratio $\xi_{2.5}$ (Eq.~\eqref{eqn:Compactness}), the parameters $M_4$ (Eq.~\eqref{eqn:M4}), and $\mu_4$ (Eq.~\eqref{eqn:mu4}),  the kinetic energy contained within $M_4$, $\mathcal{T}_{4}$, and the dimensionless spin of the compact remnant, $a_{\rm rm}$ (Eq.\eqref{eqn:a4}; in cases in which it is larger than 1, we annotate it with the number in parenthesis). Finally, the last column displays whether the compact remnant predicted according to \cite{2016-Ertl-Janka} criterion is a BH or an NS.
}
\label{table:magnetic}
\end{table*}

Most stellar models do not incorporate the magnetic field as a dynamic variable \citep[but see e.g.][]{Takahashi_2021A&A...646A..19}. Instead, the magnetic field is estimated as resulting from the saturation of the growth of either the MRI or the TS dynamo in radiative layers of the star. Furthermore, the estimated saturation fields ($\overline{B_r}$ and $\overline{B_\phi}$) do not satisfy the solenoidal constraint ($\nabla\cdot \overline{\mathbf{B}}=0$). %
As a result, both magnetic field components are extremely variable in time and in radius, and non-magnetised stellar shells exist intertwined with other magnetised layers.
Admittedly, this is not a restriction for secular stellar evolution calculations. Both the MRI and the TS dynamo yield small-scale magnetic fields, whose stochastic/turbulent nature can effectively be included as a sub-grid-scale model in the stellar evolution calculation, which plays the role of a Large Eddy Simulation. In practical terms, the unresolved scales (below the scale of the Lagrangian grid size) are accounted in the governing equations through suitable models for the diffusion coefficients (see Eqs.~\ref{eqn:D}, \ref{eqn:D2} or \ref{eqn:D-circ}). 

\begin{figure*}
    \centering
        \begin{tikzpicture}
    \pgftext{%
    \includegraphics[width=0.8\textwidth]{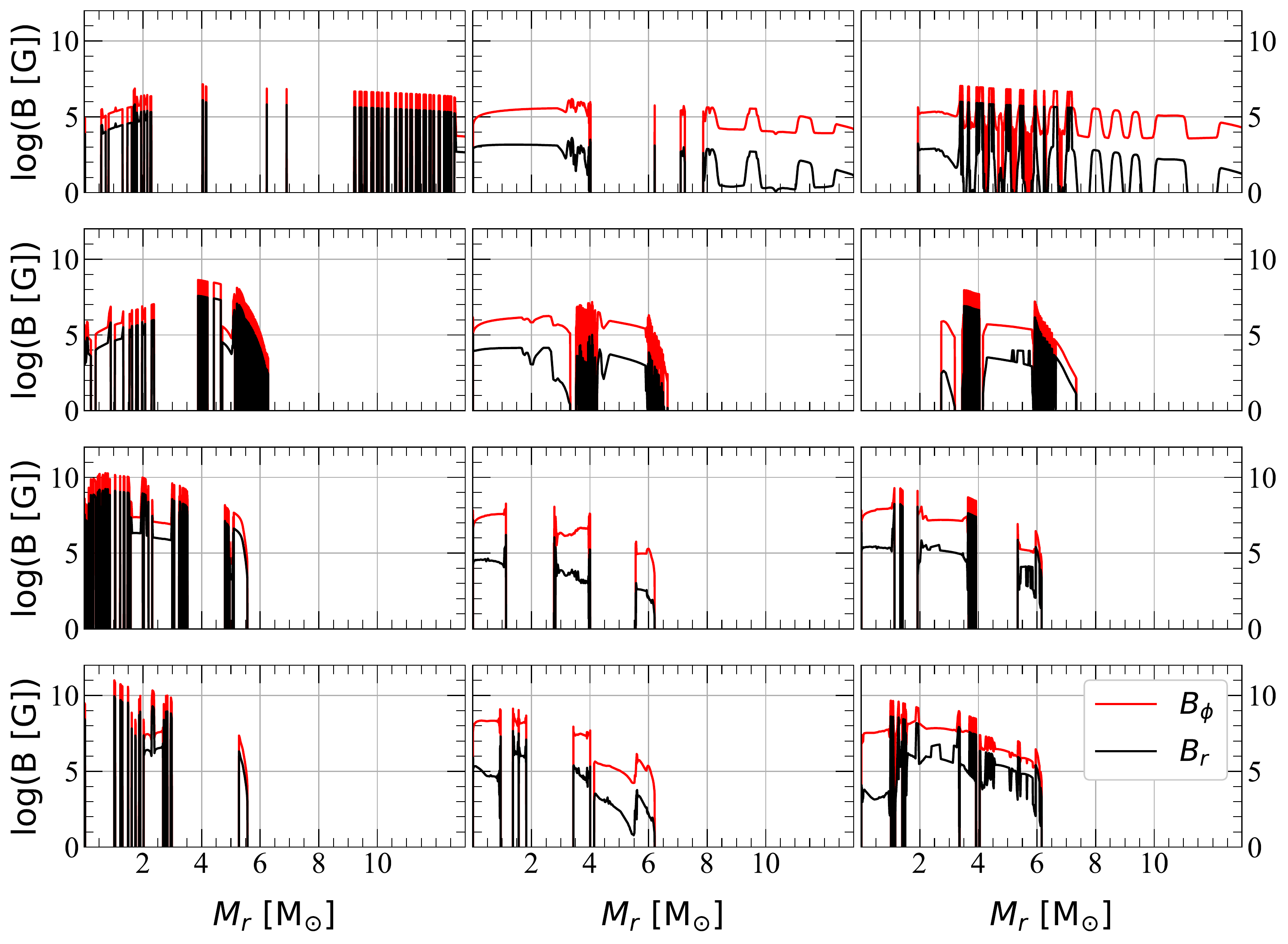}
            }%
    \node[fill=white, opacity=1, text opacity=1]  at (-5.7,4.8) {\large \modl{AM100}};
    \node[fill=white, opacity=1, text opacity=1]  at (-1.3,4.8) {\large \modl{TS100}};
    \node[fill=white, opacity=1, text opacity=1]  at (+3.2,4.8) {\large \modl{MTS100}};
    \end{tikzpicture}         

  \caption{Distribution with respect to mass of the toroidal magnetic field (in red) and the radial magnetic field (in black). We plot the magnetic field at the end of hydrogen burning (first row), at the end of helium burning (second row), at the end of carbon burning (third row), and at the end of oxygen burning (final row). Shown are model \modl{AM100} (left column), model \modl{TS100} (middle column), and \modl{MTS100} (right column). }
    \label{fig:Magnetic-fields-TS-MRI}
\end{figure*}

In \figureref{fig:Magnetic-fields-TS-MRI}  we show the saturation values
of both the azimuthal and poloidal  magnetic fields given by Eqs.~(\ref{eq:Bphi-saturation}, \ref{eq:Br-saturation}) for the MRI and Eqs.~(\ref{eqn:Bphi-TS}, \ref{eqn:Br-TS}) for the TS dynamo. 
To compute the aforementioned saturation fields, we limit the value of $|q|<1$. Due to the condition that $\omega_A \ll \Omega$ for both instabilities to be active and that at saturation of the magnetic fields we should have $\omega_A\sim q \Omega$ then for estimating the magnetic fields $|q|$ should not be larger than 1.\footnote{We note that this limitation of $q$ is only done when post-processing the magnetic field, and we do not cap the shear to 1 in the simulation.  Throughout the evolution, there are instances where (artificially) $|q|$ becomes significantly larger than 1, when a mass shell meets the triggering condition for the instability (thus reducing $\Omega$) while its neighbouring cells do not.} The fields are plotted at the end of different burning phases. When the TS dynamo is included, larger regions of the star are magnetised compared to the MRI only model, this is consistent with \citetalias{Wheeler_2015}. Most importantly, the core in \modl{TS100} and \modl{MTS100} is magnetised at the end of oxygen burning (compare the inner $\sim 1.4M_\odot$ in model \modl{AM100}, non-magnetised, with the rest of the panels in the final row of \figureref{fig:Magnetic-fields-TS-MRI}). 
At this stage, the decay timescale of the magnetic field due to reconnection is still much shorter than the evolution timescale. However, as evolution continues these timescales may become of similar orders and magnetised regions may stay as such even once the instability ceases to operate. In our numerical treatment,
the magnetic field is only computed when an instability is triggered. Thereby, as the TS dynamo is active more extensively, naturally, more regions are magnetised. In the case of the MRI, activity is more sporadic, but if the MRI triggers and then ceases to verify the instability condition - due to reduction of the shear in the zone - a strong magnetic field may still persist in the region afterwards. To assess this further, a consistent mapping of magnetic fields throughout the evolution is required, which is not implemented in stellar codes.
 
The structure of the magnetic field, and the ratio between $B_r$ and $B_{\phi}$, when the TS dynamo is included, is consistent with that in existing models in the literature including the TS dynamo \citep{Heger_et_al__2005__apj__Presupernova_Evolution_of_Differentially_Rotating_Massive_Stars_Including_Magnetic_Fields, Aguilera-Dena_2018ApJ...858..115}, 
in which the inner core of the star ($M_r\lesssim 2M_\odot$) is magnetised and, eventually, one or more extended layers around the core are magnetised as well. The magnetic field structure may have an impact on the character of the supernova explosion at the very end of evolution, which we comment upon in Sect.~\ref{sec:explosion}.  

We also note in  \figureref{fig:Magnetic-fields-TS-MRI} that neither models including the MRI nor the TS dynamo predict any surface magnetic fields after the end of the MS. This is in contrast to the fact that the outermost layers of our models are magnetised through most of the MS (see upper panels in \figureref{fig:Magnetic-fields-TS-MRI}). Indeed, we notice a significant difference in the radial magnetic field component at the surface among models that include the TS dynamo and models with only the MRI included. The former can yield r.m.s. values $\overline{B_r}\sim 10\,$G, while the latter produce $\overline{B_r}\sim 10^3\,$G. How much of the r.m.s. radial magnetic field translates into a net dipolar component ($B_\mathrm{dip}$) is difficult to predict. However, this difference in the surface values may promote the MRI as a possible ingredient to explain the relatively large dipolar magnetic fields observed in some galactic OB MS stars \citep[e.g. $0<B_\mathrm{dip}< 1.6\times 10^4\,G$][]{Aerts_2014}.

\section{Extrapolation of the results after core collapse}
\label{sec:extrapolation}
In this section, we extrapolate the results obtained after oxygen depletion to predict whether the incorporation of the MRI may change the nature of the compact remnant ensuing core collapse, and provide a forecast of the explosion type. We anticipate that our extrapolations should be taken with care, since we have not evolved our models to the brink of collapse. Also, the theory bridging from rotating and magnetised stellar cores to their compact remnants is not well developed yet.

For a better characterisation of the magneto-rotational properties of the evolved stars, we compute the total magnetic energy, $\mathcal{B}_{\rm t}$, and the total kinetic energy, $\mathcal{T}_{\rm t}$ contained in our models. The total kinetic energy is the energy that served to create the magnetic fields through the extraction of energy from differential rotation. We use the expressions\footnote{We considered that the angular average of the specific rotational energy, $\varepsilon_{\rm rot}=(1/2)r^2\Omega^2\sin^2\theta$, is $\langle \varepsilon_{\rm rot}\rangle=(1/3)r^2\Omega^2$, and that $\mathcal{T}_\mathrm{t}=\int_0^{M_t}\langle\varepsilon_{\rm rot}\rangle dm$}%
\begin{equation}
\label{eqn:Erot}
    \mathcal{T}_\mathrm{t}=\frac{1}{3}\int_0^{M_{\rm t}} \Omega^2 r^2 \mathrm{d}m\,,
\end{equation}
where $M_{\rm t}$ is the total stellar mass, and
\begin{equation}
\label{eqn:Emag}
    \mathcal{B}_\mathrm{t}=\int_0^{M_{\rm t}} \frac{B_r^2 + B_{\phi}^2}{8\pi\rho}  \mathrm{d}m\,.
\end{equation}
In the case of the magnetic energy, we assumed that the saturation magnetic fields already correspond to polar angular averages, that is,  not assumed any particular large-scale structure of the magnetic field.
Table~\ref{table:magnetic} lists the values of the total magnetic energy, rotational energy, and their ratios $\mathcal{B}_\mathrm{t}/\mathcal{T}_\mathrm{t}$. The star containing the largest magnetic energy is our standard model, \modl{AM100}, with $\alpha=0.02$ and the circulation by meridional currents being dealt with as advective. For all models, the magnetic energy is a small fraction of the rotational energy  ($\mathcal{B}_\mathrm{t}/\mathcal{T}_\mathrm{t}<4\%$), off which it feeds.\footnote{More precisely, the magnetic field feeds off the free or shear rotational energy of the star, which is $1-2$ orders of magnitude smaller than $\mathcal{T}_t$.} Models that include the MRI only display a significantly larger magnetisation ($\mathcal{B}_\mathrm{t}/\mathcal{T}_\mathrm{t}\sim 0.7\%-3\%$) than models where the TS dynamo is included ($\mathcal{B}_\mathrm{t}/\mathcal{T}_\mathrm{t}\gtrsim 0.02\%$). The smaller magnetisation of the TS dynamo models happens because of their significantly smaller magnetic energy, which is not compensated by the fact that the total rotational energy is about two orders of magnitude smaller than in models where only the MRI is included.

\subsection{MRI influence on the type of compact remnant}
\label{sec:final-fate}

Transport of AM and chemical elements caused by MRI, the TS dynamo and a combination of both, impacts the type of compact object that our models may produce after core collapse. We computed the evolution up to the end of oxygen burning and not the pre-SN link. Furthermore, the dynamics of the evolution post-bounce may change the fate of the compact remnant, turning a PNS into a BH. Thus, any predictions we make in this section on the nature of the compact remnant must be taken with due caution. However, comparing models among themselves can give an idea of how a magnetic instability may affect the outcome of the last stages of the star. In the following, we consider a number of parameters to estimate the nature of the compact remnant. 
A first parameter is the compactness. \cite{O_Connor_2011} defined it as a normalised enclosed mass-to-radius ratio computed at the time of bounce, namely,
\begin{equation}
\label{eqn:Compactness}
    \xi_M = \left.\frac{M/M_{\odot}}{R(M_{\rm b}=M)/1000\text{ km}}\right\vert_{t}
,\end{equation}
where $R(M_{\rm b}=M)$ is the radial coordinate that encloses a baryonic mass at epoch $t$. The typical mass scale chosen when employing the compactness to make a forecast of the compact remnant type is $M=2.5M_{\odot}$; therefore, we estimate $\xi_{2.5}$ for all models. 
The compactness could change (either growing or decreasing) in all of our models as they approach the point of core collapse. \cite{Sukhbold_2014} have shown that the evolution of $\xi_{2.5}$ is very moderate after oxygen depletion, for solar metallicity, non-rotating models with masses below $\sim 18M_\odot$. This is because the $2.5M_\odot$ mass shell is sufficiently far from the centre to not undergo a very substantial contraction after oxygen depletion. Singularly, for models with $15M_\odot$, \cite{Sukhbold_2014} find that most of the increase in $\xi_{2.5}$ happens after Silicon depletion, when it grows by less than $\sim 8\%$ (see their Figures 14 and 15). 
Hence, our value of $\xi_{2.5}$ after the end of oxygen burning, may underestimate by less than, say $\sim 30\%$ this parameter at the pre-SN link.

According to \cite{O_Connor_2011}, if $\xi_{2.5}>0.45$ at the time of bounce, the core collapse yields the formation of BHs without neutrino-driven explosions.  This limit was lowered by \cite{Ugliano_2012ApJ...757...69} to $\xi_{2.5}>0.35$ including the long-term effects of the fall-back. \cite{Sukhbold_2014} showed that not too much accuracy is lost when evaluating $\xi_{2.5}$ at the pre-SN link. 
At the end of oxygen burning, all our models lead to a compactness of $\xi_{2.5} <0.18$. These values are up to $\sim 27\%$ larger than in, for example, \cite{Sukhbold_2014} using the KEPLER code (cf. their Figure 14, for a model computed with solar composition, $15M_\odot$ and at the end of oxygen burning). The compactness of our models compares quite well to the results of  \cite{Sukhbold_2018ApJ...860...93} and \cite{Chieffi_2020ApJ...890...43} for  $15M_\odot$ and a solar metallicity. Even accounting for some evolution until the pre-SN link, all computed models may have values of $\xi_{2.5}$ in a range that may either produce NSs or BHs \citep[][find both explosions NS or BH formation for $0.15\le \xi_{2.5}\le 0.35$]{Ugliano_2012ApJ...757...69}. However, when including the combined effects of rotation and magnetic fields, the compactness shows some limitations to predict the remnant type \citep{Obergaulinger_2021MNRAS.503.4942}.
 
Models without magnetic fields develop the most compact cores (compare $\xi_{2.5}$ for models \modl{AN100} or \modl{DN100} with the rest in Table \ref{table:magnetic}). The action of the TS dynamo (alone or in combination with MRI) yields slightly less compact cores than models including the MRI only. 
A minuscule compactness results if one artificially lowers $f_\mu$ in GENEC models ($\xi_{2.5}^{\modl{AM005}}\approx 0.01$). 
This low value of $f_{\mu}$  allows the MRI to trigger more in the star than in other models, and thus homogenises very well the different layers of the star. Again, we recall that $q_{\rm min,MRI}\propto N_{\rm eff}^2$ and that the chemical gradient term dominates the effective buoyancy frequency. Thus, lowering the contribution of $N_\mu$ to $N_{\rm eff}$ facilitates the trigger of MRI. This model thus develops a large homogeneous region from $M_r \in [2 M_{\odot} ; 5 M_{\odot}]$, in this region the mass gradient is very small and therefore the radius $R_{2.5}$ will be larger than in other models. This explains the small compactness. We insist however that this model has only an academic and comparative value with previous work, and should not be taken as a realistic one. Indeed, the unusually low compactness adds up to the arguments against using $f_\mu<1$ in models where meridional circulation is treated as an advective process.
 
\cite{2016-Ertl-Janka} have pointed out the ambiguities of $\xi_{M}$ as a good diagnostic tool for deciding the type of compact remnant. As an alternative, they suggested using a two-parameter criterion to evaluate whether a model would explode and form an NS or fail to do so and collapse into a BH. This two-parameter criterion is computed directly from the pre-SN profiles 
(instead of computing $\xi_{2.5}$ exactly at the time of core bounce as proposed by \citealt{O_Connor_2011}).
The relevant quantities in this parameterisation are the normalised mass inside a dimensionless entropy per nucleon of $s=4k_{\rm B}$ ($k_{\rm B}$ is the Boltzmann constant\footnote{The evaluation of the entropy per baryon in our models uses the Helmholtz equation of state \citep{Timmes_2000ApJS..126..501}.}),
\begin{equation}
\label{eqn:M4}
    M_4 \equiv M_r(s=4k_{\rm B})/M_{\odot},
\end{equation}
and the mass derivative at this location,
\begin{equation}
\label{eqn:mu4}
    \mu_4 \equiv \frac{\text{dm}/M_{\odot}}{\text{dr}/1000 \text{km}}\Bigg|_{s=4k_{\rm B}}.
\end{equation}
Ertl's criterion consists of comparing $\mu_4$, which is akin to the mass accretion rate, with the product $M_4\mu_4$, 
which is a proxy for the neutrino  luminosity induced by mass accretion $\mathcal{L}_{\nu}^{\rm acc}$. This leads to a separation of models that produce NSs and models that produce BHs by finding which models successfully explode and which ones fail.
\begin{figure}[H]
    \centering
    \includegraphics[width=0.5\textwidth]{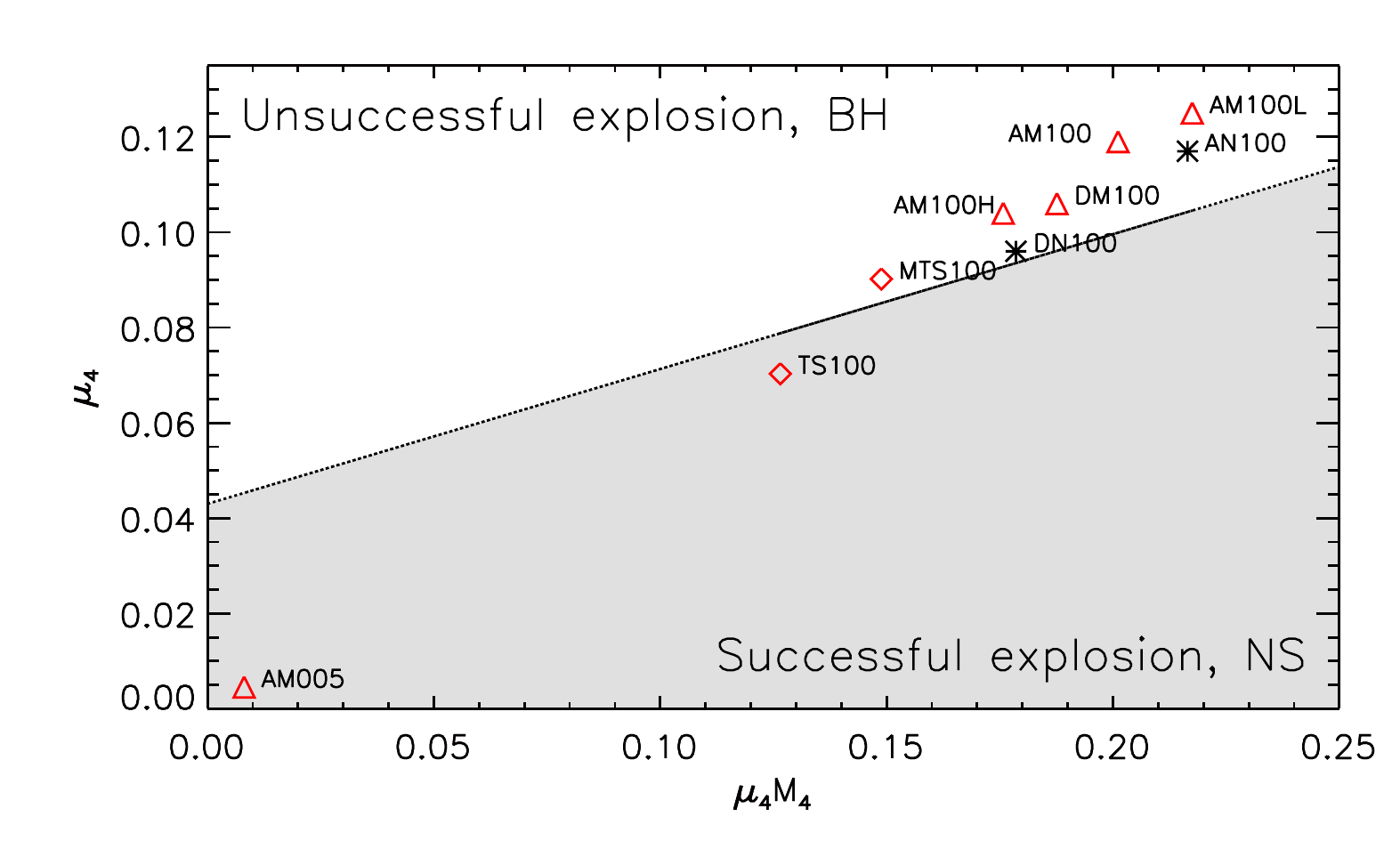}
    \caption{Representation of $\mu_4$ as a function of $M_4\mu_4$ for each of the computed models. Triangles denote models with only the MRI included, diamonds correspond to models with the TS dynamo incorporated, and asterisks show models without magnetic instabilities. The black line represents the linear relation
    $y=0.283x+0.043$, where $y=\mu_4$ and $x=\mu_4 M_4$. This line corresponds to the separation between successful and unsuccessful explosions \citep{2016-Ertl-Janka}. Models above the line will likely collapse into BHs, and those beneath are likely to form NSs.
    } %
    \label{fig:mu4-M4}%
\end{figure}
\noindent
In Fig.~\ref{fig:mu4-M4} we show our models in the phase space ($M_4\mu_4$,$\mu_4$) alongside a line that approximately  separates NS-forming from BH-forming models or, equivalently, successful and unsuccessful explosions. The separation line is not unique, but using any of the possibilities for the coefficients $k_1$ and $k_2$ in $y=k_1x+k_2$, with $y=\mu_4$ and $x=\mu_4 M_4$ from \cite{2016-Ertl-Janka} does not change whether any of our models shall form an NS or a BH. The precise values for $\mu_4$ and $M_4$ are given in Tab.~\ref{table:magnetic}. 
Models that fall above (below) the line are more likely to form BHs (NSs). 
Ertl's two-parameter criterion suggests
that almost all models will form BHs. Only models \modl{TS100} and \modl{AM005} fall below the line, and thus, may form NSs. Model \modl{AM005} is  an outlier here, and, due to its parameters being non-justified for GENEC, we will not develop further on this. 
While \modl{TS100} may form an NS, \modl{MTS100} may yield a BH. Therefore, the action of MRI could be critical to facilitate the collapse of the model into a BH. In fact, all models that include the MRI with $f_{\mu}=1$ regardless of advective or diffusive considerations for circulation by meridional currents are well above the line and will most likely form BHs according to the Ertl's two-parameter criterion. 

\begin{figure*}
    \centering
    \includegraphics[width=0.48\textwidth]{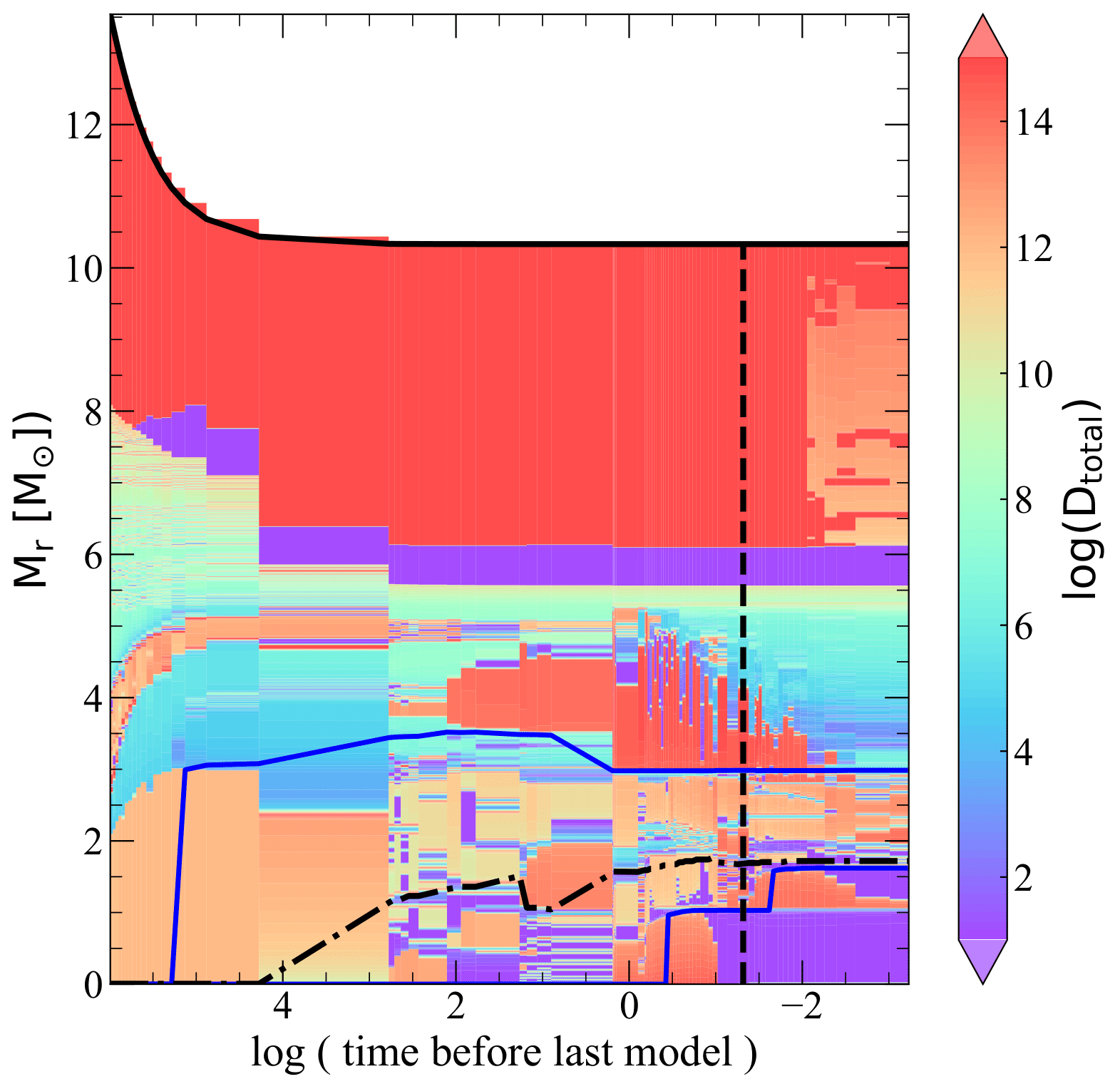}
    \includegraphics[width=0.48\textwidth]{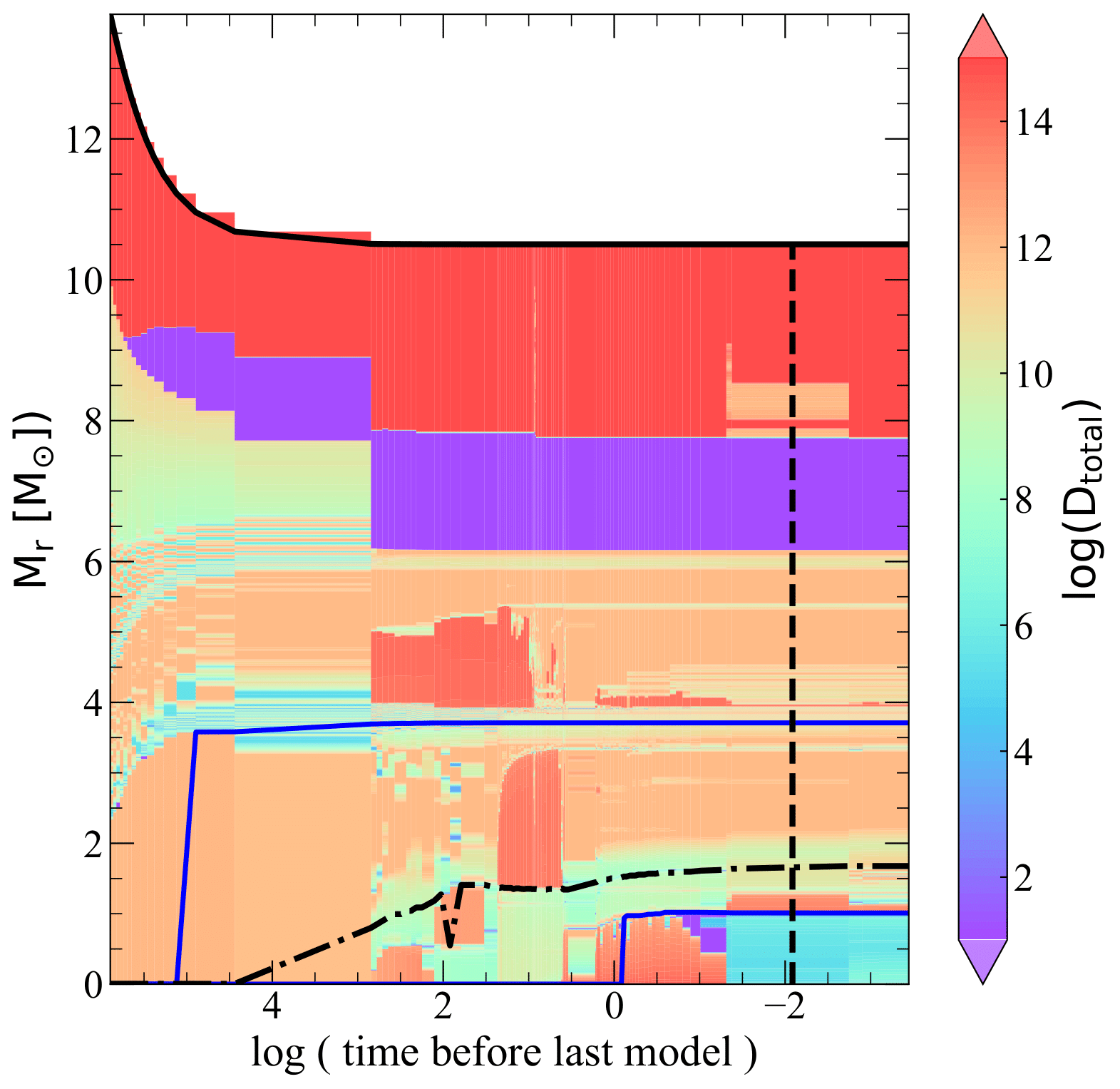}
    \caption{Kippenhahn-like diagram of the diffusion coefficient for AM transport given by Eq.\eqref{eqn:D-circ} for models \modl{AM100} (left) and \modl{MTS100} (right). The y axis is the Lagrangian mass coordinate. The x axis is the decimal logarithm of the time before the last computed model.\ We plot roughly from the end of the MS.
    As the two models are not  computed up to exactly the same time, this axis is not directly comparable. The vertical dashed black line shows the time when oxygen burning ends, which is the same time as in the plots of Fig.~\ref{fig:Composition-oxygen-burning}. The blue lines correspond to the inner and outer limits of the oxygen shell, defined as $X_{O_{16}}>0.5,$ and the dot-dash black line corresponds to the evolution of $M_4$ Eq.~\eqref{eqn:M4}. 
    }
    \label{fig:Kippenhan}
\end{figure*}

The predictions made on the basis of the compactness and the two-parameter criterion of \cite{2016-Ertl-Janka} are consistent. However, several considerations are in order here. First,
our analysis is made at the end of oxygen burning, and computing up to the pre-SN link may give different conclusions for these models. In Fig.~\ref{fig:Kippenhan} we show Kippenhahn diagrams, without the nuclear burning information, for models \modl{AM100} and \modl{MTS100}, on top of which we plot the oxygen shell boundaries in blue and $M_4$ in black dashed dotted line. The colour map tracks the evolution of the diffusion coefficient $D$ given by Eq.~\eqref{eqn:D-circ}. The zones that transport efficiently AM and chemical elements have high values of $D$. The zones that reach maximum values, in red, are convective regions of the star. Values of around $\approx$ $10^{12}$ $\rm cm^2 \ s^{-1}$ are regions that transport just as efficiently as convective regions but due to instabilities, (hydrodynamic or magnetic), in our models this is mostly due to MRI activity.
The diffusion coefficient is significant in the oxygen shell for both models so this zone behaves as a convective shell. This efficient transport impacts the extension and vigour of the oxygen-burning shell. 
The evolution of $\mu_4$ is a proxy for the oxygen-burning strength \citep{Sukhbold_2018ApJ...860...93}. As the oxygen-burning shell develops a jump in entropy \citep{Sukhbold_2014}, it also facilitates the evolution of $M_4$, as well as the compactness. The mass coordinate $M_4$ stays above the lower boundary of the oxygen shell at all times and therefore $\mu_4$ will not vary much up until the pre-SN link. So our initial extrapolation has a good chance of holding until the pre-SN link.

A second important point is that none of the two criteria employed to predict the compact remnant accounts for the action of magnetic instabilities in rotating models. The threshold value for $\xi_{2.5}$ separating BH-forming from NS-forming models may depend on the rotational energy left in the core until the pre-SN link \citep[but see][]{Obergaulinger_2022MNRAS.512.2489}. In Tab.~\ref{table:magnetic}, we observe an approximate correlation between  $\xi_{2.5}$ and the rotational energy contained within $M_4$, $\mathcal{T}_4$. The least compact cores are those with the smaller $\mathcal{T}_4$. In its turn, a small value of $\mathcal{T}_4$ is the result of an efficient transport of AM away from the core. 

To thoroughly explore whether the MRI encourages BH formation we must explore models in a larger parameter space, varying initial rotation and metallicity and compute these models up to the pre-SN link. Nonetheless, with this limited study we suggest that the MRI, by extracting AM from the core and through chemical transport making those cores more compact, tends to promote the collapse into BHs.

To further quantify the impact of the MRI on the compact remnant properties, we estimate the initial rotation period of a theoretical NS produced by the collapse of each of the models following the method of \citet{Hirschi2005}. However, on the basis of the two criteria applied above to extrapolate the type of compact remnant, it is more likely that most of our models generate BHs instead of NSs. In the former cases, the procedure of \cite{Hirschi2005} may estimate the rotational period of the PNS before collapsing to a BH.  If we suppose that the star from each model forms an NS that has a baryonic mass equal to $M_4$ (following e.g. \citealt{2016-Ertl-Janka,Sukhbold_2018ApJ...860...93}, but we note the difference with respect to \citealt{Hirschi2005}, who assume that the baryonic mass of the NS is fixed and equal to $1.56M_\odot$), then we can compute the AM contained inside this remnant mass at the end of oxygen burning with a radius of $R=12\,$km. This is defined as%
\begin{equation}
    \mathcal{L}_{4}=\frac{2}{3}\int_0^{M_{4}} \Omega r^2 dm.    
\end{equation}
This central core will lose binding energy due to the emission of neutrinos, reducing its baryonic mass until it is approximately equal to the gravitational mass of the potentially forming NS, $M_\mathrm{4,\rm g}$.
The relation between the gravitational and baryonic masses is $M_\mathrm{4}=\mathrm{BE}+M_\mathrm{4, g}$, where $\mathrm{BE} = 0.6\beta M_\mathrm{4, g}/(1-0.5\beta)$ is the PNS binding energy, and $\beta=GM_\mathrm{4,g}/Rc^2$ ($G$ and $c$ are Newton's gravitational constant and the speed of light in vacuum, respectively). These relations can be inverted to obtain $M_\mathrm{4, g}$ from a second order equation:
\begin{align}
    0.1GM^2_\mathrm{4,g}  + (Rc^2+0.5G M_\mathrm{4})M_\mathrm{4, g} -Rc^2M_\mathrm{4}=0 ,
\end{align}
where we note the typo in Eq.~(1) of  \cite{Hirschi2005} in the sign of the linear term of the previous equation. Assuming that the AM loss during the collapse and post-collapse phase is proportional to the amount of binding energy lost due to neutrinos \citep{Hirschi2005}, one can compute
the following quantities. First, the AM of the compact remnant, $\mathcal{L}_{\rm rm}$. This will be smaller than $\mathcal{L}_4$ due to the neutrino emission. Precisely \citep[cf.][]{Hirschi2005}, $\mathcal{L}_{\rm rm} = \mathcal{L}_4 (M_{4\rm ,g}/M_4)$. Second, the angular frequency of the hypothetical remnant (either an NS or a BH formed after an initial phase in which the central core forms a PNS, which collapses due to mass accretion)
\begin{equation}
    \Omega_\mathrm{rm}=\mathcal{L}_{4}/(0.36R^2M_4)
\end{equation}
with the corresponding rotational period
\begin{equation}
\label{eqn:PNS}
    \mathcal{P}_{\rm rem}=2\pi/\Omega_\mathrm{rm}.
\end{equation}
If a BH forms, possibly more mass than $M_4$ may contribute to it. Also, the reduction in the AM of the central core due to neutrino emission may not be so large (it will depend on how long the PNS lasts before it finally collapses). Thus, should the remnant be a BH, the values of $\mathcal{L}_{\rm rm}$, $\mathcal{P}_{\rm rm}$ and $\Omega_{\rm rm}$ would be poorer approximations to these quantities than if the remnant is an NS. Finally, we can also estimate the dimensionless remnant spin as
\begin{align}
    a_{\rm rm} = c \mathcal{L}_{\rm rm} / (G M_{4\rm ,g}^2). 
    \label{eqn:a4}
\end{align}


Comparing models \modl{AN100} and \modl{DN100}, we note that the choice of the implementation of the circulation by meridional currents does not drastically impact $\mathcal{P}_{\rm rm}$. The action of purely diffusive meridional currents slows down the compact core by a factor of $\sim 2$ compared to the model with advective meridional circulation (Tab.~\ref{table:magnetic}). As expected, much more important is the presence of magnetic instabilities. Models without magnetic fields (\modl{AN100} and \modl{DN100}) may produce compact remnants that rotate $\sim 3 - 4$ times faster than models that incorporate the MRI ($\sim 25-65$ times faster compared with TS dynamo models).

The effect that has the greatest impact on $\mathcal{P}_{\rm rm}$ is the TS dynamo. Its presence in the model \modl{TS100} leads to a slowdown of core rotation affecting $\mathcal{P}_{\rm rm}$ by a factor of $\sim 13$, with respect to our reference MRI model \modl{AM100} or variations thereof (models \modl{AM100H} or \modl{AM100L}). The combined action of MRI and the TS dynamo predicts a rotation rate 1.5 times slower for model \modl{MTS100} than for \modl{TS100}, showing that the MRI continues to have an impact on rotation rates in the late stages of stellar evolution.
 
Besides the relative differences between the predicted period in our models, all of them are short or too short compared with observations, \citep[see e.g.][]{Gallant2017}. 
According to these authors, classical radio pulsars should be born with a median birth period larger than $30\,$ms and no more than 10\% of pulsars could be born with birth periods inferior to $15\,$ms. 
Models implementing the TS dynamo alone or combined with the MRI are closer to the observational constraints (though their periods are still $\sim 50\%$ below them) than models including only the MRI. 
For models that may form BHs and account only for the MRI, our estimate of the dimensionless spin of the compact remnant clusters around $0.7\lesssim a_{\rm rm}\lesssim 0.85$. The special case of model \modl{MTS100} sticks out by its foreseeable tiny dimensionless spin ($a_{\rm rm}\approx 0.05$).

\subsection{MRI influence on the supernova explosion type}
\label{sec:explosion}

\begin{figure*}
    \centering
    \includegraphics[width=0.99\textwidth]{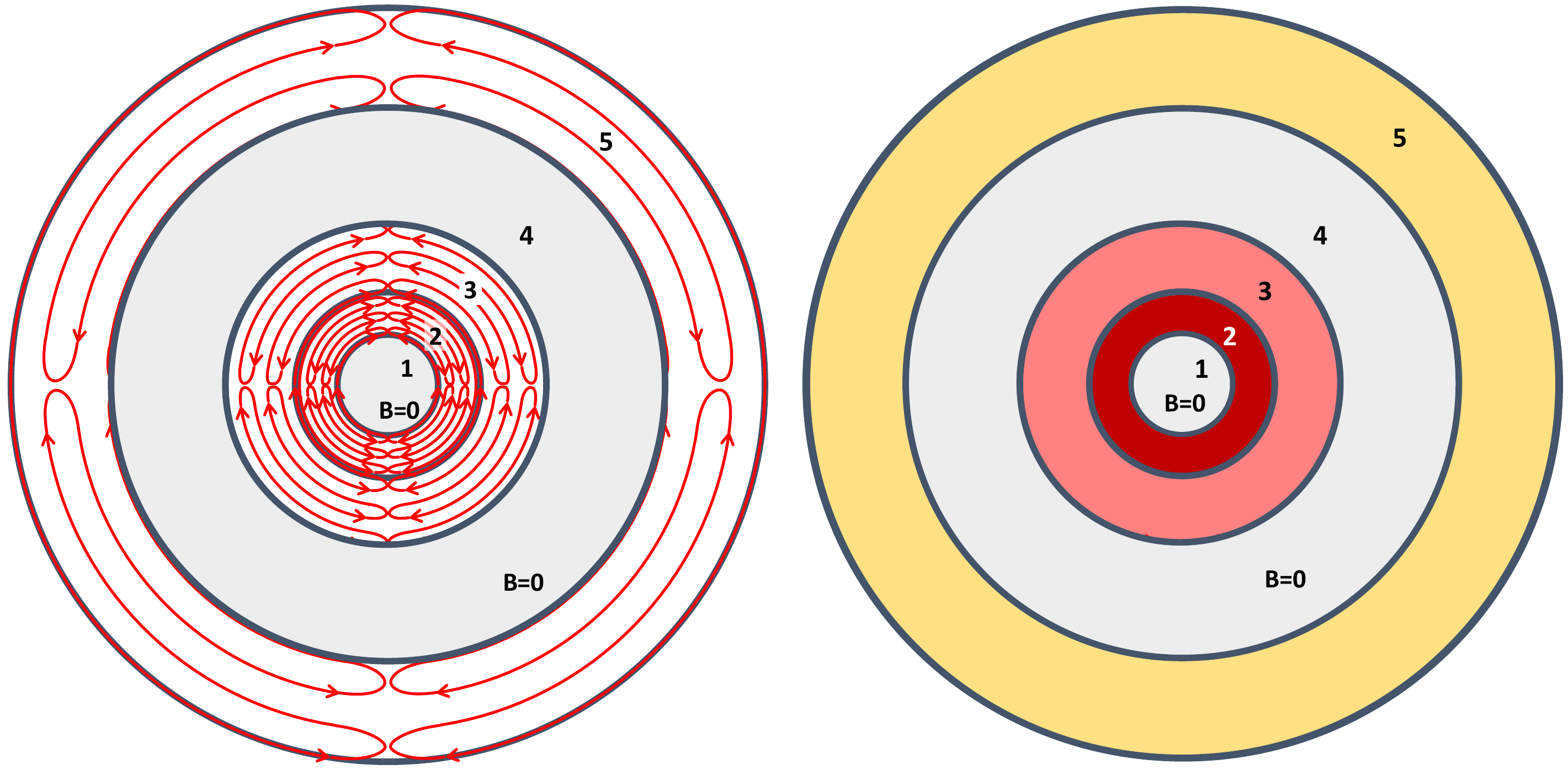}
    \caption{Schematic representation of the magnetisation of a model. \textit{Left}: Schematic structure of five stellar layers at a certain evolutionary time. Two of the layers are not magnetised (grey) and represent mass shells where the MRI or the TS dynamo cannot trigger. In layers 2, 3, and 5, magnetic instabilities have triggered. The approximate topology of the poloidal projection of the resulting magnetic field is represented by the red lines, with the direction of the field shown by the arrows. The schematic representation roughly corresponds to the field topology developed in MRI channel flows (perpendicular to the radial direction). While the magnetic field is clearly discontinuous between shells 3 and 5 (there are no magnetic field lines connecting the two regions), shells 2 and 3 may be connected. \textit{Right}: Schematic of the poloidal magnetic field in a 1D stellar evolution code. The root mean square values in shells 2, 3, and 5 replace the exact small-scale magnetic field topology shown in the left panel. Different colours denote distinct root mean square field strengths. The exact topology of the magnetic field lines is not accessible in the 1D representation.}%
    \label{fig:MRI_Bfield}%
\end{figure*}

A magnetic structure within the inner core (resulting from stellar evolution models including the TS mechanism) may trigger either magneto-rotational supernova explosions or neutrino-driven explosions \citep{Obergaulinger_2017MNRAS.469L..43, Obergaulinger_2020MNRAS.492.4613, Obergaulinger_2021MNRAS.503.4942, Obergaulinger_2022MNRAS.512.2489, Obergaulinger_2018JPhG...45h4001, Aloy_2021MNRAS.500.4365}. In the following, we assess whether the structure induced by the MRI as the main mechanism of AM transport, for \modl{AM100}, during the lifetime of the star has also the potential to drive magneto-rotational supernovae. Even if  the inner $\lesssim 1.5M_\odot$ is essentially non-magnetised and will collapse forming a barely magnetised PNS, there are still layers of the star with $1.5\lesssim M_r/M_\odot\lesssim 2.5$ that are magnetised (Fig.~\ref{fig:Magnetic-fields-TS-MRI}, lower left panel), whose accretion onto the PNS may take some seconds after core bounce. A magneto-rotational explosion requires that the poloidal magnetic field 
has a coherence length large enough to allow the connection of the PNS with the supernova shock \citep{Bugli_2020MNRAS.492...58,Aloy_2021MNRAS.500.4365}.
Following exactly the definition of \citet{Obergaulinger_2022MNRAS.512.2489}, we estimated the average coherence length of the poloidal magnetic field within the innermost $2.5M_\odot$ of each model, $\overline{l}_\mathrm{pol}$, computing the average length of the radial intervals in which $B_r \ne 0$:
\begin{align}
\overline{l}_{\rm pol}\equiv \frac{1}{N_{\rm layers}}\sum_{i=1}^{N_{\rm layers}} \Delta_i ,
\label{eqn:laver}
\end{align}
where $\Delta_i$ is the radial width of a magnetised layer and $N_{\rm
  layers}$ is the number of magnetised layers up to a mass coordinate
$M_r=2.5M_\odot$. 
We note that either MRI- or TS-dynamo-generated magnetic fields are small-scale fields, which cannot be resolved with any practical mass/radial discretisation of the stellar matter. 

Thus, the representation of the magnetic field topology in one-dimensional (1D) stellar evolution codes is necessarily only approximate. We assume that within a mass shell (with a radial extension much larger than the scale of the generated fields; see \figureref{fig:MRI_Bfield}, left) where the instabilities trigger, the magnetic field corresponds to the time and volume averaged saturation values from Eqs.~(\ref{eq:Bphi-saturation}),  (\ref{eq:Br-saturation}), (\ref{eqn:Br-TS}), and (\ref{eqn:Bphi-TS}). In other words, one represents the saturation magnetic field by a uniform distribution of magnetic fields within each mass shell (\figureref{fig:MRI_Bfield}, right). This representation is inconsequential for stellar evolution. Indeed the action of the magnetic instabilities with respect to driving AM or chemical element transport is modelled as a diffusive process. Therefore, the magnetic field is not a dynamical variable of evolution. We notice that two contiguous shells where a magnetic instability triggers may be connected by magnetic field lines crossing the (numerical) boundary separating them. The estimated coherence length does not correspond to the small-scale fields but, instead, to the time and volume averaged values. The defined quantity $\overline{l}_{\rm pol}$ aims at quantifying the `connectivity' of the magnetised regions in the inner regions of the star (precisely the regions with a larger dynamical impact on the post-collapse phase) when stellar models are mapped as initial models of neutrino-magnetohydrodynamic simulations of the collapse and post-collapse phases (see below). The obtained value may be somewhat dependent on the mass resolution used in GENEC. But exploring the effects of mass resolution on $\overline{l}_{\rm pol}$ is beyond the scope of this paper.
Furthermore, the process of mapping any vector quantity (such as the magnetic field) from 1D to higher dimensions is not unique. The most important physical restriction is that the reconstructed magnetic field must be solenoidal. Therefore, the reconstruction of the multi-dimensional magnetic field topology is a necessary step to use the 1D stellar evolution results as initial models for two-dimensional (2D) or three-dimensional (3D) neutrino-magnetohydrodynamic supernova simulations. In this reconstruction process, one starts from the magnetic field representation schematically shown in \figureref{fig:MRI_Bfield} (right). The mapping to multiple dimensions typically connects, through a continuous magnetic field representation, adjacent magnetised shells. This could be the case of mass shells 2 and 3 in \figureref{fig:MRI_Bfield} for example. However, the same reconstruction process typically assigns zero magnetic field to cells that are non-magnetised (e.g. shells 1 and 4). This means that for a neutrino-magnetohydrodynamic simulations shells 2 and 3 would be connected by the magnetic field, while they are disconnected from shell 5 (owing to the magnetic gap in shell 4).

We find typical values $\overline{l}_\mathrm{pol}\sim (2 \cdots 7)\times 10^7\,$cm in our models incorporating the MRI only. Models where the TS dynamo (alone or in combination with the MRI) operates, display a bit larger values $\overline{l}_\mathrm{pol}\sim (1 \cdots 2)\times 10^8\,$cm. To place these figures into context, we also list in Tab.~\ref{table:magnetic} the ratio, $\lambda_{2.5}\equiv \overline{l}_{\rm pol}/R_{2.5}$, which compares the average poloidal field coherence length to the radius of the inner $2.5M_\odot$, $R_{2.5}$, for each model \citep{Obergaulinger_2022MNRAS.512.2489}. This inner region has the largest impact on the post-collapse dynamics. If $\overline{l}_\mathrm{pol}$ represents a sizeable fraction of $R_{2.5}$ (i.e. if $\lambda_{2.5} \sim 1$), the energy of a subsequent supernova explosion will grow with increasing $\overline{l}_\mathrm{pol}$ \citep{Bugli_2020MNRAS.492...58}, as well as  the possibility of leaving an NS as compact remnant \citep{Aloy_2021MNRAS.500.4365, Obergaulinger_2021MNRAS.503.4942}. We find $\lambda_{2.5}\le 0.04$ for our reference model \modl{AM100} and for all models incorporating only the MRI. This value is too small to produce strong magneto-rotational explosions \citep{Obergaulinger_2022MNRAS.512.2489}. However, our models are not evolved to the brink of collapse, and a further growth of $\lambda_{2.5}$ is possible because of the contraction of the core after oxygen depletion (which will reduce $R_{2.5}$).  

\cite{2016-Ertl-Janka} uses the  two parameters ($\mu_4$, $M_4$) to study the explodability of a model by the neutrino mechanism.  On the basis of that criterion, models above the division line in Fig.~\ref{fig:mu4-M4} should not go through a neutrino driven explosion. The majority of our models may, therefore, not produce neutrino-driven explosions. Only the model implementing exclusively the TS dynamo may possibly produce such an explosion. The case of \modl{MTS100} is very close to the division line separating neutrino-driven explosions from failed supernovae. It has also the largest value of $\lambda_{2.5}$, which indeed is larger than $\sim 0.1$, the approximate boundary below which neutrinos are the main driver of the explosion \citep{Obergaulinger_2022MNRAS.512.2489}. Thus, in this model, neutrinos and magnetic effects might contribute to its successful explosion.

\section{Conclusion and perspectives}
\label{sec:conclusions}

Using the GENEC stellar evolution code we have computed rotating magnetic stars with masses of 15 $M_{\odot}$, solar metallicity, and a moderate equatorial initial rotational velocity of 206 $\rm km\, s^{-1}$ to look at the effects of the MRI and the TS dynamo. 
We implement, for the first time, the MRI with the circulation of meridional currents treated as an advective process. We find that the advective circulation builds up differential rotation, which encourages the MRI to trigger, especially during the MS. In agreement with \cite{Wheeler_2015}, we find that the MRI is a key factor in stellar evolution, impacting all stages of the star's lifetime, and we find suggestions that the MRI will impact the death of the star as well. However, where and when the MRI can trigger closely depends on other parameters of the model, such as the implementation of meridional currents or the chemical gradient weight, $f_{\mu}$.

We have shown that the MRI acts differently if the circulation by meridional currents is implemented advectively or treated diffusively. The diffusive treatment of meridional circulation destroys the rotational gradient very efficiently. Therefore, the shear does not build up and does not overcome the stable stratification of the chemical gradient; thus, the MRI fails to significantly affect the model. However, when we implement the circulation by meridional currents as an advective process, the buildup of differential rotation is sufficient to overcome the stabilising effects. The MRI sets in mainly in the first few million years of the MS and affects the transport of AM and chemical elements in equal measure. We therefore observe a slowdown of the core, an acceleration of the surface rotation rates, and more heavy materials from deep in the star being `dredged up' towards the surface compared to non-magnetic models.

The MRI is  particularly sensitive to the trigger condition, Eq.~\eqref{eqn:Instab}, and the choice of the parameter $f_{\mu}$, which acts as a weight for chemical gradients. Low values of $f_{\mu}$, commonly $0.05$, inhibit the stabilising effects of the chemical gradient on the MRI and artificially allow the MRI to trigger much deeper inside the star, significantly boosting its impact. In our models we chose to remove this parameter (i.e. setting it to $f_\mu=1$). We confirm that when we set $f_{\mu}=0.05$, the MRI can trigger much more easily due to weakened chemical gradients.

The addition of the TS dynamo to our models confirms that, during the MS, this mechanism is extremely efficient at transporting AM, totally flattening the rotation profile, as was found by previous studies. The TS dynamo can easily trigger over large spatial extents and is much less sensitive to the chemical stratification compared to the MRI. In cases where the MRI and the TS dynamo may operate simultaneously, we find that it is  difficult for the MRI to trigger. The TS dynamo consistently triggers first in most of the star and dissipates the differential rotation before the MRI can develop. At the end of the MS, the chemical structure and AM distribution of models with both instabilities closely resemble those of models with just the TS dynamo.

Our study of the MRI during the MS highlights that there is an intricate dependence of the MRI on the other properties of the model, such as the implementation of other hydrodynamic or magnetic instabilities, and that care is needed in the implementation of the MRI in a stellar model. 
During the post-MS evolution of our models (up to the end of oxygen burning), the MRI remains relevant to stellar evolution by extracting AM from the core and efficiently mixing regions where it is active. 

 The slowdown of the core due to the MRI is progressive and is observed during all burning stages of stellar evolution. The dominance of the TS dynamo on AM transport observed during the MS continues during post-MS evolution, where it is the main agent slowing down the core. Nonetheless, it is when both of the magnetic instabilities, the TS dynamo and the MRI, are included together that we observe the slowest core rotation rates.  Our model with both the MRI and the TS dynamo predicts a compact remnant with a rotation period of the order of 10\,ms, values that are the closest to the expected rates of rapidly rotating NSs at birth (should this model form an NS; see below). 

The activity of both instabilities leads to relatively strong magnetised stars at the end of oxygen burning. Specifically, the TS dynamo leads to strongly magnetised cores where the poloidal field reaches saturation values of $10^8\,$G, 
estimated with Eqs.~\eqref{eqn:Bphi-TS} and \eqref{eqn:Br-TS}, 
whereas the MRI tends to magnetise smaller regions farther out from the core, but the field strengths are higher, $10^{10}\,$G. Both of these factors could be extremely relevant to the collapse dynamics of these models. As the simulations are only run up to the end of oxygen burning, still far from the pre-SN link, it is hard to predict the subsequent dynamics of the collapse, but using the two-parameter criterion of \citet{2016-Ertl-Janka}, we find that almost all of our models at this stage tend to go through an unsuccessful explosion and form BHs. 
Notably, the TS-dynamo-only model may not form a BH  but rather an NS after going through a successful explosion. The type of compact remnant left behind by the model that includes both the MRI and the TS dynamo is difficult to predict. Depending on its location in the $(\mu_4, \mu_4M_4)$ plane, it may collapse into a BH. Nevertheless, the radial coherence length of this model compared to the size of the inner core is the largest among all computed models ($\lambda_{2.5}\gtrsim 0.1$). This may render the magnetic field effects important in the post-collapse evolution, letting them contribute to a successful explosion, which could prevent the collapse into a BH. 
All these facts suggest that if the MRI can develop during stellar evolution, the star will be more prone to collapse into a moderately rotating BH (with dimensionless spin $\sim 0.8$). However, a larger parametric space in mass and initial equatorial rotational velocity needs to be explored to support this suggestion.

The MRI, as we have shown, has the potential to transport chemical elements very efficiently during the early stages of the MS, dredging heavy elements up to the surface. The next question to ask is whether the difference between the MRI and non-MRI models presented would be distinguishable through the observation of surface abundances. We have seen from Fig.~\ref{fig:Surface-Abund}a that the largest differences occur at the very beginning of the MS phase. Thus, detailed surface abundances of massive stars early in their MS phase may provide some clues for differentiating between models if the MRI developed or not, and if so to constrain the efficiency of the transport. At the moment, the observations of $[N/H]_{\rm surf}$ ratios in MS B-type stars (see the grey-shaded zones in Fig.~\ref{fig:Surface-Abund}a) are more compatible with models that do not include the MRI than with models that include it  halfway through the MS. Anyway, both cases are broadly compatible with observations at the end of the MS (if we were to vary initial velocity and metallicity). Boron might be an interesting element to study in this context. This element can be used as a probe of very progressive mixing at the surface of massive stars. Indeed, a process active only in the outer layer of a massive star can be sufficient to destroy boron at its surface (by dredging it down to layers sufficiently hot to destroy it, with temperatures of around $6\times 10^6\,$K ) while not affecting the abundances of elements that are only modified at higher temperatures (such as nitrogen, whose abundance can change if the temperature rises above a few tens of millions of degrees). This question has been discussed by \citet{Pro1999}, \citet{Venn2002}, \citet{Mendel2006}, \citet{Frisch2010}, and \citet{Pro2016}, but in the frame of models that do not account for the MRI instability.  Furthermore, since MRI models enrich their surface rapidly with elements produced in their core, such models may boost the ejection of the radioisotope $^{26}$Al by stellar winds \citep[e.g.][]{2005A&A...429..613P}. This would also be an interesting future area of application for our models.

As the MRI is strongly inhibited by chemical gradients, further studies should explore its role in (quasi-)chemical homogeneous evolution  \cite[e.g.][]{Aguilera-Dena_2018ApJ...858..115}. Stars following (quasi-)chemical homogeneous evolution have little to no chemical gradients in their interior, and thus the MRI could be easily triggered. This in turn could change the AM content of the resulting compact objects. The MRI could help foster conditions for creating an even more homogeneous evolution than that modelled already.

\begin{acknowledgements}
We thank the anonymous referee for their careful revision of the original manuscript, as a result of which, the final paper has greatly improved. We acknowledge the use of the Helmholtz equation of state as implemented in the web page of F. Timmes (\url{https://cococubed.com/code_pages/eos.shtml}).
AG, PE and GM have received funding from the European Research Council (ERC) under the European Union's Horizon 2020 research and innovation program
(grant agreement No 833925, project STAREX). MAA and AG thank the support of the Spanish Ministry of Science, Education and Universities (PGC2018-095984-B-I00) and of the Valencian Community (PROMETEU/2019/071).
\end{acknowledgements}
\bibliographystyle{aa}
\bibliography{biblio.bib} 

\begin{thebibliography}{102}
\expandafter\ifx\csname natexlab\endcsname\relax\def\natexlab#1{#1}\fi

\bibitem[{Aerts {et~al.}(2014)Aerts, Molenberghs, Kenward, \&
  Neiner}]{Aerts_2014}
Aerts, C., Molenberghs, G., Kenward, M.~G., \& Neiner, C. 2014, The
  Astrophysical Journal, 781, 88

\bibitem[{{Aguilera-Dena} {et~al.}(2018){Aguilera-Dena}, {Langer}, {Moriya}, \&
  {Schootemeijer}}]{Aguilera-Dena_2018ApJ...858..115}
{Aguilera-Dena}, D.~R., {Langer}, N., {Moriya}, T.~J., \& {Schootemeijer}, A.
  2018, \apj, 858, 115

\bibitem[{{Akiyama} {et~al.}(2003){Akiyama}, {Wheeler}, {Meier}, \&
  {Lichtenstadt}}]{Akiyama_2003ApJ...584..954}
{Akiyama}, S., {Wheeler}, J.~C., {Meier}, D.~L., \& {Lichtenstadt}, I. 2003,
  \apj, 584, 954

\bibitem[{{Aloy} \& {Obergaulinger}(2021)}]{Aloy_2021MNRAS.500.4365}
{Aloy}, M.~{\'A}. \& {Obergaulinger}, M. 2021, \mnras, 500, 4365

\bibitem[{{Balbus}(1995)}]{Balbus_1995ApJ...453..380}
{Balbus}, S.~A. 1995, \apj, 453, 380

\bibitem[{{Balbus} \& {Hawley}(1991)}]{1991BandH}
{Balbus}, S.~A. \& {Hawley}, J.~F. 1991, \apj, 376, 214

\bibitem[{Balbus \& Hawley(1998)}]{RevModPhys.70.1}
Balbus, S.~A. \& Hawley, J.~F. 1998, Rev. Mod. Phys., 70, 1

\bibitem[{{Bavera} {et~al.}(2020){Bavera}, {Fragos}, {Qin}, {Zapartas},
  {Neijssel}, {Mandel}, {Batta}, {Gaebel}, {Kimball}, \&
  {Stevenson}}]{Bavera2020}
{Bavera}, S.~S., {Fragos}, T., {Qin}, Y., {et~al.} 2020, \aap, 635, A97

\bibitem[{{Belczynski} {et~al.}(2020){Belczynski}, {Klencki}, {Fields},
  {Olejak}, {Berti}, {Meynet}, {Fryer}, {Holz}, {O'Shaughnessy}, {Brown},
  {Bulik}, {Leung}, {Nomoto}, {Madau}, {Hirschi}, {Kaiser}, {Jones}, {Mondal},
  {Chruslinska}, {Drozda}, {Gerosa}, {Doctor}, {Giersz}, {Ekstrom}, {Georgy},
  {Askar}, {Baibhav}, {Wysocki}, {Natan}, {Farr}, {Wiktorowicz}, {Coleman
  Miller}, {Farr}, \& {Lasota}}]{Bel2020}
{Belczynski}, K., {Klencki}, J., {Fields}, C.~E., {et~al.} 2020, \aap, 636,
  A104

\bibitem[{{Braithwaite}(2006)}]{Braithwaite_2006A&A...449..451}
{Braithwaite}, J. 2006, \aap, 449, 451

\bibitem[{{Brott} {et~al.}(2011){Brott}, {de Mink}, {Cantiello}, {Langer}, {de
  Koter}, {Evans}, {Hunter}, {Trundle}, \& {Vink}}]{Brott2011}
{Brott}, I., {de Mink}, S.~E., {Cantiello}, M., {et~al.} 2011, \aap, 530, A115

\bibitem[{{Bugli} {et~al.}(2020){Bugli}, {Guilet}, {Obergaulinger},
  {Cerd{\'a}-Dur{\'a}n}, \& {Aloy}}]{Bugli_2020MNRAS.492...58}
{Bugli}, M., {Guilet}, J., {Obergaulinger}, M., {Cerd{\'a}-Dur{\'a}n}, P., \&
  {Aloy}, M.~A. 2020, \mnras, 492, 58

\bibitem[{{Chaboyer} \& {Zahn}(1992)}]{1992Chaboyer-Zahn}
{Chaboyer}, B. \& {Zahn}, J.~P. 1992, \aap, 253, 173

\bibitem[{{Chieffi} \& {Limongi}(2020)}]{Chieffi_2020ApJ...890...43}
{Chieffi}, A. \& {Limongi}, M. 2020, \apj, 890, 43

\bibitem[{{Choi} {et~al.}(2017){Choi}, {Conroy}, \& {Byler}}]{Choi2017}
{Choi}, J., {Conroy}, C., \& {Byler}, N. 2017, \apj, 838, 159

\bibitem[{{Choplin} {et~al.}(2018){Choplin}, {Hirschi}, {Meynet},
  {Ekstr{\"o}m}, {Chiappini}, \& {Laird}}]{Choplin2018}
{Choplin}, A., {Hirschi}, R., {Meynet}, G., {et~al.} 2018, \aap, 618, A133

\bibitem[{{Deheuvels} {et~al.}(2020){Deheuvels}, {Ballot}, {Eggenberger},
  {Spada}, {Noll}, \& {den Hartogh}}]{Deheuvels2020}
{Deheuvels}, S., {Ballot}, J., {Eggenberger}, P., {et~al.} 2020, \aap, 641,
  A117

\bibitem[{{den Hartogh} {et~al.}(2019{\natexlab{a}}){den Hartogh},
  {Eggenberger}, \& {Hirschi}}]{Hart2019}
{den Hartogh}, J.~W., {Eggenberger}, P., \& {Hirschi}, R. 2019{\natexlab{a}},
  \aap, 622, A187

\bibitem[{{den Hartogh} {et~al.}(2019{\natexlab{b}}){den Hartogh},
  {Eggenberger}, \& {Hirschi}}]{Hartogh2019}
{den Hartogh}, J.~W., {Eggenberger}, P., \& {Hirschi}, R. 2019{\natexlab{b}},
  \aap, 622, A187

\bibitem[{{Eddington}(1933)}]{Eddington1933}
{Eddington}, A.~S. 1933, \nat, 132, 639

\bibitem[{{Eggenberger} {et~al.}(2019{\natexlab{a}}){Eggenberger}, {Buldgen},
  \& {Salmon}}]{2019A&A...626L...1EggenbergerSun}
{Eggenberger}, P., {Buldgen}, G., \& {Salmon}, S.~J.~A.~J. 2019{\natexlab{a}},
  \aap, 626, L1

\bibitem[{{Eggenberger} {et~al.}(2019{\natexlab{b}}){Eggenberger}, {Deheuvels},
  {Miglio}, {Ekstr{\"o}m}, {Georgy}, {Meynet}, {Lagarde}, {Salmon}, {Buldgen},
  {Montalb{\'a}n}, {Spada}, \& {Ballot}}]{Egg2019a}
{Eggenberger}, P., {Deheuvels}, S., {Miglio}, A., {et~al.} 2019{\natexlab{b}},
  \aap, 621, A66

\bibitem[{{Eggenberger} {et~al.}(2019{\natexlab{c}}){Eggenberger}, {den
  Hartogh}, {Buldgen}, {Meynet}, {Salmon}, \& {Deheuvels}}]{Egg2019b}
{Eggenberger}, P., {den Hartogh}, J.~W., {Buldgen}, G., {et~al.}
  2019{\natexlab{c}}, \aap, 631, L6

\bibitem[{{Eggenberger} {et~al.}(2008){Eggenberger}, {Meynet}, {Maeder},
  {Hirschi}, {Charbonnel}, {Talon}, \&
  {Ekstr{\"o}m}}]{2008Ap&SS.316...43Eggenberger}
{Eggenberger}, P., {Meynet}, G., {Maeder}, A., {et~al.} 2008, \apss, 316, 43

\bibitem[{{Ekstr{\"o}m} {et~al.}(2012){Ekstr{\"o}m}, {Georgy}, {Eggenberger},
  {Meynet}, {Mowlavi}, {Wyttenbach}, {Granada}, {Decressin}, {Hirschi},
  {Frischknecht}, {Charbonnel}, \& {Maeder}}]{Ekstrom2012}
{Ekstr{\"o}m}, S., {Georgy}, C., {Eggenberger}, P., {et~al.} 2012, \aap, 537,
  A146

\bibitem[{{Endal} \& {Sofia}(1981)}]{1981Endal}
{Endal}, A.~S. \& {Sofia}, S. 1981, \apj, 243, 625

\bibitem[{{Ertl} {et~al.}(2016){Ertl}, {Janka}, {Woosley}, {Sukhbold}, \&
  {Ugliano}}]{2016-Ertl-Janka}
{Ertl}, T., {Janka}, H.~T., {Woosley}, S.~E., {Sukhbold}, T., \& {Ugliano}, M.
  2016, \apj, 818, 124

\bibitem[{{Fricke}(1969)}]{1969Fricke}
{Fricke}, K. 1969, \aplett, 3, 219

\bibitem[{{Frischknecht} {et~al.}(2010){Frischknecht}, {Hirschi}, {Meynet},
  {Ekstr{\"o}m}, {Georgy}, {Rauscher}, {Winteler}, \&
  {Thielemann}}]{Frisch2010}
{Frischknecht}, U., {Hirschi}, R., {Meynet}, G., {et~al.} 2010, \aap, 522, A39

\bibitem[{{Fuller} \& {Ma}(2019)}]{Fuller2019c}
{Fuller}, J. \& {Ma}, L. 2019, \apjl, 881, L1

\bibitem[{{Fuller} {et~al.}(2019){Fuller}, {Piro}, \& {Jermyn}}]{Fuller2019b}
{Fuller}, J., {Piro}, A.~L., \& {Jermyn}, A.~S. 2019, \mnras, 485, 3661

\bibitem[{{Gallant} {et~al.}(2017){Gallant}, {Bandiera}, {Bucciantini}, \&
  {Amato}}]{Gallant2017}
{Gallant}, Y.~A., {Bandiera}, R., {Bucciantini}, N., \& {Amato}, E. 2017, in
  Supernova 1987A:30 years later - Cosmic Rays and Nuclei from Supernovae and
  their Aftermaths, ed. A.~{Marcowith}, M.~{Renaud}, G.~{Dubner}, A.~{Ray}, \&
  A.~{Bykov}, Vol. 331, 63--68

\bibitem[{{Georgy} {et~al.}(2012){Georgy}, {Ekstr{\"o}m}, {Meynet}, {Massey},
  {Levesque}, {Hirschi}, {Eggenberger}, \& {Maeder}}]{Georgy2012}
{Georgy}, C., {Ekstr{\"o}m}, S., {Meynet}, G., {et~al.} 2012, \aap, 542, A29

\bibitem[{{Goldreich} \& {Schubert}(1967)}]{Goldreich1967}
{Goldreich}, P. \& {Schubert}, G. 1967, \apj, 150, 571

\bibitem[{{Groh} {et~al.}(2013){Groh}, {Meynet}, {Georgy}, \&
  {Ekstr{\"o}m}}]{Groh2013}
{Groh}, J.~H., {Meynet}, G., {Georgy}, C., \& {Ekstr{\"o}m}, S. 2013, \aap,
  558, A131

\bibitem[{{Hawley} {et~al.}(2011){Hawley}, {Guan}, \& {Krolik}}]{Hawley2011}
{Hawley}, J.~F., {Guan}, X., \& {Krolik}, J.~H. 2011, \apj, 738, 84

\bibitem[{Heger {et~al.}(2000)Heger, Langer, \& Woosley}]{Heger_2000}
Heger, A., Langer, N., \& Woosley, S.~E. 2000, The Astrophysical Journal, 528,
  368–396

\bibitem[{{Heger} {et~al.}(2005{\natexlab{a}}){Heger}, {Woosley}, \&
  {Spruit}}]{Heger2005}
{Heger}, A., {Woosley}, S.~E., \& {Spruit}, H.~C. 2005{\natexlab{a}}, \apj,
  626, 350

\bibitem[{{Heger} {et~al.}(2005{\natexlab{b}}){Heger}, {Woosley}, \&
  {Spruit}}]{Heger_et_al__2005__apj__Presupernova_Evolution_of_Differentially_Rotating_Massive_Stars_Including_Magnetic_Fields}
{Heger}, A., {Woosley}, S.~E., \& {Spruit}, H.~C. 2005{\natexlab{b}}, \apj,
  626, 350

\bibitem[{{Hirose} {et~al.}(2009){Hirose}, {Krolik}, \&
  {Blaes}}]{Hirose_2009ApJ...691...16}
{Hirose}, S., {Krolik}, J.~H., \& {Blaes}, O. 2009, \apj, 691, 16

\bibitem[{{Hirschi} {et~al.}(2005){Hirschi}, {Meynet}, \&
  {Maeder}}]{Hirschi2005}
{Hirschi}, R., {Meynet}, G., \& {Maeder}, A. 2005, \aap, 443, 581

\bibitem[{{Jouve} {et~al.}(2015){Jouve}, {Gastine}, \&
  {Ligni{\`e}res}}]{Jouve_2015A&A...575A.106}
{Jouve}, L., {Gastine}, T., \& {Ligni{\`e}res}, F. 2015, \aap, 575, A106

\bibitem[{{Kagan} \& {Wheeler}(2014)}]{Kagan_2014ApJ...787...21}
{Kagan}, D. \& {Wheeler}, J.~C. 2014, \apj, 787, 21

\bibitem[{{Levan} {et~al.}(2016){Levan}, {Crowther}, {de Grijs}, {Langer},
  {Xu}, \& {Yoon}}]{Levan2016}
{Levan}, A., {Crowther}, P., {de Grijs}, R., {et~al.} 2016, \ssr, 202, 33

\bibitem[{{Limongi} \& {Chieffi}(2018)}]{Limongi2018}
{Limongi}, M. \& {Chieffi}, A. 2018, \apjs, 237, 13

\bibitem[{{Maeder}(1997)}]{Maeder1997-2}
{Maeder}, A. 1997, \aap, 321, 134

\bibitem[{Maeder(2009)}]{Maeder2009}
Maeder, A. 2009, Physics, Formation And Evolution Of Rotating Stars, ed.
  H.~Kienle (Springer-Verlag Berlin Heidelberg)

\bibitem[{{Maeder} \& {Meynet}(2004)}]{Maeder_2004A&A...422..225}
{Maeder}, A. \& {Meynet}, G. 2004, \aap, 422, 225

\bibitem[{{Maeder} \& {Meynet}(2014)}]{2014ApJ...793..123Maederbraking}
{Maeder}, A. \& {Meynet}, G. 2014, \apj, 793, 123

\bibitem[{{Maeder} \& {Zahn}(1998)}]{MZ1998}
{Maeder}, A. \& {Zahn}, J.-P. 1998, \aap, 334, 1000

\bibitem[{{Marchant} \& {Moriya}(2020)}]{Marchant2020}
{Marchant}, P. \& {Moriya}, T.~J. 2020, \aap, 640, L18

\bibitem[{{Masada} {et~al.}(2006){Masada}, {Sano}, \&
  {Takabe}}]{Masada_2006ApJ...641..447}
{Masada}, Y., {Sano}, T., \& {Takabe}, H. 2006, \apj, 641, 447

\bibitem[{{Masada} {et~al.}(2012){Masada}, {Takiwaki}, {Kotake}, \&
  {Sano}}]{Masada_2012ApJ...759..110}
{Masada}, Y., {Takiwaki}, T., {Kotake}, K., \& {Sano}, T. 2012, \apj, 759, 110

\bibitem[{{Mendel} {et~al.}(2006){Mendel}, {Venn}, {Proffitt}, {Brooks}, \&
  {Lambert}}]{Mendel2006}
{Mendel}, J.~T., {Venn}, K.~A., {Proffitt}, C.~R., {Brooks}, A.~M., \&
  {Lambert}, D.~L. 2006, \apj, 640, 1039

\bibitem[{{Menou} {et~al.}(2004){Menou}, {Balbus}, \&
  {Spruit}}]{Menou_2004ApJ...607..564}
{Menou}, K., {Balbus}, S.~A., \& {Spruit}, H.~C. 2004, \apj, 607, 564

\bibitem[{{Mestel}(1966)}]{Mestel1966}
{Mestel}, L. 1966, \zap, 63, 196

\bibitem[{{Meynet}(2009)}]{Meynet2009}
{Meynet}, G. 2009, {Physics of Rotation in Stellar Models}, Vol. 765
  (Springer), 139--169

\bibitem[{{Meynet} \& {Maeder}(1997)}]{MM1997}
{Meynet}, G. \& {Maeder}, A. 1997, \aap, 321, 465

\bibitem[{{Meynet} \& {Maeder}(2000)}]{MandM2000}
{Meynet}, G. \& {Maeder}, A. 2000, \aap, 361, 101

\bibitem[{{Meynet} \& {Maeder}(2017)}]{MM2017}
{Meynet}, G. \& {Maeder}, A. 2017, {Supernovae from Rotating Stars}, ed. A.~W.
  {Alsabti} \& P.~{Murdin} (Springer), 601

\bibitem[{{Murphy} {et~al.}(2021){Murphy}, {Groh}, {Farrell}, {Meynet},
  {Ekstr{\"o}m}, {Tsiatsiou}, {Hackett}, \& {Martinet}}]{Murphy2021}
{Murphy}, L.~J., {Groh}, J.~H., {Farrell}, E., {et~al.} 2021, \mnras, 506, 5731

\bibitem[{{Obergaulinger} \& {Aloy}(2017)}]{Obergaulinger_2017MNRAS.469L..43}
{Obergaulinger}, M. \& {Aloy}, M.~{\'A}. 2017, \mnras, 469, L43

\bibitem[{{Obergaulinger} \& {Aloy}(2020)}]{Obergaulinger_2020MNRAS.492.4613}
{Obergaulinger}, M. \& {Aloy}, M.~{\'A}. 2020, \mnras, 492, 4613

\bibitem[{{Obergaulinger} \& {Aloy}(2021)}]{Obergaulinger_2021MNRAS.503.4942}
{Obergaulinger}, M. \& {Aloy}, M.~{\'A}. 2021, \mnras, 503, 4942

\bibitem[{{Obergaulinger} \& {Aloy}(2022)}]{Obergaulinger_2022MNRAS.512.2489}
{Obergaulinger}, M. \& {Aloy}, M.~{\'A}. 2022, \mnras, 512, 2489

\bibitem[{{Obergaulinger} {et~al.}(2009){Obergaulinger}, {Cerd{\'a}-Dur{\'a}n},
  {M{\"u}ller}, \& {Aloy}}]{Obergaulinger_2009A&A...498..241}
{Obergaulinger}, M., {Cerd{\'a}-Dur{\'a}n}, P., {M{\"u}ller}, E., \& {Aloy},
  M.~A. 2009, \aap, 498, 241

\bibitem[{{Obergaulinger} {et~al.}(2018){Obergaulinger}, {Just}, \&
  {Aloy}}]{Obergaulinger_2018JPhG...45h4001}
{Obergaulinger}, M., {Just}, O., \& {Aloy}, M.~A. 2018, Journal of Physics G
  Nuclear Physics, 45, 084001

\bibitem[{O’Connor \& Ott(2011)}]{O_Connor_2011}
O’Connor, E. \& Ott, C.~D. 2011, The Astrophysical Journal, 730, 70

\bibitem[{{Palacios} {et~al.}(2005){Palacios}, {Meynet}, {Vuissoz},
  {Kn{\"o}dlseder}, {Schaerer}, {Cervi{\~n}o}, \&
  {Mowlavi}}]{2005A&A...429..613P}
{Palacios}, A., {Meynet}, G., {Vuissoz}, C., {et~al.} 2005, \aap, 429, 613

\bibitem[{{Paxton}(2019)}]{2019zndo...3473377Paxton}
{Paxton}, B. 2019, {Modules for Experiments in Stellar Astrophysics (MESA)}

\bibitem[{{Paxton} {et~al.}(2011){Paxton}, {Bildsten}, {Dotter}, {Herwig},
  {Lesaffre}, \& {Timmes}}]{2011ApJS..192....3Paxton}
{Paxton}, B., {Bildsten}, L., {Dotter}, A., {et~al.} 2011, \apjs, 192, 3

\bibitem[{{Paxton} {et~al.}(2013){Paxton}, {Cantiello}, {Arras}, {Bildsten},
  {Brown}, {Dotter}, {Mankovich}, {Montgomery}, {Stello}, {Timmes}, \&
  {Townsend}}]{2013ApJS..208....4Paxton}
{Paxton}, B., {Cantiello}, M., {Arras}, P., {et~al.} 2013, \apjs, 208, 4

\bibitem[{{Petrovic} {et~al.}(2005){Petrovic}, {Langer}, \& {van der
  Hucht}}]{Petrovic_2005A&A...435.1013}
{Petrovic}, J., {Langer}, N., \& {van der Hucht}, K.~A. 2005, \aap, 435, 1013

\bibitem[{{Prantzos} {et~al.}(2020){Prantzos}, {Abia}, {Cristallo}, {Limongi},
  \& {Chieffi}}]{Prantzos2020}
{Prantzos}, N., {Abia}, C., {Cristallo}, S., {Limongi}, M., \& {Chieffi}, A.
  2020, \mnras, 491, 1832

\bibitem[{{Proffitt} {et~al.}(1999){Proffitt}, {J{\"o}nsson}, {Litz{\'e}n},
  {Pickering}, \& {Wahlgren}}]{Pro1999}
{Proffitt}, C.~R., {J{\"o}nsson}, P., {Litz{\'e}n}, U., {Pickering}, J.~C., \&
  {Wahlgren}, G.~M. 1999, \apj, 516, 342

\bibitem[{{Proffitt} {et~al.}(2016){Proffitt}, {Lennon}, {Langer}, \&
  {Brott}}]{Pro2016}
{Proffitt}, C.~R., {Lennon}, D.~J., {Langer}, N., \& {Brott}, I. 2016, \apj,
  824, 3

\bibitem[{{Qin} {et~al.}(2019){Qin}, {Marchant}, {Fragos}, {Meynet}, \&
  {Kalogera}}]{Qin2019}
{Qin}, Y., {Marchant}, P., {Fragos}, T., {Meynet}, G., \& {Kalogera}, V. 2019,
  \apjl, 870, L18

\bibitem[{{Reboul-Salze} {et~al.}(2021){Reboul-Salze}, {Guilet}, {Raynaud}, \&
  {Bugli}}]{Reboul-Salze_2021A&A...645A.109}
{Reboul-Salze}, A., {Guilet}, J., {Raynaud}, R., \& {Bugli}, M. 2021, \aap,
  645, A109

\bibitem[{{Rembiasz} {et~al.}(2016{\natexlab{a}}){Rembiasz}, {Guilet},
  {Obergaulinger}, {Cerd{\'a}-Dur{\'a}n}, {Aloy}, \&
  {M{\"u}ller}}]{Rembiasz_et_al__2016__mnras__Onthemaximummagneticfieldamplificationbythemagnetorotationalinstabilityincore-collapsesupernovae}
{Rembiasz}, T., {Guilet}, J., {Obergaulinger}, M., {et~al.} 2016{\natexlab{a}},
  \mnras, 460, 3316

\bibitem[{{Rembiasz} {et~al.}(2016{\natexlab{b}}){Rembiasz}, {Obergaulinger},
  {Cerd{\'a}-Dur{\'a}n}, {M{\"u}ller}, \&
  {Aloy}}]{Rembiasz_et_al__2016__mnras__Terminationofthemagnetorotationalinstabilityviaparasiticinstabilitiesincore-collapsesupernovae}
{Rembiasz}, T., {Obergaulinger}, M., {Cerd{\'a}-Dur{\'a}n}, P., {M{\"u}ller},
  E., \& {Aloy}, M.~A. 2016{\natexlab{b}}, \mnras, 456, 3782

\bibitem[{{Shi} {et~al.}(2010){Shi}, {Krolik}, \&
  {Hirose}}]{Shi_2010ApJ...708.1716}
{Shi}, J., {Krolik}, J.~H., \& {Hirose}, S. 2010, \apj, 708, 1716

\bibitem[{{Shi} {et~al.}(2016){Shi}, {Stone}, \&
  {Huang}}]{Shi_2016MNRAS.456.2273}
{Shi}, J.-M., {Stone}, J.~M., \& {Huang}, C.~X. 2016, \mnras, 456, 2273

\bibitem[{Spitzer(2006)}]{Spitzer2006}
Spitzer, L. 2006, The Physics of Fully Ionized Gases (Dover Publications;
  Second Edition, Revised (July 7, 2006))

\bibitem[{{Spruit}(1999)}]{1999Spruit}
{Spruit}, H.~C. 1999, \aap, 349, 189

\bibitem[{{Spruit}(2002)}]{Spruit2002}
{Spruit}, H.~C. 2002, A\&A, 381, 923

\bibitem[{{Suijs} {et~al.}(2008){Suijs}, {Langer}, {Poelarends}, {Yoon},
  {Heger}, \& {Herwig}}]{Suijs2008}
{Suijs}, M.~P.~L., {Langer}, N., {Poelarends}, A.~J., {et~al.} 2008, \aap, 481,
  L87

\bibitem[{Sukhbold \& Woosley(2014)}]{Sukhbold_2014}
Sukhbold, T. \& Woosley, S.~E. 2014, The Astrophysical Journal, 783, 10

\bibitem[{{Sukhbold} {et~al.}(2018){Sukhbold}, {Woosley}, \&
  {Heger}}]{Sukhbold_2018ApJ...860...93}
{Sukhbold}, T., {Woosley}, S.~E., \& {Heger}, A. 2018, \apj, 860, 93

\bibitem[{{Sweet}(1950)}]{Sweet1950}
{Sweet}, P.~A. 1950, \mnras, 110, 548

\bibitem[{{Takahashi} \& {Langer}(2021)}]{Takahashi_2021A&A...646A..19}
{Takahashi}, K. \& {Langer}, N. 2021, \aap, 646, A19

\bibitem[{{Tassoul}(1978)}]{tassoul1978}
{Tassoul}, J.-L. 1978, {Theory of rotating stars} (Princeton: Univ. Press)

\bibitem[{{Timmes} \& {Swesty}(2000)}]{Timmes_2000ApJS..126..501}
{Timmes}, F.~X. \& {Swesty}, F.~D. 2000, \apjs, 126, 501

\bibitem[{{Ugliano} {et~al.}(2012){Ugliano}, {Janka}, {Marek}, \&
  {Arcones}}]{Ugliano_2012ApJ...757...69}
{Ugliano}, M., {Janka}, H.-T., {Marek}, A., \& {Arcones}, A. 2012, \apj, 757,
  69

\bibitem[{{Venn} {et~al.}(2002){Venn}, {Brooks}, {Lambert}, {Lemke}, {Langer},
  {Lennon}, \& {Keenan}}]{Venn2002}
{Venn}, K.~A., {Brooks}, A.~M., {Lambert}, D.~L., {et~al.} 2002, \apj, 565, 571

\bibitem[{Wheeler {et~al.}(2015)Wheeler, Kagan, \& Chatzopoulos}]{Wheeler_2015}
Wheeler, J.~C., Kagan, D., \& Chatzopoulos, E. 2015, The Astrophysical Journal,
  799, 85

\bibitem[{{Woosley} \& {Heger}(2006)}]{Woosley_2006ApJ...637..914}
{Woosley}, S.~E. \& {Heger}, A. 2006, \apj, 637, 914

\bibitem[{{Yoon} {et~al.}(2012){Yoon}, {Dierks}, \& {Langer}}]{Yoon2012}
{Yoon}, S.~C., {Dierks}, A., \& {Langer}, N. 2012, \aap, 542, A113

\bibitem[{{Yoon} {et~al.}(2006){Yoon}, {Langer}, \&
  {Norman}}]{Yoon_2006A&A...460..199}
{Yoon}, S.~C., {Langer}, N., \& {Norman}, C. 2006, \aap, 460, 199

\bibitem[{{Zahn}(1992{\natexlab{a}})}]{JPZahn1992}
{Zahn}, J.~P. 1992{\natexlab{a}}, \aap, 265, 115

\bibitem[{{Zahn}(1992{\natexlab{b}})}]{Zahn1992}
{Zahn}, J.~P. 1992{\natexlab{b}}, \aap, 265, 115

\bibitem[{{Zahn} {et~al.}(2007){Zahn}, {Brun}, \&
  {Mathis}}]{Zahn_2007A&A...474..145}
{Zahn}, J.~P., {Brun}, A.~S., \& {Mathis}, S. 2007, \aap, 474, 145

\bibitem[{Zeipel(1924)}]{Zeipel1924}
Zeipel, H.~v. 1924, Zum Strahlungsgleichgewicht der Sterne, ed. H.~Kienle
  (Berlin, Heidelberg: Springer Berlin Heidelberg), 144--152

\end{thebibliography}

\end{document}